\documentclass[a4paper,twoside,12pt,titlepage]{article}

\usepackage[cm]{fullpage}
\usepackage[section]{placeins}	
\usepackage{amsmath,bm,mathrsfs,yfonts,commath}
\numberwithin{equation}{section}
\numberwithin{figure}{section}
\usepackage{amssymb,wasysym}
\usepackage{setspace} \allowdisplaybreaks
\usepackage{cite}
\usepackage[hidelinks]{hyperref}
\usepackage[noabbrev]{cleveref}

\usepackage{tikz,tikz-3dplot}
\usetikzlibrary{patterns,calc,3d}
\usepackage{subcaption,wrapfig}

\let\oldsection\section
\def\section{\cleardoublepage\oldsection}

\begin{document}

\pagenumbering{gobble}
\thispagestyle{empty}
\begin{titlepage}
\begin{center}
\vspace*{\fill}

\includegraphics[width=0.4\textwidth]{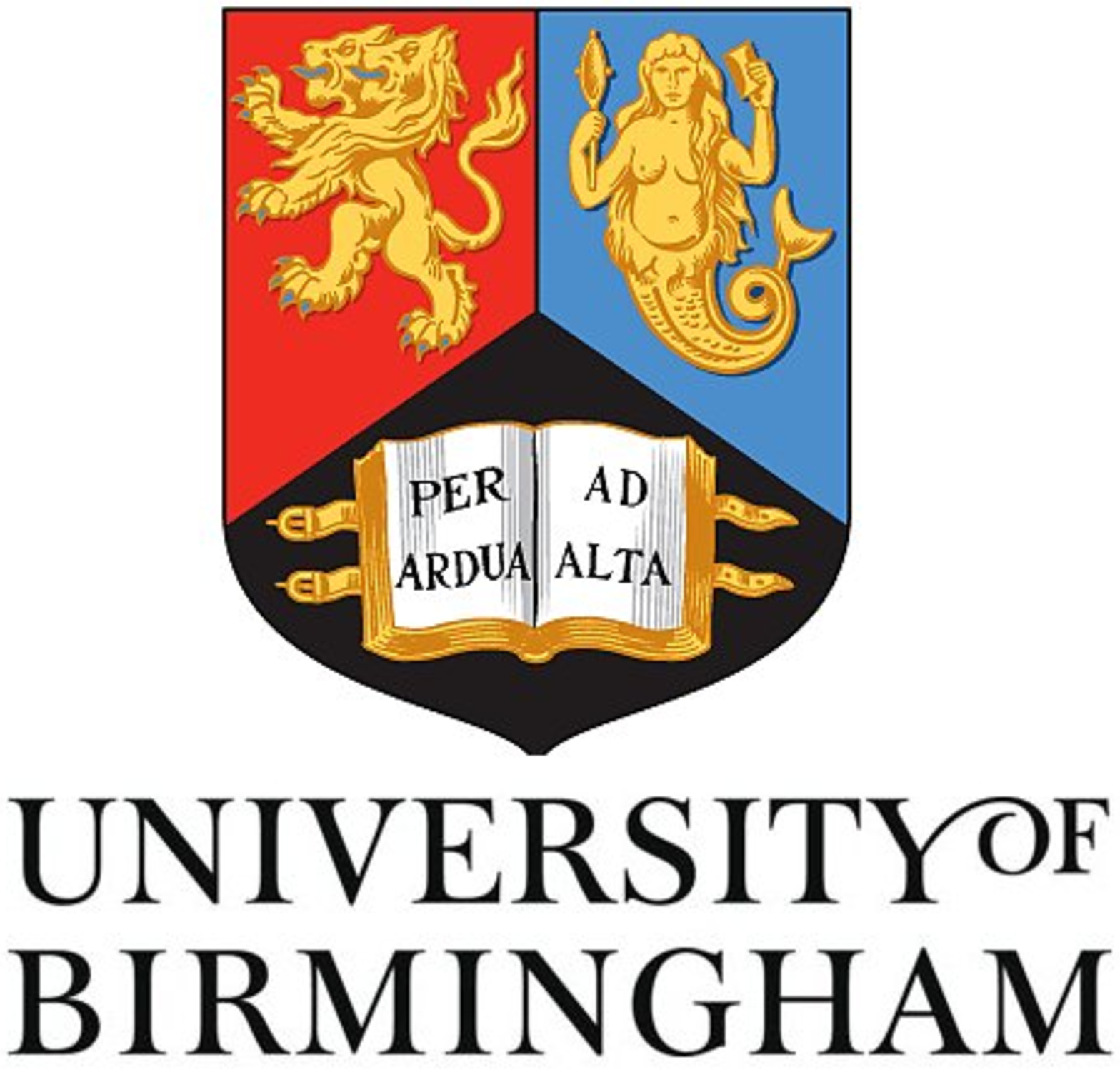}

\vspace{1cm}

\Large
\textsc{School of Mathematics}

\vspace{2cm}

\textbf{Wetting Fronts in Porous Media}

\vspace{1cm}

\large
An MSci Research Project

\vspace{3cm}

\begin{minipage}{0.8\textwidth}
\begin{minipage}[t]{0.49\textwidth}
\begin{flushleft} \large
\emph{Author:}\\
Edward W.G. Skevington\\
1071929
\end{flushleft}
\end{minipage}
\hfill
\begin{minipage}[t]{0.49\textwidth}
\begin{flushright} \large
\emph{Supervisor:} \\
Prof. Y.D. Shikhmurzaev
\end{flushright}
\end{minipage}
\end{minipage}

\vspace{3cm}

May 29, 2014

\vspace*{\fill}
\end{center}
\end{titlepage}

\normalsize

\clearpage

\begin{abstract}
The dynamics of the wetting front are considered during the imbibition of a fluid into a porous substrate through a circular drawing area. A mathematical model of this process, assuming incompressible Darcy flow, is presented, before the full finite element scheme for solving this set of equations is given allowing the reader to reproduce all presented results. Asymptotic analysis is performed revealing contradictions between the assumptions of Darcy's equation and the solutions it produces, along with qualitative results for the behaviour of the wetting front and macroscopic contact angles. Velocity and pressure distributions across the wetted region are presented, as well as plots of the evolution of the wetting front and parameters with discussion.
\end{abstract}

\pagenumbering{gobble}
\thispagestyle{empty}
\tableofcontents
\thispagestyle{empty}

\cleardoublepage
\pagenumbering{arabic}
\section{Introduction}

\begin{figure}[ptb]
	\centering
	\begin{tikzpicture}
	\draw (-5,0) -- (5,0);
	\fill[pattern=dots,pattern color=gray] (-5,0) rectangle (5,-4);
	
	\draw (-2,0) -- (-2,4) (2,0) -- (2,4);
	\fill[pattern=north east lines] (-2.5,0) rectangle (-2,4);
	\fill[pattern=north east lines] (2,0) rectangle (2.5,4);
	
	\begin{scope}
		\clip (-5,0) rectangle (5,-4);
		\draw (0,0) ellipse (4.5 and 3);
	\end{scope}
	
	\draw (0,2.5) node {External Reservoir};
	\draw[->] (0,0.5) node {Drawing Area} +(0,-0.2) -- (0,0);
	\draw (-4.5,2.5) node {Atmosphere};
	\draw (0,-1.5) node {Wetted Region};
	\draw (-3.5,-3.1) node {Dry region};
	\draw[->] (3.5,-3.1) node {Wetting front} +(0,0.3) -- (-35:3.8);
	
	\fill (2,0) circle (0.05);
	\fill (-2,0) circle (0.05);
	
	\fill (4.5,0) circle (0.05);
	\fill (-4.5,0) circle (0.05);
\end{tikzpicture}
	\caption{Illustration of the imbibition of a wetting fluid, labelling the various regions. The external reservoir depicted is a vertical column of fluid supported by a solid cylinder.}
	\label{f:intro_terms}
\end{figure}
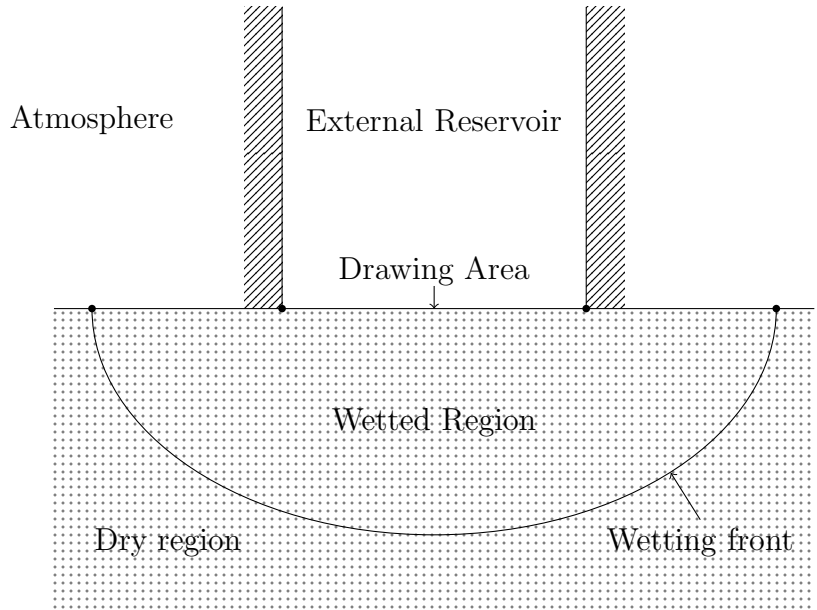

The flow of fluids through porous media is present in a vast variety of natural phenomena and industrial applications. Some examples are oil recovery, carbon-dioxide sequestration, hydro-geology, fuel cells, ink-jet and 3D printing, and the creation of ceramics. Porous media are materials such as sandstone, paper or packed beads, which have small voids in their bulk, called \textit{pores}, connected together to form a network of thin passageways on a microscopic scale. The connectivity of the pores allows fluids to flow through them. When more than one fluid occupies the porous medium there will be pores in which the two fluids meet, causing an interfacial surface to form where surface tension will act. The fluids on either side of this surface may be part of a large \textit{bulk} which occupies the pore space on a length scale much larger than that of the pores, such as an aquifer or oil reservoir, or be in the form of \textit{ganglia} only occupying a few pores at most. When the fluid is flowing rapidly into a porous medium, or \textit{wetting} it, a sharp interface may form on the macroscopic scale between the bulk phases of the wetting and displaced fluids called a \textit{wetting front}. Whether or not a clear wetting front is formed depends on the characteristics of the two fluids and the porous solid. If the porous medium is initially saturated with and surrounded by one fluid, and is then brought into contact with another fluid which then wets it, this is called \textit{imbibition}. The body of fluid that has been introduced shall be called the \textit{external reservoir}, the area of contact between this and the porous medium the \textit{drawing area}, the resulting bulk phase of the wetting fluid the \textit{wetted region}, and the bulk phase of the displaced fluid the \textit{dry region}. The terminology that we employ is illustrated in figure \ref{f:intro_terms}.

The most important parameter characterising the porous medium itself is the \textit{porosity}, which is the volume fraction of the material that is pore space. That is, if we consider a volume $V$ within the porous medium, then $V_1$ of this total volume will be made up of the pore voids and $V_2$ of the solid matrix itself, such that $V=V_1+V_2$. The porosity is $V_1/V$. This is what we mean by a volume fraction, the terms length fraction and area fraction shall also be used in this work.

The field of flows in porous media has been under investigation for many years now and progress has been made in the mathematical description and conceptual understanding of all the topics above. However, a full theoretical model is still in wanting.

We aim to investigate mathematical models of the wetting front. We shall do this by theoretically studying the imbibition of a liquid through a horizontal surface into a porous substrate. The external reservoir may either be a column of liquid or a droplet. This will produce theoretical predictions which can then be tested empirically. In the present work explicit modelling of the fluid exterior to the porous medium shall not be undertaken, instead we will model and begin to study the bulk region of the imbibed fluid, the subject of interest being the propagation of the wetting front into the porous medium as time progresses. In addition we shall only investigate only a very simple model of the wetting process, but put forward a scheme that can be easily enhanced to investigate much more complicated models.

In the literature review that follows we will first overview in broad terms the approaches to modelling fluid flows in porous media, followed by a closer look at the continuum models.  Another example of imbibition shall then be considered, where boundary conditions for the wetting front have already been proposed and tested, of particular interest is that of Shikhmurzaev and Sprittles'. Finally, we shall examine the progress made into imbibition through a horizontal surface, especially that of droplets since much progress has been made in this area.

\subsection{Approaches to Modelling}

 The main problem in this area is to change the scale of the description of the flow from that of the pore to that of the macroscopic domain, which may be, for example, an oil field or a piece of paper. On the pore scale the standard equations for macroscopic fluids (such as the Navier-Stokes equations) are valid, and the domain of the flow is the pores. Performing an order of magnitude estimate, the length scale of a pore may be $\sim 10^{-5} \mathrm{m}$ \cite{natural_rock}, having a volume $\sim 10^{-15} \mathrm{m}^3$. A rain drop has a length scale of $\sim 10^{-2} \mathrm{m}$, thus if a rain drop imbibes into a porous medium it will pass into $\sim 10^{9}$ pores, the precise dynamics of the flow being required in every one. It is not only impractical to attempt to calculate the solution in such a domain, but also unwise to require detailed knowledge of the pore structure in the sample, which would render impossible the modelling of flows without sophisticated apertures to scan the sample first. Therefore other methods have to be devised.

\begin{figure}[ptb]
	\centering
	\includegraphics[width=0.35\textwidth]{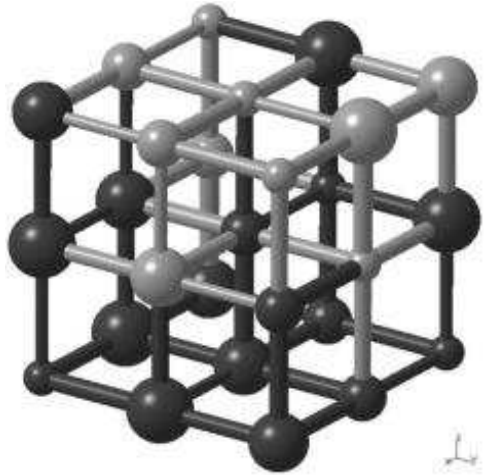}
	\caption{Capillary network from \cite{spread_sessile_numerics}, the pores and throats connecting them are arranged in a regular rectangular grid, the diameter of both being randomly generated}
	\label{f:intro_capil_net}
\end{figure}

Adler and Brenner \cite{review_adler} review various methodologies still present in the field. The more recent review by Alava \textit{et al.} in 2004 \cite{disordered_alava} discusses many of the more modern (and advanced) forms of these methods, which broadly speaking can be classified into two types. 

Firstly there are continuum descriptions. Here we consider the case where the pores are on a much smaller scale than the bulk region of fluid, and the time scales characterising the flow in the pores is much shorter than that of the macroscopic flow we are investigating. Thus we can model the flow using averaged quantities on intermediate scales. These approaches have the advantage that they provide a macroscopic description of a macroscopic phenomena. This is what is ultimately desired from any model; even of we could use the Navier-Stokes equation to describe the flow in every pore, the desired results would be the concentrations of fluids in different regions, their flux and their averaged stress or pressure. If these can be calculated directly then this is to great advantage analytically and intuitively.

Secondly there are the lattice or particle models. These typically operate by considering the porous network to be regular in some sense. For example the capillary network in figure \ref{f:intro_capil_net}, where a rectangular grid of spherical pores with throats connecting them is used to represent the porous network. The flow is then modelled using some algorithm dictating which fluid each pore, throat or other small region is occupied by. The algorithm is deduced from assumptions about the behaviour at each modelling point to approximate when one fluid will displace the other. Of course the porous network in a rock will not resemble the figure, it will be much more disordered, and one fluid does not suddenly displace another, it takes time if only a very small amount. We see that these models operate in the same regime as the continuum models, requiring the separation of scales.

Contrasting the two approaches, continuum models have the advantage of giving direct access to the macroscopic parameters that will ultimately be of interest, and are analytically tractable to provide asymptotic information in limiting cases. Another consideration is topology, since the pores are modelled directly in the lattice model, and their orientation cannot be guaranteed to be (and often isn't intended to be) the orientation of the true pores, a huge number of pores must be used in the model to hide the inaccuracies produced, and more than can be feasibly simulated. Continuum models do not face this obstacle. The lattice models must be proven to have some advantage over a continuum model, which can only be that they have unsurpassed accuracy and precision when describing a range of phenomena. This has not been achieved so far. In what follows an overview of continuum models is presented.

\subsection{Continuum Mechanical Models}

The assumptions involved in continuum mechanics  shall now be stated more formally. In general, continuum mechanics assumes a separation of the length and time scales between the macroscopic behaviour of interest and the microscopic processes that drive it. Thus the macroscopic behaviour can be modelled using spatio-temporally averaged quantities on intermediate scales (which are almost always the quantities of interest). The equations used can be thought of as the dominant terms in the asymptotic expansion as the ratio of microscopic to macroscopic scales tends to zero. Within an individual pore a primary continuum limit\footnote{In this limit, the microscopic behaviour is that of atoms and molecules, the macroscopic behaviour is that of the fluid flow in a single pore.}
is used to model the fluid, yielding such equations as the Navier-Stokes equation. The flow in a pore and the flow of the bulk regions of fluid are assumed to be on scales separated by orders of magnitude, thus we can model the macroscopic flow using a secondary continuum limit, which shall be used unless otherwise stated. The scales characterising the macroscopic region shall hereafter be referred to as Darcy scales.

Under this secondary continuum limit, the porosity can be viewed as the continuum average of a function that takes the value 1 in the pores and 0 in the solid matrix. Using this definition porosity is clearly, in general, a function of position, and if the porous medium is homogeneous the the porosity is a constant.

When developing continuum models the behaviour under the primary continuum limit is sometimes required, and the behaviour under the second is calculated as a result. However, we do not wish to consider a specific porous network, and instead choose to represent it using cylindrical pores. The flow in these representative pores is assumed to approximate well the flow that occurs in the real pores once the secondary continuum limit is applied. The representative pores have an effective pore radius which is not only a function of position but also of the direction of the pore, and in isotropic and homogeneous porous media becomes a constant. Calculating the effective pore radius that will best describe a particular material is subtle, a method for doing so is presented in \cite{cap_rise_powders} and tested in \cite{wetting_quartz}.

In our study we will require equations that describe the macroscopic flow of the fluid through the wetted region. Examples of these equations will now be discussed and an appropriate equation chosen.

The simplest continuum description was discovered empirically by Darcy in 1856, and is explained in \cite{darcy_intro,natural_rock}. It has been well tested and is used extensively in engineering applications. That is not to say that it is the best equation, but it certainly is adequate for most situations. If gravity is the only applied body force then Darcy's equation is, denoting the velocity $\boldsymbol{u}$ and pressure $p$,
\begin{equation*}
	\boldsymbol{u}=-\frac{k}{\mu}(\nabla p -\rho \boldsymbol{g}),
\end{equation*}
where $\mu$ and $\rho$ are the viscosity and density of the fluid, $k$ the permeability and $\boldsymbol{g}$ the free fall acceleration due to gravity. The permeability characterises the resistance of the porous medium to the motion of the fluid. Interpreting this equation, the fluid only experiences forces due to the pressure gradient and body force, convection and viscous diffusion having negligible effect. Also, since the acceleration occurs on a time scale much shorter than that of the macroscopic flow, it is the velocity that responds to these forces (in the continuum limit). Darcy's equation applies to the flow in a region saturated with one fluid phase. To apply as-is to imbibition, the wetting fronts between the phases must be surfaces and there must be no ganglia. We will discuss shortly the ways in which Darcy's equation is modified to model more complicated flow scenarios.

Darcy's equation can be derived by explicitly volume averaging the equations of motion within the individual pores, as in \cite{darcy_whitaker}. The assumptions that must be made in this derivation give insight into the equations conditions of validity. The most important conditions are that the pore size is much smaller than the domain of the flow  and that the macroscopic acceleration of the fluid is small (as should be expected). The paper then goes on to derive alternative equations which include some correction terms for small effects. The equations developed are the Navier-Stokes equation with perturbing terms, and not Darcy's equation with corrections, since the mathematical technique applies the correction of including the porous matrix to the free flow. An equation produced in this manner may well be valid for particle suspension phenomena, since there the flow is indeed perturbed by the presence of solid particles. However, it has not been shown that any equation derived in this manner is more accurate than Darcy's, nor that they give any advantages for describing flows in porous media where the effects of the solid matrix dominate.

Other equations have been produced that are corrections to Darcy's equation. One of these is Brinkman's equation, which includes a correction for long range viscous effects. This equation has often been justified (see \cite{bc_haber}) by the claim that it allows for the Beavers and Joseph boundary condition \cite{bc_beavers} and the experimental results that accompany it in the paper. This boundary condition states that, at the edge of the porous medium where the fluid transitions into free flow, the components of velocity tangential to the boundary change rapidly in the direction normal to the boundary. However, as demonstrated in \cite{about_bj}, the condition itself does not show the separation of scales required for a valid continuum mechanical model, nor is their experimental data of true porous flow and free flow, but rather the `free flow' is in a region of a similar scale to the pores. This does not invalidate Brinkman's equation, but does show that we have no reason to believe in its validity. Many more examples of corrections do exist (the other classic example is the Forchheimer equation \cite{forcheimer_derivation}), but it is not clear if any of them are valid and in what regime, and they all reduce to Darcy's equation in the continuum limit.

A more complete description would include the modelling of ganglia, as well as intertwined percolating bulk phases. In a continuum model with mixed phases we must introduce saturations of the different fluids as functions of position and time, as described in \cite[Ch. 5]{natural_rock}. Of course this makes the modelling of the interactions between fluids much more difficult, since we do not know the size and extent of each region of fluid, nor the geometry of the surfaces that separate them. Typically the interaction is modelled via a constitutive equation specifying a pressure difference between the phases, which will likely be a function of the saturation. If Darcy's equation is used for each fluid phase then the permeability may be altered by a factor known as the relative permeability, which will also be a function of the saturation. In some formulations even terms involving the direct effect of the pressure in other phases are included into a modified Darcy's equation.

Hilfer has attempted to create a very general model of multiphase fluid flows. In his recent paper \cite{hilfer_theory} divulging all theoretical development he starts with general statements of mass and momentum conservation. He also models the bulk phases and ganglia as different phases, such that each possesses its own saturation and can be modelled using its own constitutive equations. These constitutive equations are then proposed characterising the behaviour of one of the fluid phases, or the interaction between two fluid phases, or between a fluid and the solid matrix. However, the constitutive equations proposed are of forms that are unjustified and have so many free parameters that the resulting model is simply unusable in its most general form. This is well demonstrated by what happens when he applies sufficient restrictions are applied to the model to produce Darcy flow in the two bulk phases. The pressure difference between them is a function of one variable with \textit{ten} arbitrary parameters. It is no wonder that the model fits well to a small number of empirical curves, it would be a surprise if it didn't. The model is also simulated numerically in \cite{hilfer_numerics} in a one dimensional situation, however no empirical evidence is provided. For this model to be validated, it needs to be shown that it can predict experimental results in a manner that is not indicative of its vast number of free parameters, but that the parameters are constants for the materials in the system.

In this study we do not intend to include the effects of ganglia in our model. Of the models that do not include these effects, Darcy's equation is the only one that has been extensively verified. All others that have been developed have not been been sufficiently well tested or have been shown to be inaccurate. Since we do not intend to test bulk equations, Darcy's equation will be used.

\subsection{Capillary Rise in a Porous Column}

The mathematical modelling of the interfaces between different fluid phases is a difficult topic in its own right, and thus a simple situation is required in which it can be studied. This can be achieved by considering a vertical column of a porous material initially saturated with one fluid. The base of this column is then immersed in an external reservoir that imbibes into it, rising up against gravity. This process is known as \textit{capillary rise}, and is a simplification since the wetting front will be approximately horizontal and propagating in the vertical direction which makes it reasonable to model it as a one-dimensional phenomenon. The behaviour of interest is that of the menisci at the wetting front as the fluid propagates, and the boundary conditions required to describe it. Of these we are especially interested in that of Shikhmurzaev and Sprittles, which has recently been shown to accurately describe this phenomenon. First we shall briefly discuss the relevant bulk equations and then move onto the boundary conditions.

The equation that is used to describe the bulk flow may be Darcy's, but often Washburn's equation \cite{washburn_original} is used. The flow along a long thin tube or \textit{capillary} of constant circular cross section, that in general may be curved, is assumed to follow Poiseuille's law for locally unidirectional flow. The only coordinate for this one dimensional flow is the distance along the tube, and the only variables of interest are the velocity and pressure averaged over the cross-section. The velocity in Poiseuille flow is a function of the distance from the centre of the tube and time, therefore the cross-sectionally averaged velocity will only be a function of time. The porous medium is modelled as a bundle of these capillaries, aligned in the vertical direction. The assumption of unidirectional flow is invalidated at the inlet, leading to the development of corrections to this equation such as \cite{washburn_applicability}. Another improvement that has been made is the inclusion of pore doublets \cite{extended_washburn}. These improvements are of little interest here, since  Washburn's equation, or preferably Darcy's equation since this is what is used in a general flow in a porous material, are sufficient to examine boundary conditions that may be applied at the wetting front.

The simplest assumption that may be made about the menisci in the pores (or capillaries) on the wetting front is that they form spherical caps that, at the edge of the capillary, subtend a prescribed constant angle to the solid boundary known as the \textit{contact angle}. Across each meniscus surface tension acts, causing a bulk pressure difference between the imbibing and displaced fluids. If the contact angle is less than $\pi/2$ then the pressure in the imbibing fluid is less than that of the of the displaced fluid. This decrease in pressure will cause a pressure gradient in the imbibing fluid, since the pressure at the base of the porous column will be less than that at the wetting front, and if the force of the pressure gradient is greater than the force of gravity then the fluid will be driven upwards.

\begin{figure}[ptb]
	\centering
	\includegraphics[width=0.6\textwidth]{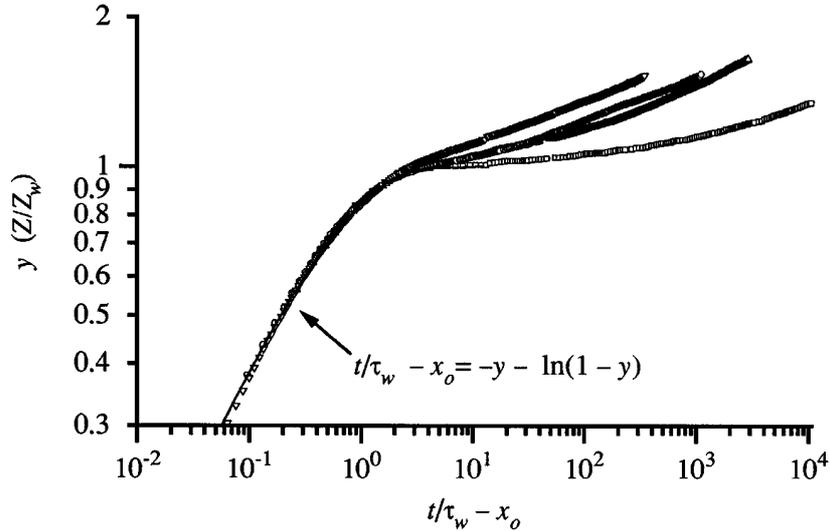}
	\caption{Four sets of empirical data for capillary rise from \cite{interface_pinning}, the x-axis showing our $t$ and y-axis our $h$. For early times the solution to Darcy's equation for constant contact angle in the pores of the wetting front fits well. However, for later times the flow rate reduces dramatically below what is predicted. The experimental data is for packed beads of diameter 180$\mu$m ($\triangledown$), 253$\mu$m ($\fullmoon$), 359$\mu$m ($\vartriangle$), and 510$\mu$m ($\square$).}
	\label{f:intro_capil_rise_delker}
\end{figure}

Delker \textit{et al.} \cite{interface_pinning} model the vertical porous material using Darcy's equation and the assumption of a constant contact angle. They show analytically that $h(t)-h_0 \propto e^{t/\tau}$, where $h(t)$ is the current height, $h_0$ is the equilibrium height and $\tau$ is the characteristic time scale for the imbibition. They then go on to present experimental data that is included here in figure \ref{f:intro_capil_rise_delker}, along with a plot of the analytic solution. It is observed that the analytic solution fits well for small times, but that for large times the flow is much slower.

\begin{figure}[ptb]
	\centering
	\includegraphics[width=0.6\textwidth]{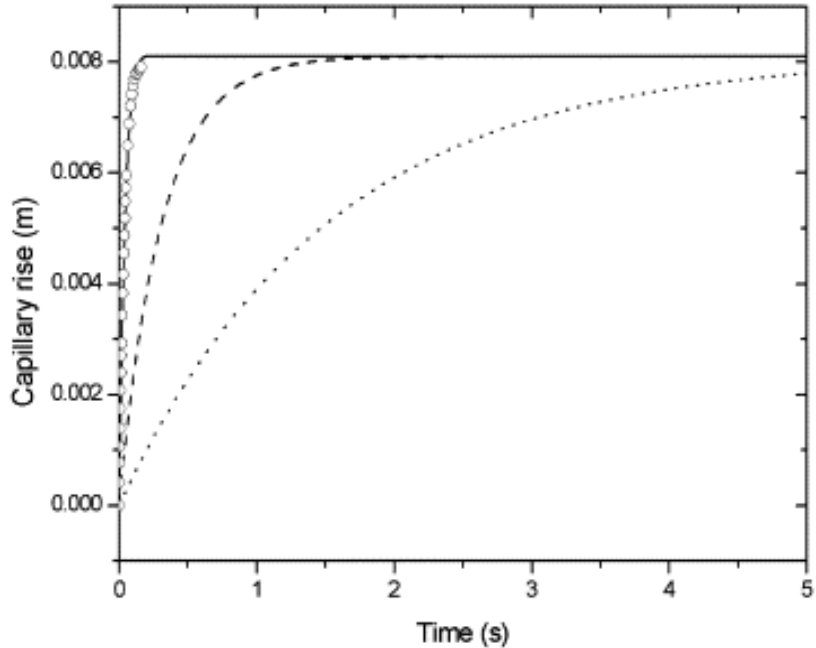}
	\caption{Numerical simulations of capillary rise from \cite{cap_rise_martic}, the x-axis showing our $t$ and y-axis our $h$. The graph demonstrates that by increasing the variation of the dynamic contact angle the equilibrium state takes longer to reach. The white circles represent empirical data from \cite{cap_rise_quere}.}
	\label{f:intro_capil_rise_martic}
\end{figure}

A possible solution to this problem is to allow the contact angle to vary as a \textit{dynamic contact angle}. In any propagation of a fluid, the contact angle is a functional\footnote{A functional is a mapping from a function to a number, this is usually an integral of the function. In this case it would likely be an integral involving the velocity field and some weight function.} 
of the local velocity field \cite[\S 3.2.3.3]{shk_capillary_flows}. Since capillary rise is modelled in one dimension, all of the local velocities are characterised by a single scalar velocity which is equal to the velocity of the meniscus itself. Therefore, we assume that there is an equation that relates the velocity of the meniscus and the contact angle, preferably such that one is a function of the other. Martic \textit{et al.} \cite{cap_rise_martic} used Washburn's equation to model capillary rise.  At the wetting front the meniscus velocity was restricted to be a monotonically increasing function of contact angle for the range of contact angles involved in the process, with a parameter to govern the magnitude of contact angle variation. A larger contact angle will lead to a flatter meniscus and lower pressure difference across it, thus a lower velocity, which is what is shown by their simulations in figure \ref{f:intro_capil_rise_martic}. To describe the results in figure \ref{f:intro_capil_rise_delker}, we could employ a model of contact angle variation that is almost constant for the range of velocities encountered at early times, and smoothly increases for the lower velocities encountered near the end.

\begin{figure}[ptb]
	\centering
	\includegraphics[width=0.7\textwidth]{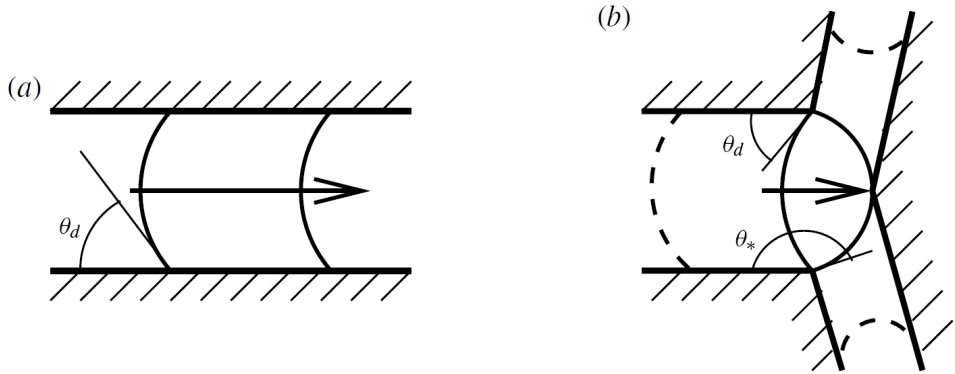}
	\caption{Illustration of the two modes proposed in \cite{wetting_dynamics_shk}, the wetting mode (a) and the threshold mode (b).}
	\label{f:intro_modes}
\end{figure}

The model developed by Shikhmurzaev and Sprittles in \cite{wetting_dynamics_shk} slows the advancement using a different method, involving two distinct modes as illustrated in figure \ref{f:intro_modes}. These modes are modelled in a representative cylindrical pore that (in an isotropic medium) is perpendicular to the wetting front, and itself modelled in the one-dimensional manner using velocities and pressures averaged over the cross-section. In mode 1 the meniscus is advancing along the pore freely, as illustrated by figure \ref{f:intro_modes}a, its free surface forming dynamic contact angle $\theta_d$ with the pore wall. In mode 2 the contact line is pinned until the contact angle reaches $\theta_*$, as illustrated by figure \ref{f:intro_modes}b. The length fraction along the pore traversed in mode $i$ is $s_i$. If $\theta_d\geq \theta_*$ then pinning does not occur and $s_1=1$, otherwise it takes the value $s_1=s_{10}$ where $s_{10}$ is the representative length fraction over which pinning cannot occur. From these length fractions and the velocity of the meniscus in each of the modes, the area fraction of the wetting front in mode $i$ is calculated. The pressure and normal velocity of the wetting front are equal to the mean weighted by area fraction of the values of the representative menisci.

The pressure in mode 1 is calculated relative to the pressure in the displaced fluid using the surface tension across the spherical cap, as usual. The proposed function for the dynamic contact angle is that from the theory of capillary flows with forming interfaces \cite{shk_capillary_flows}. Thus, in mode 1, the condition is a non-linear relationship between pressure and normal velocity. In mode 2, the stagnation pressure is defined as the pressure that builds up on the meniscus when it is prevented from deforming. This is then used to derive the pressure and velocity at the meniscus as it deforms, averaged over time. The resulting boundary condition is a non-linear relationship between the normal velocity of the wetting front, the pressure and the stagnation pressure.

\begin{figure}[ptb]
	\centering
	\includegraphics[width=0.6\textwidth]{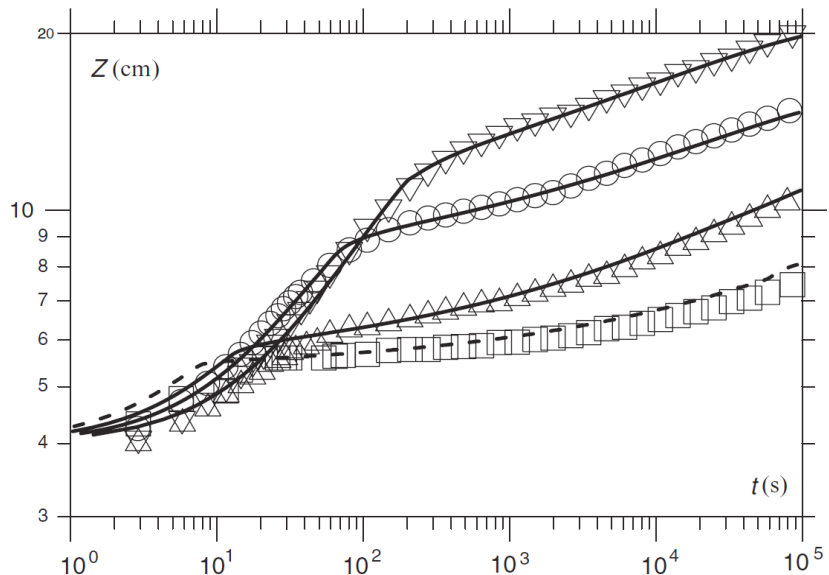}
	\caption{Solid and dashed lines are numerical simulations of capillary rise from \cite{cap_rise_shk}, with the empirical data from \cite{interface_pinning} that is also plotted in figure \ref{f:intro_capil_rise_delker}, the x-axis showing our $t$ and y-axis our $h$. The experimental data is for packed beads of diameter 180$\mu$m ($\triangledown$), 253$\mu$m ($\fullmoon$), 359$\mu$m ($\vartriangle$), and 510$\mu$m ($\square$).}
	\label{f:intro_modes_rise}
\end{figure}

Numerical simulations were performed to compare the results of Shikhmurzaev and Sprittles' model with the empirical results of Delker \textit{et al.}, and are included in figure \ref{f:intro_modes_rise}. Qualitatively, the plots show the same behaviour. However, there does seem to be some discrepancy in the results, especially for the beads with a diameter of 510$\mu$m. Denoting the diameter of the beads as $d$ and the distance moved in the vertical direction as $h$, continuum mechanics is valid in the limit $d/h \rightarrow 0$, and averaged quantities being defined on a scale $\sqrt{d/h}\,h$. For the largest beads the separation of scales is $\sim 1/6$, which is nowhere near zero as required. For the smallest beads the separation is $\sim1/30$, which is acceptable. The most likely explanation for the increase in accuracy as the bead diameter decreases is that the experiments were not sufficiently well within the continuum regime.

Now that a theoretical model has been shown to describe otherwise unexplained phenomena in a simple situation, its effects should be investigated in a more complicated environment. Our aim is to start an investigation into modelling the phenomena discussed below.

\subsection{Imbibition into a Porous Substrate}

An important topic of research is the dynamics of imbibition when we cannot model the phenomenon as one dimensional. These flows reveal more complicated behaviours across the wetted region and wetting front, as we discover in our study. We consider a fluid imbibing into the flat horizontal top of a porous substrate from a reservoir of fluid that has been placed on it. This is a three-dimensional process, or in the axisymmetric case where the drawing area is circular, two-dimensional. In Shikhmurzaev and Sprittles' \cite{wetting_dynamics_shk} model the multi-dimensional wetting front allows different regions of the wetting front to have different area fractions in each mode. The pressure of the fluid in the external reservoir is of little importance, since it is insignificant in comparison to the Darcy pressure \cite{dynam_angle_shk}, thus the wetted region draws in any fluid it requires through this drawing area with no resistance from the reservoir. Therefore, the only parameter from the reservoir that affects imbibition is the radius of the drawing area. If the reservoir is a cylindrical column of fluid then this radius will be constant (or possibly a known function of time), if it is a droplet then it may be a constant, a function of time or a function of the volume of imbibed fluid for simple cases.

The phenomenon that we will be considering is imbibition through a circular drawing area of constant radius, whilst the main topic of research in this area is the imbibition of liquid droplets into porous substrates. This phenomenon is the most common subject for multi-dimensional imbibition processes. Despite our research not being on this subject specifically since we will not be modelling the droplet, the area of research is important due to its presence in the literature and its applications in ink-jet printing, 3D printing and the manufacture of ceramics. It is relevant since, in the simplest case, the drawing area of the droplet is constant. In addition, our model of the wetted region could easily be expanded to use a simple model of the droplet to vary the radius of the drawing area. The remainder of this subsection shall be devoted to analytical, experimental and numerical progress in this area.

\begin{figure}[ptb]
	\centering
	\begin{tikzpicture}
	\draw (-5,0) -- (5,0);
	\fill[pattern=dots,pattern color=gray] (-5,0) rectangle (5,-4);
	
	\begin{scope}
		\clip (-5,0) rectangle (5,4);
		\draw (0,0) ellipse (3 and 2.6);
	\end{scope}
	
	\begin{scope}
		\clip (-5,0) rectangle (5,-4);
		\draw (0,0) ellipse (4.5 and 3);
	\end{scope}
	
	\draw (0,1) node {Droplet};
	\draw (-3.5,2.5) node {Atmosphere};
	\draw (0,-1.5) node {Wetted Region};
	\draw (-3.5,-3.1) node {Dry region};
	\draw[->] (3.5,-3.1) node {Wetting front} +(0,0.3) -- (-35:3.8);
	
	\draw (3,0.5) arc (90:180:0.5);
	\draw[dashed] (3,0) -- (3,0.7);
	\draw (2.2,0.5) node {CA1};
	
	\draw (4.5,-0.5) arc (-90:-180:0.5);
	\draw[dashed] (4.5,0) -- (4.5,-0.7);
	\draw (3.8,-0.5) node {CA2};
	
	\fill (3,0) circle (0.05);
	\fill (-3,0) circle (0.05);
	\draw (2.9,-0.3) node {CL1};
	
	\fill (4.5,0) circle (0.05);
	\fill (-4.5,0) circle (0.05);
	\draw (4.5,0.3) node {CL2};
\end{tikzpicture}
	\caption{Illustration of droplet imbibition, labelling the contact lines and contact angles.}
	\label{f:intro_CA_CL}
\end{figure}
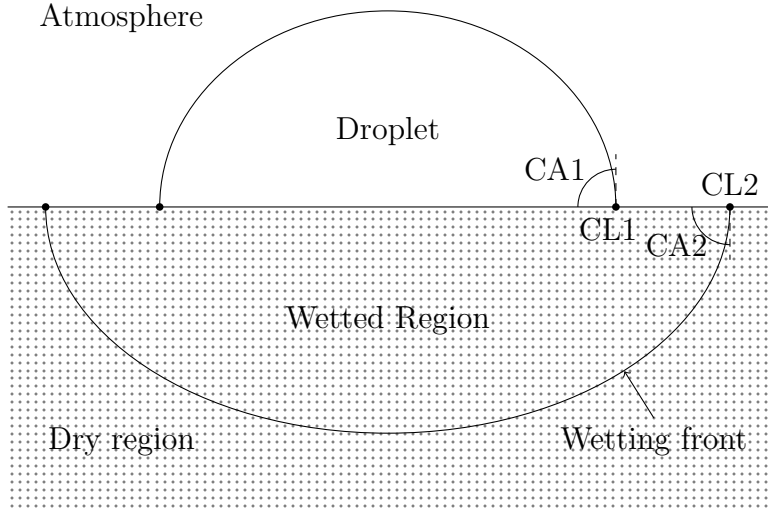

It is helpful to define two contact lines, which are lines at which three different materials meet, and contact angles, which are the angles subtended through one of the materials at the contact line. The contact lines and angles discussed are labelled in figure \ref{f:intro_CA_CL}. Let CL1 be the contact line between the droplet, the wetted region and whatever `atmosphere' the droplet is surrounded by. Let CL2 be the contact line at which the wetting front and solid surface meet. Let CA1 be the contact angle subtended by the droplet at CL1, and CA2 be the angle subtended by the wetted region at CL2. This terminology shall also be used for a column of fluid. Of course CL1 and CL2 could meet at the same line, as is investigated by Shikhmurzaev in \cite{dynam_angle_shk} for droplet imbibition. He also shows that, as CA1 and CA2 tend to $\pi/2$, the contact lines split with CL2 advancing ahead.

\begin{figure}[ptb]
	\centering
	\includegraphics[width=0.6\textwidth]{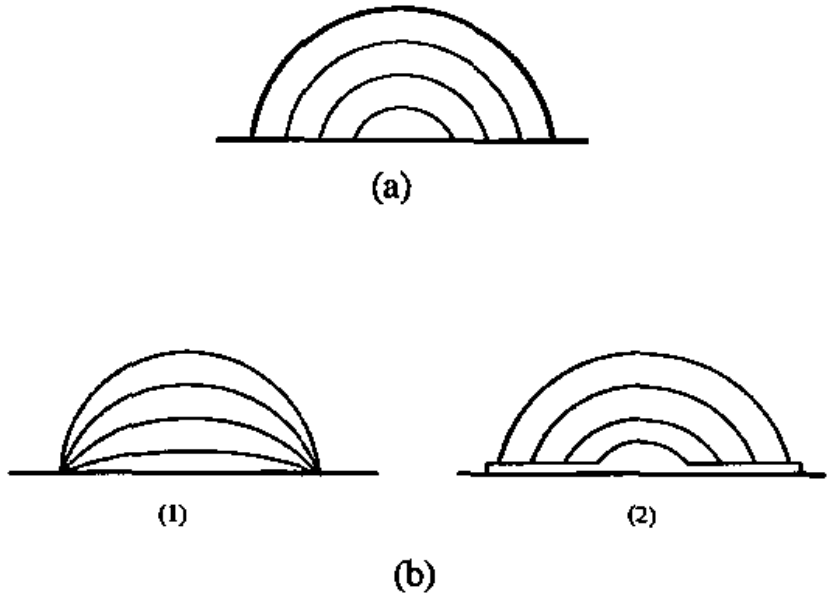}
	\caption{Illustration of the different cases of droplet depletion from \cite{penetration_denesuk}, described in the text.}
	\label{f:intro_denesuk_cases}
\end{figure}

Denesuk \textit{et al.} \cite{dynamics_denesuk} define three regimes of behaviour for the spread of a liquid droplet over a porous solid. Let the time scale of spreading be $\tau_s$ and the time scale of imbibition (or, as it is called in their paper, depletion) be $\tau_d$. If $\tau_d \gg \tau_s$ then the droplet will spread out in a similar manner to spreading over a non-porous substrate, before slowly imbibing in a semi-static manner. If $\tau_d \ll \tau_s$ then the fluid will imbibe into the solid before any significant spreading can occur. If $\tau_d \approx \tau_s$ then the droplet will imbibe whilst the fluid spreads, but the imbibition itself is only affected by the radius of the drawing area, therefore imbibition controls (in part) the dynamics of spread. In our investigation, since we shall not be modelling the droplet, we will only be able to consider cases where the droplet moves in a semi-static manner. That is for $\tau_d \gg \tau_s$, and possibly late times for $\tau_d \approx \tau_s$, once the droplet has already spread out and the behaviour of the droplet is driven by imbibition in such a manner that inertial effects of the droplet are negligible. In the earlier paper by Denesuk \textit{et al.} \cite{penetration_denesuk} they consider the imbibition of a droplet that has already spread out, specifying three cases that occur as the droplets volume depletes (see figure \ref{f:intro_denesuk_cases}). Case (a) is that of decreasing drawing area (DDA), where CL1 recedes, decreasing the radius of the drawing area to zero for a droplet of zero volume. In case (b) the drawing area remains constant, CL1 being pinned in place, proving a constant drawing area (CDA). This can occur in two ways that are experimentally distinct: (b1) where the drawing area maintains the appearance of having a constant radius; (b2) where the drawing area appears to decrease in radius, but a thin film remains that can supply the pores with fluid from the bulk of the droplet. Both cases of (b) produce the same behaviour within the porous material, thus we consider the distinction no further. In our work we will only model the case of CDA. It is likely that neither of DDA or CDA are commonplace, and that as droplets imbibe their drawing area decreases but not to zero. Denesuk \textit{et al.} then perform theoretical analysis of the two cases, using a Washburn type model for the porous solid. From this they deduce that the time for imbibition with DDA, and constant contact angle CA1, is nine times greater than that of CDA.

Experiments have been performed in a variety of the cases and limits described by Denesuk \textit{et al.} \cite{dynamics_denesuk}. Holman \textit{et al.} \cite{spread_holman} perform experiments for droplets with $\tau_d \approx \tau_s$, using materials: HPA 0.5 with porosity $0.549$ and representative pore radius $0.07\mu$m; HPA 1 with porosity $0.575$ and representative pore radius $0.17\mu$m. Droplets of diameter $54\mu$m are placed onto the substrate. Performing a best fit for their data, the radius of the drawing area at short times is approximately $R(t) = 54.1(0.04+t)^{0.176}$. At later times it is assumed to follow the model presented by Denesuk \textit{et al.} \cite{penetration_denesuk} for DDA, but this is not plotted for a comparison.

Hapgood \textit{et al.} \cite{drop_penetration} perform experiments of imbibition into various powders and packed beads. The photographs they provide are informative as to the dynamics of the process and the time scales involved, but no data on the radii of the drawing area is provided. 

Popovich \textit{et al.} \cite{popovich_rise_spread} experimentally investigate the spread of various fluids over carbon black, reporting initial and maximal radii, the rate of spread and the time for imbibition. However the porous substrate did fracture during the experiments, thus it is unclear as to the quality of the results.

Chandra and Avedisan \cite{droplet_experiment_chandra} perform experiments into the spread of droplets over a ceramic substrate, including images of the droplets spreading in their paper. 

To investigate the level of agreement between theory and experiment, numerical simulations have been performed. Reis \textit{et al.} in \cite{droplet_reis} produced numerical simulations of both the flow in a droplet imbibing into the solid and the flow within the solid. They then compared them to empirical results, which show a good level of agreement for some of the simulations. They chose to use a spatially averaged Navier-Stokes equation, which is appropriate for particle suspension phenomena and has not been shown to be valid for flow in a porous material, as has already been discussed. Equivalent simulations need to be performed using Darcy's law for a fair comparison to be made as to the merits of their choice of bulk equation. They also use a constant contact angle CA1 as a boundary condition, which they justify with results from \cite{wetting_fukai}, which is for a droplet rapidly spreading on a non-porous substrate. The assumption may also be valid for spreading on a porous substrate, but it is expected that (unless we have DDA) the contact angle will initially be some finite value and zero when all the fluid has been imbibed. This is what is shown in their plots in \cite{droplet_reis_2} which do not maintain the contact angle they specify, although this may be because the method of approximating the boundary that they use does not produce a smooth curve as it should. Finally, the contact angle that they use in the pores is constant, which may or may not be a good approximation for droplet imbibition, this is yet to be tested. Considering all of these questionable elements, the results produced are remarkably similar to the empirical results which does suggest that their mathematical model may be largely correct, but without many alternatives to compare it to we cannot yet draw this conclusion.

Another relevant study has been done by Markicevec \textit{et al.} \cite{spread_sessile_numerics}. In this study a capillary network model is used, producing numerical results with around 20\% accuracy. The final example is that by Alleborn and Razillier \cite{spread_porous}, considering a very wide flat droplet using lubrication theory, in which motion can only occur in the vertical direction, producing surprisingly conical wetted regions. The validity of the lubrication approximation used shall be discussed later.

\subsection{The Present Work}

Our purpose is to investigate the dynamics of the wetting front by modelling and simulating imbibition into a porous substrate. In the present work the boundary condition on the wetting front that is used is for a constant contact angle within the pores, but the numerical scheme developed is easily expandable to include dynamic contact angles and even the modes proposed by Shikhmurzaev and Sprittles in \cite{wetting_dynamics_shk}. The numerical scheme is for axisymmetric imbibition obeying Darcy's equation and incompressibility.

In  section \ref{s:ProbForm} we will formulate a model of imbibition through a circular region of constant radius. Then in section \ref{s:Numerics} we describe the numerical model that will be used to produce solutions to the equations, and simulate the imbibition process. In section \ref{s:simple} we investigate the velocity and pressure distributions across the wetted region for particular wetting fronts, both using our numerical solutions and asymptotic analysis in regions of interest. Following this we produce numerical simulations of the wetting fronts evolution for various initial conditions. Finally we summarise the results and propose future work in section \ref{s:conc}.

During our study we discover problems with the solutions to Darcy's law that are unexpected and reveal it to be an invalid equation when modelling a range of flows. This motivates the existence of the improvements we discussed earlier, although none of these have been proposed to solve problems like those that we discover.

\section{Problem Formulation}\label{s:ProbForm}

Consider a non-deformable isotropic homogeneous porous solid initially filled with a gas, which in the process to be studied will be regarded as dynamically passive. We assume the solid is large enough to ignore all of its faces other than its flat horizontal top, through which an incompressible fluid is imbibed over a circular region of radius $R$. Outside the solid, we call the region of fluid the \textit{external reservoir} and the rest the \textit{atmosphere}. Within the solid the region of fluid is called the \textit{wetted region}, and the rest is the \textit{dry region}. Here is set out the modelling of the dynamics of the wetted region under the secondary continuum limit, i.e. the limit as the ratio of the pore scale to the Darcy scale tends to zero, which shall be used unless otherwise stated.

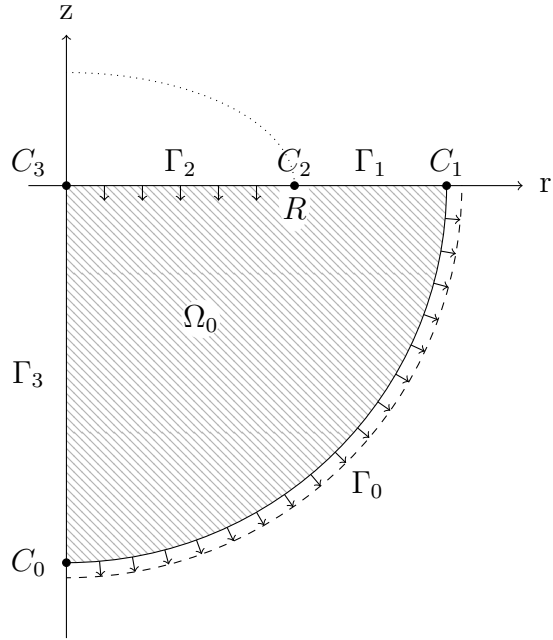
\begin{figure}[t]
	\centering
	\begin{tikzpicture}
	\newcommand\Rmin{-0.5};
	\newcommand\Rmax{6};
	\newcommand\Zmin{-6};
	\newcommand\Zmax{2};
	\begin{scope}
		\clip (0,0) rectangle (6,-6);
		\draw[pattern=north west lines, pattern color=black!30!white] (0,0) circle [radius=5cm];
		\filldraw[white] (3,-0.2)	ellipse (2mm and 4mm);
		\draw[dashed] (0,0) circle [radius=5.2cm];
	\end{scope}
	\begin{scope}
		\clip (0,0) rectangle (3,3);
		\draw[dotted] (0,0) ellipse (3cm and 1.5cm);
	\end{scope}
	\draw[->]	(\Rmin,0) -- (\Rmax,0);
	\draw[->]	(0,\Zmin) -- (0,\Zmax);
	\draw		($(\Rmax,0)+(0.3,0)$) node {r}
				($(0,\Zmax)+(0,0.3)$) node {z};
	\filldraw
		(0,-5)	circle [radius=0.5mm]
		(5,0)	circle [radius=0.5mm]
		(0,0)	circle [radius=0.5mm]
		(3,0)	circle [radius=0.5mm];
	\filldraw[white]
		(-45:2.5)	circle [radius=2.5mm];
	\draw
		(-0.5,-5)	node {$C_0$}
		(5,0.3)		node {$C_1$}
		(3,0.3)		node {$C_2$}
		(-0.5,0.3)	node {$C_3$}
		(-45:5.6)	node {$\Gamma_0$}
		(4,0.3)		node {$\Gamma_1$}
		(1.5,0.3)	node {$\Gamma_2$}
		(-0.5,-2.5)	node {$\Gamma_3$}
		(-45:2.5)	node {$\Omega_0$};
	\draw
			(3,-0.3)	node {$R$};
	\foreach \n in {5,10,...,85} {
		\draw[->] (-\n:5) -- (-\n:5.2);
	}
	\foreach \n in {0.5,1,...,2.51} {
		\draw[->] (\n,0) -- (\n,-0.2);
	}
\end{tikzpicture}
	\caption{Illustration of the axisymmetric imbibition process, with moving free surface $\Gamma_0$ and a droplet (or any appropriate external reservoir) resting on the solid being imbibed through $\Gamma_2$.}
	\label{f:imbib}
\end{figure}

We assume that the velocity, pressure and wetted region are axisymmetric, thus we choose to use cylindrical polar coordinates. The cylindrical axis is placed on the axis of symmetry with its coordinate $z$ such that $z<0$ in the solid and $z=0$ on its top, as shown in figure \ref{f:imbib}. The radial coordinate shall be $r$, the azimuth $\phi$, the time $t$ and the position $\boldsymbol{r}$. Our model will be developed in the $r$-$z$ plane, which contains all the information of the problem. Figure \ref{f:imbib} illustrates an example configuration. In it $\Omega_0$ is the wetted region, $\Gamma_1$ and $\Gamma_2$ are the boundaries to the atmosphere and external reservoir respectively, $\Gamma_3$ is on the axis of symmetry, and $C_0$, $C_1$, $C_2$ and $C_3$ are defined by the figure. $\Gamma_0$ is the boundary to the dry region, known as the \textit{wetting front}, that moves as the fluid imbibes. All other regions may also evolve with time.

For later convenience, we define the total boundary as $\partial \Omega_0 = \Gamma_0 \cup \Gamma_1 \cup \Gamma_2 \cup \Gamma_3 \cup C_0 \cup C_1 \cup C_2 \cup C_3$, and $\hat{\boldsymbol{n}}$ to be the outward pointing unit normal to $\partial \Omega_0$.

Let us use the notation $\boldsymbol{u}(\boldsymbol{r},t)=u(r,z,t)\,\hat{\boldsymbol{r}}(\phi)+v(r,z,t)\,\hat{\boldsymbol{z}}$ to be the velocity and $p(\boldsymbol{r},t)=p(r,z,t)$ to be the pressure of the averaged flow on the Darcy scale. Using the assumptions of incompressibility, isotropy and homogeneity the continuity equation can be written as
\begin{equation}\label{eq:cont_dimfull}
	\nabla \cdot \boldsymbol{u}=0 \hspace{1cm} \forall \: \boldsymbol{r} \in \Omega_0.
\end{equation}
The momentum balance in the wetted region is given by Darcy's equation
\begin{equation} \label{eq:darcy_dimfull}
	\boldsymbol{u}=-\frac{k}{\mu}\nabla (p+\rho g z) \hspace{1cm} \forall \: \boldsymbol{r} \in \Omega_0,
\end{equation}
where $k$ is the permeability of the porous solid, $\mu$ and $\rho$ are the dynamic viscosity and density of the imbibing fluid respectively, and $g$ the magnitude of free-fall acceleration due to gravity, all being constant. Combining \eqref{eq:cont_dimfull} and \eqref{eq:darcy_dimfull} we see that $\nabla^2 p=0$ so that, if the boundary isn't moving, we require one boundary condition at every boundary point, and for a moving boundary we require two conditions.

In general, fluid could pass through $\Gamma_1$ to form a new region of fluid above the surface or be drawn down creating a new de-wetting front. This would require the modelling of the process of creating new boundaries, as well as the formulation of boundary conditions that allow for the de-wetting process. For simplicity we assume that these processes do not occur and thus
\begin{equation}	\label{eq:bc_gamma1_dimful}
	\boldsymbol{u}\cdot\hat{\boldsymbol{n}}=0 \hspace{1cm} \forall \: \boldsymbol{r} \in \Gamma_1.
\end{equation}

The boundary $\Gamma_2$ must have a condition that matches the solution in the wetted region to the external reservoir. We consider the scales of pressure in the regions, using the same technique as in \cite{dynam_angle_shk}, measuring the pressure relative to that of the dynamically passive gas. Note that variables with a tilde represent those of the external reservoir. Define the surface tension to be $\sigma$, the representative pore radius to be $a$, and the velocity and length scales to be $U$ and $L$ respectively. Note that $\tilde{L}=L$. The scale of pressure in the wetted region is $P=2\sigma / a$ from the assumption that the pores are cylinders and the menisci are spherical caps, as shall be discussed later. The scale of pressure in the external reservoir is $\tilde{P}=\mu \tilde{U}/\tilde{L}$, from the Navier-Stokes equation in the bulk at Reynolds numbers that are small or approximately one. The pressure is continuous across the boundary, $p=\tilde{p}$ on $\Gamma_2$, marking dimensionless parameters with a prime this is
\begin{equation*}
	p'=\frac{\mu \tilde{U}}{2 \sigma}\frac{a}{L}\tilde{p}' \hspace{1cm} \forall \: \boldsymbol{r} \in \Gamma_2.
\end{equation*}
The secondary continuum limit is the limit that $a/L\rightarrow0$, and hence the pressure in the reservoir is negligible compared to that of the wetted region. Therefore the continuum mechanical boundary condition is
\begin{equation}	\label{eq:bc_gamma2_dimfull}
	p=0 \hspace{1cm} \forall \: \boldsymbol{r} \in \Gamma_2.
\end{equation}
In a physical situation the external pressure can of course be chosen to be of the same order of magnitude as the Darcy pressure, but in most circumstances this requires significant engineering to achieve and would almost certainly not be the case in droplet imbibition.

On $\Gamma_3$, we have the condition of axisymmetry
\begin{equation}	\label{eq:bc_gamma3_dimful}
	\boldsymbol{u}\cdot\hat{\boldsymbol{n}}=0 \hspace{1cm} \forall \: \boldsymbol{r} \in \Gamma_3.
\end{equation}

Considering the boundary $\Gamma_0$, it is first assumed that the wetting front moves with the velocity of the fluid. Denoting the normal velocity of the wetting front by $v_s$, this assumption is stated mathematically as $v_s=\boldsymbol{u}\cdot\hat{\boldsymbol{n}}$. We define a function $F(\boldsymbol{r},t)$ such that $F=0$ on $\Gamma_0$, in our case this equation can be written in differential form as the kinematic boundary condition
\begin{equation}	\label{eq:bc_timestep_F_dimfull}
	\frac{\partial F}{\partial t} + \boldsymbol{u}\cdot\nabla F =0.
\end{equation}
For the dynamic boundary condition we use the standard model of wetting, which is mode 1 of Shikhmurzaev and Sprittles' model  \cite{wetting_dynamics_shk}. Under the primary continuum limit the wetting front consists of the menisci within the pores. In this model representative pores are used, aligned normal to the surface, containing a representative meniscus that is a spherical cap forming the contact angle $\theta$ with the wall. The meniscus is advancing along the pore with velocity $u_1$ and pressure $p_1$ (both averaged across the pore cross section). The variables in the representative pore and of the secondary continuum limit are related by the equations
\begin{alignat}{2}
	p&=p_1 \hspace{1cm} &\forall \: \boldsymbol{r} &\in \Gamma_0,	\label{eq:bc_p1_dimfull}\\
	\boldsymbol{u}\cdot\hat{\boldsymbol{n}}&=u_1 \hspace{1cm} & \forall \: \boldsymbol{r} &\in \Gamma_0,	\label{eq:bc_u1_dimfull}
\end{alignat}
As discussed in the introduction, there is a function that relates the dynamic contact angle $\theta_d$ and the velocity of the meniscus, $G(\theta_d,u_1)=0$. Due to the spherical cap approximation for the meniscus shape, in a pore with representative radius $a$ and surface tension $\sigma$ the fluid has a pressure relative to the constant pressure of the dynamically passive gas given by
\begin{equation}
	p_1 = -\frac{2\sigma}{a} \cos(\theta_d).	\label{eq:bc_gamma0_dimfull}
\end{equation}

Finally we require an initial condition for \eqref{eq:bc_timestep_F_dimfull}. This initial condition must specify the shape of the wetting front, i.e.  $F(\boldsymbol{r},0)=0$, although it is much easier to provide the curve along which it is zero. Thus we shall require functions $r(s)$ and $z(s)$ such that $F(\hat{\boldsymbol{r}}r(s)+\hat{\boldsymbol{z}}z(s),0)=0 \: \forall s\in[0,s_\mathrm{max}]$ where $s_\mathrm{max}$ is the end point of the wetting front. We also require that $\hat{\boldsymbol{r}}r(0)+\hat{\boldsymbol{z}}z(0)$ is the point $C_1$ and $\hat{\boldsymbol{r}}r(s_\mathrm{max})+\hat{\boldsymbol{z}}z(s_\mathrm{max})$ is the point $C_0$ at time $t=0$.

The equations we have discussed are
\begin{alignat}{2}
	\nabla \cdot \boldsymbol{u}&=0 \hspace{1cm} \hspace{1cm} &\forall \: \boldsymbol{r} &\in \Omega_0,	\tag{\ref{eq:cont_dimfull}}	\\
	\boldsymbol{u}&=-\frac{k}{\mu}\nabla (p+\rho g z) \hspace{1cm} \hspace{1cm} &\forall \: \boldsymbol{r} &\in \Omega_0,	\tag{\ref{eq:darcy_dimfull}}	\\
	\frac{\partial F}{\partial t} + \boldsymbol{u}\cdot\nabla F &=0,		\tag{\ref{eq:bc_timestep_F_dimfull}}	\\
	\boldsymbol{u}\cdot\hat{\boldsymbol{n}}&=0 \hspace{1cm} \hspace{1cm} &\forall \: \boldsymbol{r} &\in \Gamma_1	\tag{\ref{eq:bc_gamma1_dimful} and \ref{eq:bc_gamma3_dimful}}	\\
	p&=0 \hspace{1cm} &\forall \: \boldsymbol{r} &\in \Gamma_2	\tag{\ref{eq:bc_gamma2_dimfull}}	\\
	p&=p_1 \hspace{1cm} &\forall \: \boldsymbol{r} &\in \Gamma_0,	\tag{\ref{eq:bc_p1_dimfull}}	\\
	\boldsymbol{u}\cdot\hat{\boldsymbol{n}}&=u_1 \hspace{1cm} & \forall \: \boldsymbol{r} &\in \Gamma_0	\tag{\ref{eq:bc_u1_dimfull}},	\\
	p_1 &= -\frac{2\sigma}{a} \cos(\theta_d) \tag{\ref{eq:bc_gamma0_dimfull}}	\\
	G(\theta_d,u_1)&=0.
\end{alignat}
In this work we will only consider the simplest of wetting processes, that of constant contact angle. We enforce $\theta_d=\theta_s$ where $\theta_s \in (0,\pi)$, therefore $G(\theta_d,u_1)=\theta_d-\theta_s$. The equations are now written in dimensionless form, where the scales of pressure, length, velocity and time are $P=2\sigma\cos(\theta_s)/a$, $L=R$, $U=(k/\mu L) P$ and $T=L/U$ respectively, using the same symbols for the dimensionless functions as we did for the dimensional ones. The only dimensionless parameter of the system is $\gamma=k \rho g/\mu U$.
\begin{subequations}\label{seq:dimless_system}\begin{alignat}{2}
	\nabla \cdot \boldsymbol{u}&=0 \hspace{1cm} &\forall \: \boldsymbol{r} &\in \Omega_0	\label{eq:continuity}\\
	\boldsymbol{u}&=-\nabla (p +\gamma z) \hspace{1cm} & \forall \: \boldsymbol{r} &\in \Omega_0	\label{eq:darcy}\\
	\frac{\partial F}{\partial t} + \boldsymbol{u}\cdot\nabla F&=0 \label{eq:bc_timestep_F}\\
	\boldsymbol{u}\cdot\hat{\boldsymbol{n}}&=0 \hspace{1cm} &\forall \: \boldsymbol{r} &\in \Gamma_1 \cup \Gamma_3	\label{eq:bc_gamma13}\\
	p&=-1 \hspace{1cm} &\forall \: \boldsymbol{r} &\in \Gamma_0	\label{eq:bc_gamma0}	\\
	p&=0 \hspace{1cm} &\forall \: \boldsymbol{r} &\in \Gamma_2	\label{eq:bc_gamma2}
\end{alignat}\end{subequations}

The equations in \eqref{seq:dimless_system} along with specifying the initial conditions $r(s)$ and $z(s)$ form the closed set of equations to solve.

\section{Discrete form of the Equations} \label{s:Numerics}

In the set of equations to solve, \eqref{seq:dimless_system}, it is important to observe that the only time dependence is in the advancing of the wetting front, \eqref{eq:bc_timestep_F}. Thus the equations can be solved at each instant of time for the velocity and pressure distribution independently of temporal evolution. First we shall present the scheme for numerical solution to the spatial problem, which shall then be tested, before giving the method for time stepping.

\subsection{Interpolation Functions}\label{ss:Interp}

The numerical simulations are performed using the finite element method, described in \cite{intro_fem,fem_framework_shk}. A finite set of nodes are chosen at positions $\boldsymbol{r}_i(t)=r_i(t)\hat{\boldsymbol{r}}+z_i(t)\hat{\boldsymbol{z}} \in \overline{\Omega_0}$, arranged into triangles with curved sides, one node at each vertex and one on each side, as shown in figure \ref{f:2d_element}. These triangles are known as quadratic triangular elements, the domain of the $e$th element being denoted $\Omega^e$. We define continuous interpolation functions $\psi_i(\boldsymbol{r},t)$ such that $\psi_i(\boldsymbol{r}_j(t),t)=\delta_{ij}$, where $\delta_{ij}$ is the Kronecker delta, and $\sum_i \psi_i=1$. Note that this definition does not uniquely specify the interpolation functions.

\begin{figure}[t]
	\centering
	\begin{tabular}{c c}
		\begin{subfigure}[t]{0.47\textwidth}
			\centering
			\begin{tikzpicture}
	\def\varTension{1};
	\def\varRadius{2mm};
	\coordinate (C0) at (1.5,4);
	\coordinate (C1) at (1.5,1);
	\coordinate (C2) at (5,1);
	\coordinate (C3) at (3,3);
	\coordinate (C4) at (0.9,2.3);
	\coordinate (C5) at (3,1.2);
	\draw[->] (-0.5,0) -- (5,0);
	\draw[->] (0,-0.5) -- (0,5);
	\draw 
		(5,-0.3) node {$r$}
		(-0.3,5) node {$z$};
	\draw plot[smooth,tension=\varTension] coordinates{(C0) (C4) (C1)};
	\draw plot[smooth,tension=\varTension] coordinates{(C1) (C5) (C2)};
	\draw plot[smooth,tension=\varTension] coordinates{(C2) (C3) (C0)};
	\fill[fill=white, draw=black]
		(C0) circle (\varRadius)
		(C1) circle (\varRadius)
		(C2) circle (\varRadius)
		(C3) circle (\varRadius)
		(C4) circle (\varRadius)
		(C5) circle (\varRadius);
	\draw
		(C0) node {$0$}
		(C1) node {$1$}
		(C2) node {$2$}
		(C3) node {$3$}
		(C4) node {$4$}
		(C5) node {$5$};
	\draw
		($(C3)!0.5!(C4)!0.3333333!(C5)$) node {$\Omega^e$}
		($(C3)+(45:0.6)$) node {$\Gamma^{e,2}$}
		($(C4)+(180:0.5)$) node {$\Gamma^{e,0}$}
		($(C5)+(-90:0.5)$) node {$\Gamma^{e,1}$};
\end{tikzpicture}
			\caption{An example element in the mesh. This is the $e$th element with domain $\Omega^e$ and nodes at the numbered locations.}
			\label{f:2d_element}
		\end{subfigure}
		&
		\begin{subfigure}[t]{0.47\textwidth}
			\centering
			\begin{tikzpicture}
	\def\varTension{1};
	\def\varRadius{2mm};
	\coordinate (C0) at (-2,2);
	\coordinate (C1) at (-2,-2);
	\coordinate (C2) at (2,-2);
	\coordinate (C3) at (0,0);
	\coordinate (C4) at (-2,0);
	\coordinate (C5) at (0,-2);
	\draw[->] (-2.5,0) -- (2.5,0);
	\draw[->] (0,-2.5) -- (0,2.5);
	\draw 
		(2.5,-0.3) node {$\xi$}
		(-0.3,2.5) node {$\eta$};
	\draw plot[smooth,tension=\varTension] coordinates{(C0) (C4) (C1)};
	\draw plot[smooth,tension=\varTension] coordinates{(C1) (C5) (C2)};
	\draw plot[smooth,tension=\varTension] coordinates{(C2) (C3) (C0)};
	\fill[fill=white, draw=black]
		(C0) circle (\varRadius)
		(C1) circle (\varRadius)
		(C2) circle (\varRadius)
		(C3) circle (\varRadius)
		(C4) circle (\varRadius)
		(C5) circle (\varRadius);
	\draw
		(C0) node {$0$}
		(C1) node {$1$}
		(C2) node {$2$}
		(C3) node {$3$}
		(C4) node {$4$}
		(C5) node {$5$};
	\draw ($(C3)!0.5!(C4)!0.3333333!(C5)$) node {$\Omega^M$}
	($(C3)+(40:0.6)$) node {$\Gamma^{M,2}$}
	($(C4)+(-0.5,0.5)$) node {$\Gamma^{M,0}$}
	($(C5)+(0.5,-0.5)$) node {$\Gamma^{M,1}$};
	\draw
		($(-2,0)+(-135:0.45)$) node {$-1$}
		(-0.3,-0.3) node {$0$}
		(2,-0.3) node {$1$}
		($(0,-2)+(-135:0.45)$) node {$-1$}
		(-0.3,2) node {$1$};
\end{tikzpicture}
			\caption{The master element in master coordinates $\xi$ and $\eta$, with domain $\Omega^M$ and nodes at the numbered locations.}
			\label{f:2d_master}
		\end{subfigure}
		\\
		\begin{subfigure}[t]{0.47\textwidth}
			\centering
			\begin{tikzpicture}
	\def\varTension{1};
	\def\varRadius{2mm};
	\coordinate (C0) at (1,2.5);
	\coordinate (C1) at (3,2.5);
	\coordinate (C2) at (4,0.5);
	\draw[->] (-0.5,0) -- (5,0);
	\draw[->] (0,-0.5) -- (0,3);
	\draw 
		(5,-0.3) node {$r$}
		(-0.3,3) node {$z$};
	\draw plot[smooth,tension=\varTension] coordinates{(C0) (C1) (C2)};
	\fill[fill=white, draw=black]
		(C0) circle (\varRadius)
		(C1) circle (\varRadius)
		(C2) circle (\varRadius);
	\draw
		(C0) node {$0$}
		(C1) node {$1$}
		(C2) node {$2$};
	\draw ($(C1)+(0,0.5)$) node {$\Omega^{eb}$};
\end{tikzpicture}
			\caption{An example boundary element in the mesh. This is the $b$th boundary of the $e$th element with domain $\Omega^{eb}$ and nodes at the numbered locations.}
			\label{f:1d_element}
		\end{subfigure}
		&
		\begin{subfigure}[t]{0.47\textwidth}
			\centering
			\begin{tikzpicture}
	\def\varTension{1};
	\def\varRadius{2mm};
	\coordinate (C0) at (-2,0);
	\coordinate (C1) at (0,0);
	\coordinate (C2) at (2,0);
	\draw[->] (-2,0) -- (2.5,0);
	\draw 
		(2.5,-0.3) node {$\omega$};
	\draw plot[smooth,tension=\varTension] coordinates{(C0) (C1) (C2)};
	\fill[fill=white, draw=black]
		(C0) circle (\varRadius)
		(C1) circle (\varRadius)
		(C2) circle (\varRadius);
	\draw
		(C0) node {$0$}
		(C1) node {$1$}
		(C2) node {$2$};
	\draw ($(C1)+(0,0.5)$) node {$\Omega^{B}$};
	\draw
		(-2,-0.4) node {$-1$}
		(0,-0.4) node {$0$}
		(2,-0.4) node {$1$};
\end{tikzpicture}
			\caption{The master boundary element in master coordinate $\omega$, with domain $\Omega^{B}$ and nodes at the numbered locations.}
			\label{f:1d_master}
		\end{subfigure}
	\end{tabular}
	\caption{}
\end{figure}

The bulk variables $u$, $v$ and $p$ are interpolated using their values at all the nodes, which is the scheme used in \cite{stable_darcy}. Using the same notation for the approximations as for the true solutions, we have
\begin{subequations}\label{seq:function_discrete}\begin{align}
	u(\boldsymbol{r},t)&=\sum_i u_i(t) \psi_i(\boldsymbol{r},t),	\\
	v(\boldsymbol{r},t)&=\sum_i v_i(t) \psi_i(\boldsymbol{r},t),	\\
	p(\boldsymbol{r},t)&=\sum_i p_i(t) \psi_i(\boldsymbol{r},t).
\end{align}\end{subequations}
Note that $u(\boldsymbol{r}_i(t),t)=u_i(t)$, etc. thus the new variables are the values of the unknown functions at the nodes.

To obtain unique interpolation functions, we first define global node numbers to be the italicised indices used so far, and local node numbers over the $e$th element that have the values 0 to 5, as shown in figure \ref{f:2d_element}, will be denoted by Roman indices and a superscript $e$ index. Local node numbers only exist for the nodes that are part of the element, and there is an arbitrary choice of three configurations of the node numbers corresponding to rotating the definition of the numbering heuristic in figure \ref{f:2d_element}. The global node number $i$ is a function of the element number $e$ and the local node number $\mathrm{i}$, such a function is represented as a connectivity matrix $\boldsymbol{M}$, such that $i(e,\mathrm{i})=M^e_\mathrm{i}$. Local interpolation functions are defined as $\psi_i^e(\boldsymbol{r},t)=\psi_i(\boldsymbol{r},t)\:\forall \boldsymbol{r} \in \Omega^e(t)$.

Next, we define the master element to have domain $\Omega^M$ in a master coordinate system $(\xi,\eta)$. Its local node numbers are defined in figure \ref{f:2d_master} with coordinates $(\xi_\mathrm{i},\eta_\mathrm{i})$, and its sides are straight. The master interpolation functions $\psi^M_{\mathrm{i}}(\xi,\eta)$ are uniquely defined by the condition $\psi^M_\mathrm{i}(\xi_\mathrm{j},\eta_\mathrm{j})=\delta_{\mathrm{i}\mathrm{j}}$ and the requirement that they be quadratics in the master coordinates, explicitly
\begin{equation}
	\begin{array}{r@{=}l@{\hspace{1cm}}r@{=}l@{\hspace{1cm}}r@{=}l}	\vspace{1mm}
		\psi_0^M	&	\frac{1}{2}\eta(\eta+1)				,&
		\psi_1^M	&	\frac{1}{2}(\xi+\eta)(\xi+\eta+1)	,&
		\psi_2^M	&	\frac{1}{2}\xi(\xi+1)				,\\
		\psi_3^M	&	(\xi+1)(\eta+1)						,&
		\psi_4^M	&	-(\xi+\eta)(\eta+1)					,&
		\psi_5^M	&	-(\xi+\eta)(\eta+1)					.
	\end{array}
\end{equation}
We define an isoparametric coordinate transformation between $\Omega^M$ and $\Omega^e$
\begin{equation}	\label{eq:elem_iso_coord_trans}
	\boldsymbol{r}(\xi,\eta;e)=\sum_{\mathrm{i}} \boldsymbol{r}_\mathrm{i}^e(t) \, \psi^M_\mathrm{i}(\xi,\eta),
\end{equation}
which uniquely specifies the curve of the elemental boundaries, and thereby $\Omega^e$. Finally the interpolation functions are uniquely defined by
\begin{equation}	\label{eq:local_interp_master_def}
	\psi^e_\mathrm{i}(\boldsymbol{r}(\xi,\eta;e))=\psi^M_\mathrm{i}(\xi,\eta).
\end{equation}
and $\psi_i=0$ in any element that does not contain node $i$.

Boundary elements and interpolation functions are also needed. The elemental boundaries are identified by a parameter $b$: the boundary from node $0$ anticlockwise to node $1$ corresponds to $b=0$; from $1$ to $2$ has $b=1$; from $2$ to $0$ has  $b=2$. The domain of the boundary is denoted $\Gamma^{eb}$ in an element and $\Gamma^{Mb}$ in the master element, which are illustrated in figures \ref{f:2d_element} and \ref{f:2d_master} respectively. The master boundary element is defined in the master coordinate $\omega$ to have domain $\Omega^{B}$, and is shown in figure \ref{f:1d_master}. Its boundary node numbers as shown are denoted by a fraktur index $\textfrak{i}$ and a superscript $B$. A linear transformation can be defined between any of the master elements three boundaries onto the master boundary element which means that $\textfrak{i}=\textfrak{i}(b,\mathrm{i})$. Under any of these transformations the interpolation functions become what we shall call the master boundary interpolation functions
\begin{equation}
	\begin{array}{r@{=}l@{\hspace{1cm}}r@{=}l@{\hspace{1cm}}r@{=}l}	\vspace{1mm}
		\psi_0^{B}(\omega)	&	\frac{1}{2} \omega (\omega-1)	,&
		\psi_1^{B}(\omega)	&	(1+\omega)(1-\omega)			,&
		\psi_2^{B}(\omega)	&	\frac{1}{2} \omega (\omega+1)	.
	\end{array}
\end{equation}
Under the coordinate transformation \eqref{eq:elem_iso_coord_trans}, the chosen boundary of the master element transforms into a boundary of the element $e$, so we define the local boundary node number to be denoted with an index $\textfrak{i}$ and superscript indices $e$ and $b$. Since the master boundary interpolation functions are only master interpolation functions for a restricted domain, the boundary interpolation functions are defined as
\begin{align}
	\psi^{eb}_\textfrak{i}(\boldsymbol{r}(\omega;e,b))&=\psi^{B}_\textfrak{i}(\omega),	\label{eq:boundary_interp_master_def}
\intertext{where}
	\boldsymbol{r}(\omega;e,b)&=\sum_{\textfrak{i}} \boldsymbol{r}_\textfrak{i}^{eb}(t) \, \psi^{B}_\textfrak{i}(\omega). \label{eq:bound_iso_coord_trans}
\end{align}
The approximated solutions can therefore be expressed over the elemental boundaries as
\begin{subequations}\label{seq:function_discrete_boundary}\begin{align}
	u(\boldsymbol{r},t)&=\sum_{i} u_i^{eb}(t) \psi_i^{eb}(\boldsymbol{r},t),	\\
	v(\boldsymbol{r},t)&=\sum_{i} v_i^{eb}(t) \psi_i^{eb}(\boldsymbol{r},t),	\\
	p(\boldsymbol{r},t)&=\sum_{i} p_i^{eb}(t) \psi_i^{eb}(\boldsymbol{r},t),
\end{align}\end{subequations}
for appropriate $e$ and $b$.

Schemes that have the same degree of interpolation for pressure and velocity are used to approximate solutions to Darcy's equation elsewhere, for example \cite{stable_darcy}, which we use to justify the choice of interpolation outlined above. Schemes which have the interpolation of pressure one degree higher than that for velocity can also be used, for example that in \cite{roberts_FEM}. The most convenient scheme of this nature for our purposes is to have velocity interpolated linearly using only the corner nodes in each element. However, when this was used the discrete form of the bulk equations broke down at the corner nodes in each element, so this has not been used.

\subsection{Numerical Integration}

When we construct the finite element method for our problem, we shall need to be able to evaluate integrals over both the domain and its boundary. First we shall consider integrals over the domain of the form
\begin{equation*}
	I=\int_{\Omega_0} f(\boldsymbol{r}) \dif r \dif z.
\end{equation*}
We notice that the integral over the entire domain is the sum of the parts over the elements, thus
\begin{equation*}
	I=\sum_e\int_{\Omega^e} f(\boldsymbol{r}) \dif r \dif z.
\end{equation*}
Next we transform the integrals into the master element. For this we require the Jacobian of the transformation defined in \eqref{eq:elem_iso_coord_trans}
\begin{equation}	\label{eq:jacobian}
	\boldsymbol{J}^e=
	\left(\begin{array}{cc}\vspace{1mm}
		\dfrac{\partial r}{\partial \xi}	&	\dfrac{\partial z}{\partial \xi}	\\
		\dfrac{\partial r}{\partial \eta}	&	\dfrac{\partial z}{\partial \eta}
	\end{array}\right)
	=
	\left(\begin{array}{cccccc}\vspace{1mm}
		\dfrac{\partial \psi_0^M}{\partial \xi}	&
		\dfrac{\partial \psi_1^M}{\partial \xi}	&
		\dfrac{\partial \psi_2^M}{\partial \xi}	&
		\dfrac{\partial \psi_3^M}{\partial \xi}	&
		\dfrac{\partial \psi_4^M}{\partial \xi}	&
		\dfrac{\partial \psi_5^M}{\partial \xi}	\\
		\dfrac{\partial \psi_0^M}{\partial \eta}	&
		\dfrac{\partial \psi_1^M}{\partial \eta}	&
		\dfrac{\partial \psi_2^M}{\partial \eta}	&
		\dfrac{\partial \psi_3^M}{\partial \eta}	&
		\dfrac{\partial \psi_4^M}{\partial \eta}	&
		\dfrac{\partial \psi_5^M}{\partial \eta}	\\
	\end{array}\right)
	\left(\begin{array}{cc}
		r_0^e	&	z_0^e	\\
		r_1^e	&	z_1^e	\\
		r_2^e	&	z_2^e	\\
		r_3^e	&	z_3^e	\\
		r_4^e	&	z_4^e	\\
		r_5^e	&	z_5^e	\\
	\end{array}\right).
\end{equation}
Thus we have, using $J^e \equiv |\boldsymbol{J}^e|$,
\begin{equation}	\label{eq:2d_master_integration}
	\int_{\Omega_0} f(\boldsymbol{r}) \dif r \dif z=\sum_e\int_{\Omega^m} f(\boldsymbol{r}(\xi,\eta)) J^e \dif \xi \dif \eta.
\end{equation}
To evaluate the integrals over the master element, we use the quadrature set out in \cite{fem_framework_shk} which uses nine points and exactly integrates polynomials of order five. The integrands will be polynomials up to order eight, but as the size of the elements decrease the result of the numerical approximation will tend towards the true value. The scheme is
\begin{equation*}
	\int_{\Omega^M}g(\xi,\eta) \dif \xi \dif \eta \sim \sum_{j=1}^9 g(\xi_j, \eta_j) W_j
\end{equation*}
where
\begin{gather*}
	\begin{array}{c@{=}c@{\hspace{1cm}}c@{=}c@{\hspace{1cm}}c@{=}c}
		\xi_1	&	+0.00000	\,	00000	\,	00000,	&
		\xi_4	&	+0.77459	\,	66692	\,	41483,	\\
		\eta_1	&	-0.88729	\,	83346	\,	20741,	&
		\eta_2	&	-0.50000	\,	00000	\,	00000,	&
		\eta_3	&	-0.11270	\,	16653	\,	79258,	\\
		\eta_4	&	-0.97459	\,	66692	\,	41483,	&
		\eta_6	&	-0.80000	\,	00000	\,	00000,	&
		\eta_9	&	+0.57459	\,	66692	\,	41483,	\\
		W_1		&	+0.24691	\,	35802	\,	46913,	&
		W_2		&	+0.39506	\,	17283	\,	95061,	&
		W_4		&	+0.03478	\,	44646	\,	23227,	\\
		W_5		&	+0.05565	\,	51433	\,	97164,	&
		W_7		&	+0.27385	\,	75106	\,	85414,	&
		W_8		&	+0.43817	\,	20170	\,	96662,	
	\end{array}
	\\
	\xi_{2,3}=\xi_1,
	\quad
	\xi_{5,6}=-\xi_{7,8,9}=\xi_4,
	\quad
	\eta_5=\eta_1,
	\quad
	\eta_7=\eta_6,
	\quad
	\eta_8=\eta_3,
	\quad
	W_3=W_1,
	\quad
	W_6=W_4,
	\quad
	W_9=W_7.
\end{gather*}

Next we consider a boundary integral over $\partial \Omega_0$ of the form
\begin{equation*}
	I=\int_{\partial \Omega_0} f(\boldsymbol{r}) \dif s
\end{equation*}
where $s$ is the arc-length along the boundary. We write this as a sum over the elemental boundaries that are part of $\partial \Omega_0$,
\begin{equation*}	\label{eq:integral_length}
	I=\sum_{e,b : \Gamma^{eb}\subseteq\partial\Omega_0}\int_{\Gamma^{eb}} f(\boldsymbol{r}) \dif s.
\end{equation*}
These integrals can now be transformed onto the master boundary, using \eqref{eq:bound_iso_coord_trans} we see that
\begin{equation}	\label{eq:dsdomega}
	\frac{\dif s}{\dif \omega}=\sqrt{\left(\frac{\dif r}{\dif \omega}\right)^2+\left(\frac{\dif z}{\dif \omega}\right)^2}
	=\sqrt{\left(\sum_{\textfrak{i}} r_\textfrak{i}^{eb}\frac{\dif \psi_\textfrak{i}^{eb}}{\dif \omega}\right)^2+\left(\sum_{\textfrak{i}} z_\textfrak{i}^{eb}\frac{\dif \psi_\textfrak{i}^{eb}}{\dif \omega}\right)^2}
\end{equation}
thus
\begin{equation}	\label{eq:1d_master_integration}
	\int_{\partial \Omega_0} f(\boldsymbol{r}) \dif s
	=
	\sum_{e,b : \Gamma^{eb}\subseteq\partial\Omega_0}\int_{\Omega^{B}} f(\boldsymbol{r}(\omega)) \frac{\dif s}{\dif \omega}\dif \omega.
\end{equation}
To evaluate the integrals over the master boundary, the standard eight point Gaussian quadrature is used. This is exact for polynomials of order fifteen and will converge for any of the integrals we consider as the element size decreases. In fact, if $1/x$ and $\sqrt{x}$ can be accurately approximated by quadratic Taylor expansions for any given integral, the integrals will be exact. The scheme is
\begin{equation*}
	\int_{\Omega^{B}} g(\omega) \dif \omega \sim \sum_{j=1}^8 g(\omega_j) W_j
\end{equation*}
where
\begin{gather*}
	\begin{array}{c@{=}c@{\hspace{1cm}}c@{=}c}
		\omega_1	&	0.18343	\,	46424	\,	95649	\,	8,	&
		\omega_3	&	0.52553	\,	24099	\,	16329	\,	0,	\\
		\omega_5	&	0.79666	\,	64774	\,	13626	\,	7,	&
		\omega_7	&	0.96028	\,	98564	\,	97536	\,	3,	\\
		W_1	&	0.36268	\,	37833	\,	78362	\,	0,	&
		W_3	&	0.31370	\,	66458	\,	77887	\,	3,	\\
		W_5	&	0.22238	\,	10344	\,	53374	\,	5,	&
		W_7	&	0.10122	\,	85362	\,	90376	\,	3	\\
	\end{array}
	\\
	\omega_2=-\omega_1,
	\quad
	\omega_4=-\omega_3,
	\quad
	\omega_6=-\omega_5,
	\quad
	\omega_8=-\omega_7,
	\\
	W_2=W_1,
	\quad
	W_4=W_3,
	\quad
	W_6=W_5,
	\quad
	W_8=W_7.
\end{gather*}

\subsection{The Gradient Operator and Normals}

We will require the gradient of differentiable axisymmetric scalar functions, let us denote a generic such function by $f(r,z)$. This is the gradient in cylindrical coordinates, but $f$ is not a function of $\phi$, thus
\begin{equation*}
	\nabla f = \frac{\partial f}{\partial r} \hat{\boldsymbol{r}} + \frac{\partial f}{\partial z} \hat{\boldsymbol{z}}
	=\left(\begin{array}{c} \displaystyle \vspace{1mm}
		\frac{\partial f}{\partial r}	\\ \displaystyle
		\frac{\partial f}{\partial z}
	\end{array}\right)
\end{equation*}
Defining
\begin{equation*}
	\boldsymbol{\partial_\xi} f = \frac{\partial f}{\partial \xi} \hat{\boldsymbol{\xi}} + \frac{\partial f}{\partial \eta} \hat{\boldsymbol{\eta}}
	=\left(\begin{array}{c} \displaystyle \vspace{1mm}
		\frac{\partial f}{\partial \xi}	\\ \displaystyle
		\frac{\partial f}{\partial \eta}
	\end{array}\right)
\end{equation*}
and using \eqref{eq:elem_iso_coord_trans} we see that
\begin{align}\notag
	\left(\begin{array}{c} \displaystyle \vspace{1mm}
		\frac{\partial f}{\partial \xi}	\\ \displaystyle
		\frac{\partial f}{\partial \eta}
	\end{array}\right)
	&=
	\left(\begin{array}{cc}\vspace{1mm}
			\dfrac{\partial r}{\partial \xi}	&	\dfrac{\partial z}{\partial \xi}	\\
			\dfrac{\partial r}{\partial \eta}	&	\dfrac{\partial z}{\partial \eta}
		\end{array}\right)
	\left(\begin{array}{c} \displaystyle \vspace{1mm}
		\frac{\partial f}{\partial r}	\\ \displaystyle
		\frac{\partial f}{\partial z}
	\end{array}\right)
	\\ \Rightarrow \qquad
	\nabla f &= (\boldsymbol{J}^e)^{-1} \boldsymbol{\partial_\xi} f.
\end{align}
In this manner spatial derivatives of scalar functions are calculated.

To obtain the outward unit normal on $\partial \Omega_0$ we impose some restrictions on $f$. We require $f=0$ and $|\nabla f| \neq 0$ on the boundary $\partial \Omega_0$, $f<0$ in the region $\Omega_0$ and $f>0$ otherwise. Under these conditions an outward normal is $\boldsymbol{n}=\nabla f$, and since the transformation \eqref{eq:elem_iso_coord_trans} takes the boundary to the master elements boundary, $\boldsymbol{n_\xi}=\boldsymbol{\partial_\xi} f$ is an outward normal to the master element at the transformed point. Therefore
\begin{equation}	\label{eq:normal}
	\boldsymbol{n}=(\boldsymbol{J}^e)^{-1} \boldsymbol{n_\xi},
\end{equation}
and the outward unit normal can be obtained by normalising the transformation of a sensible choice of outward normal in the master coordinates, for example
\begin{equation*}
	\boldsymbol{n_\xi}=-\hat{\boldsymbol{\xi}} \text{ on } \Gamma^{M0},
	\qquad
	\boldsymbol{n_\xi}=-\hat{\boldsymbol{\eta}} \text{ on } \Gamma^{M1},
	\qquad
	\boldsymbol{n_\xi}=\hat{\boldsymbol{\xi}}+\hat{\boldsymbol{\eta}} \text{ on } \Gamma^{M2}.
\end{equation*}

\subsection{Mesh Design and the Method of Spines} \label{ss:Spines}

From section \ref{ss:Interp} we are left with five unknowns for every node at any given instant of time: its position and the values of the functions at the node. The method of constructing a mesh of nodes shall be discussed here, first describing how to construct the elements from spines, then how to position the spines, and finally how to refine around a point.

\subsubsection{Constructing Elements from Spines} \label{sss:spines_to_elements}

\begin{figure}[t]
	\centering
	\begin{tabular}{c c}
		\begin{subfigure}[t]{0.47\textwidth}
			\centering
			\begin{tikzpicture}
	\draw[->] (-0.5,0) -- (5,0);
	\draw (5,-0.3) node {$r$};
	\draw[->] (0,-4.5) -- (0,1);
	\draw (-0.3,1) node {$z$};
	
	\filldraw
		(0,-4)	circle [radius=0.5mm]
		(4,0)	circle [radius=0.5mm]
		(0,0)	circle [radius=0.5mm]
		(2,0)	circle [radius=0.5mm];
		
	\begin{scope}
		\clip (0,0) rectangle (4.1,-4.1);
		\draw (0,0) circle (4);
	\end{scope}
		
	\foreach \n in {1,2,...,7} {
		\begin{scope}
			\clip (0,0) rectangle (4.1,-4.1);
			\clip (0,0) circle (4);
			\pgfmathsetmacro{\chi}{ln(abs((-0.5*\n+8)/(-0.5*\n)))};
			\draw[domain=0:-90,smooth,variable=\zeta] plot ({(4*sinh(\chi))/(cosh(\chi)+cos(\zeta))},{(4*sin(\zeta))/(cosh(\chi)+cos(\zeta))});
		\end{scope}
		\draw (4-0.5*\n,0.3) node {\n};
	}
	\draw (4,0.3) node {0};
	\draw (0,0.3) node {8};
	
	\draw (4,-0.3) node {$r_f$};
\end{tikzpicture}
			\caption{Example spines for mesh construction. In a simulation the spines would be much more densely packed to produce a high resolution on the solution. The circles show the location of the corner points $C_0$, $C_1$, $C_2$, and $C_3$.}
			\label{f:spines}
		\end{subfigure}
		&
		\begin{subfigure}[t]{0.47\textwidth}
			\centering
			\begin{tikzpicture}
	\begin{scope}
		\clip (-2,-4) rectangle (2.5,0);
		\draw[dashed] (5,0) circle (7);
		\draw[dashed] (8.5,0) circle (7);
		\draw[dotted] (6.75,0) circle (7);
	\end{scope}
	
	\coordinate (TR) at ($(8.5,0)+(-175:7)$);
	\coordinate (TL) at ($(5,0)+(-175:7)$);
	\coordinate (BR) at ($(8.5,0)+(-155:7)$);
	\coordinate (BL) at ($(5,0)+(-155:7)$);
	\draw (TL) -- (TR) -- (BR) -- (BL) -- (TL) -- (BR);
	
	\filldraw
		(TL)	circle (0.05)
		(TR)	circle (0.05)
		(BL)	circle (0.05)
		(BR)	circle (0.05)
		($(TL)!0.5!(TR)$)	circle (0.05)
		($(TR)!0.5!(BR)$)	circle (0.05)
		($(BR)!0.5!(BL)$)	circle (0.05)
		($(BL)!0.5!(TL)$)	circle (0.05)
		($(TL)!0.5!(BR)$)	circle (0.05);
\end{tikzpicture}
			\caption{A block of elements between two spines, the standard way of generating elements. The circles show the location of the nodes that are part of the elements depicted.}
			\label{f:spines_block}
		\end{subfigure}
		\\
		\begin{subfigure}[t]{0.47\textwidth}
			\centering
			\begin{tikzpicture}
	\begin{scope}
		\clip (-2,-4) rectangle (2.5,0);
		\draw[dashed] (5,0) circle (7);
		\draw[dashed] (8.5,0) circle (7);
		\draw[dotted] (6.75,0) circle (7);
	\end{scope}
	
	\coordinate (TL) at ($(5,0)+(-175:7)$);
	\coordinate (BL) at ($(5,0)+(-155:7)$);
	\coordinate (R) at ($(8.5,0)+(-165:7)$);
	\draw (TL) -- (R) -- (BL) -- (TL);
	
	\filldraw
		(TL)	circle (0.05)
		(BL)	circle (0.05)
		(R)		circle (0.05)
		($(TL)!0.5!(R)$)	circle (0.05)
		($(R)!0.5!(BL)$)	circle (0.05)
		($(BL)!0.5!(TL)$)	circle (0.05);
\end{tikzpicture}
			\caption{An increasing wedge between two spines, for when the next spine has more intervals to fill then the previous. The circles show the location of the nodes that are part of the element depicted.}
			\label{f:spines_wedge_up}
		\end{subfigure}
		&
		\begin{subfigure}[t]{0.47\textwidth}
			\centering
			\begin{tikzpicture}
	\begin{scope}
		\clip (-2,-4) rectangle (2.5,0);
		\draw[dashed] (5,0) circle (7);
		\draw[dashed] (8.5,0) circle (7);
		\draw[dotted] (6.75,0) circle (7);
	\end{scope}
	
	\coordinate (TR) at ($(8.5,0)+(-175:7)$);
	\coordinate (BR) at ($(8.5,0)+(-155:7)$);
	\coordinate (L) at ($(5,0)+(-165:7)$);
	\draw (TR) -- (L) -- (BR) -- (TR);
	
	\filldraw
		(TR)	circle (0.05)
		(BR)	circle (0.05)
		(L)		circle (0.05)
		($(TR)!0.5!(L)$)	circle (0.05)
		($(L)!0.5!(BR)$)	circle (0.05)
		($(BR)!0.5!(TR)$)	circle (0.05);
\end{tikzpicture}
			\caption{A decreasing wedge between two spines, for when the next spine has fewer intervals to fill then the previous. The circles show the location of the nodes that are part of the element depicted.}
			\label{f:spines_wedge_down}
		\end{subfigure}
	\end{tabular}
	\caption{}
\end{figure}

Spines are curves which are used to generate elements, the elements are positioned in between the spines such that the base of the triangle is along one spine and the point opposite is on an adjacent spine. We first construct the spines and then position elements in between them, the spines to be used are shown graphically in figure \ref{f:spines}. These spines are good because they are centred around $C_1$ which will allow us to refine the mesh around this point, tend towards straight lines at $\Gamma_3$ which makes aligning the elements with the axis trivial, and are approximately perpendicular to the wetting front if it is a simple arc, thus a significant degree of distortion will have to occur for the wetting front to become parallel to them and the mesh unusable. They are isoclines of the bipolar coordinate system, specifically the coordinate 
\begin{equation}	\label{eq:spines_chi}
	\chi=\ln\left(\frac{\sqrt{(r+r_f)^2+z^2}}  {\sqrt{(r-r_f)^2+z^2}}\right)
\end{equation}
where $r_f$ is the radial coordinate of the focus, in our case $C_1$. The spines satisfy $\chi=\chi_n$, where $n$ is the index of the spine, the numbering starting at $C_1$ and increasing for decreasing $r$. This causes the spines to be circles with centre $(R_n,0)$ and radius $\rho_n$, where
\begin{equation}	\label{eq:spines_Rrho}
	R_n=\frac{r_f}{\tanh\left(\chi_n\right)},	\qquad	\rho_n=\frac{r_f}{\sinh\left(\chi_n\right)}.
\end{equation}
If $(r,z)$ is the point of intersection of the spine with the wetting front $\Gamma_0$, the angle subtended along the spine is
\begin{equation}	\label{eq:spines_theta}
	\theta_n=\arctan\left(\frac{-z}{r_f-r}\right)
\end{equation}
Let $r_n$ be the point of intersection of the spine with the $r$-axis. It is $r_n$ that we shall calculate first to position the spine, this process shall be described in the next section. For this section it will suffice to imagine that they are evenly distributed along $0 \leq r \leq r_f$. From $r_n$, we can use \eqref{eq:spines_chi} to calculate $\chi_n $ and then \eqref{eq:spines_Rrho} and \eqref{eq:spines_theta} to calculate $R_n$, $\rho_n$ and $\theta_n$.

When the spines are generated, we shall ensure that $r_{n-1}-r_n \approx r_{n}-r_{n+1}$, since it is important that small elements and large elements are not too close to each other for solution accuracy. Define for the spines that have two neighbours
\begin{equation}	\label{eq:spines_h}
	h_n=(r_{n-1}-r_{n+1})/2,
\end{equation}
this is the mean distance from this spine to its two neighbours. For the spines at $C_1$ and $C_3$, $h_n$ is the distance to the single adjacent spine. The elements generated must not be overly distorted, an element that is long and thin will induce error, so we should divide the spine up into intervals approximately of length $h_n$, each interval being an elemental boundary. Define
\begin{equation}	\label{eq:spines_J}
	J_n=\lceil \rho_n \theta_n/h_n\rceil,
\end{equation}
this shall be the number of intervals the spine is divided into, each interval being of equal length as measured along the arc of the spine. 

\paragraph{Element Generation}To generate elements, we run between two spines from the $r$-axis to the wetting front generating elements that span between an interval on one side, and the point between two intervals on the other, as shown in figures \ref{f:spines_block}, \ref{f:spines_wedge_up} and \ref{f:spines_wedge_down}. The usual method is to create a block that advances along one interval for each spine, as shown in figure \ref{f:spines_block}. The four corner nodes are places at the ends of the intervals, and the remaining nodes are placed at the midpoints of the sides they are on. However, this method will only be able to generate all the elements if $J_n=J_{n+1}$. If $J_n<J_{n+1}$ then there will be left over intervals on the next spine, which can be filled by single elements known as \textit{increasing wedges} as shown in figure \ref{f:spines_wedge_up}. If $J_n>J_{n+1}$ then there will be left over intervals on the current spine, which can be filled by single elements known as \textit{decreasing wedges} as shown in figure \ref{f:spines_wedge_down}. These extra elements should be spread out evenly along the spine to minimise the amount of distortion in the elements, for example in the current implementation if there are two elements to be added these will be added at $1/4$ and $3/4$ of the way along the spine. This is achieved by setting a counter to $0.5$ at the start of a run between two spines. Each time elements are going to be added, the counter is increased by $|J_n-J_{n+1}|/\max\{J_n,J_{n+1}\}$, if the counter exceeds $1$ then a wedge is added next and the counter decreased by $1$, otherwise a block is added.

It should be noted that each spine must carry information about its $\chi_n$ and endpoint at the wetting front. I.e. when programming this algorithm the spines should be stored in such a way that, knowing the value $n$, the values $\chi_n$ and the coordinates of the endpoint of the spine can be accessed. In the above discussion it has not been mentioned how the elements at the wetting font will be constructed. The centre points of the elemental boundaries that lie along $\Gamma_0$ must be on $\Gamma_0$, and not the midpoint of the endpoints of the spines, otherwise the solution will be inaccurate since we will not have approximated the domain as well as we can. Thus we include `pseudo-spines' that will be placed in between each pair of spines such that $r_{n+(1/2)}=(r_n+r_{n+1})/2$ is the point of intersection of the pseudo-spine with the $r$-axis. These will be used purely to hold their point of intersection with the wetting front, such that the last element generated between every pair of spines can use this point and have its boundary along the wetting front.

To find the point of intersection of a spine with the wetting front, we use the notation that the wetting front is parametrically $r=\tilde{r}(s)$, $z=\tilde{z}(s)$, where $s=0$ is $C_1$ and $0<s<s_{\textrm{max}}$ is the wetting front (this is the form in which the initial conditions are given). This means that $r(0)=r_f$ and $z(0)=0$. The point of intersection will occur at the root of the function
\begin{equation}
	f(s)=\ln\left(\frac{\sqrt{(\tilde{r}(s)+r_f)^2+\tilde{z}(s)^2}}  {\sqrt{(\tilde{r}(s)-r_f)^2+\tilde{z}(s)^2}}\right)-\chi_n
\end{equation}
which satisfies $0\leq \tilde{r}(s) \leq r_f$, $\tilde{z}(s) \leq 0$, and we must have that there is only one solution to be able to generate the mesh. We solve this equation by using the Newton-Raphson method, where an initial guess $s_0$ is produced (the arbitrary nature of this guess is why we require the solution to be unique), and refinements on this guess are produced by
\begin{equation}
	s_{m+1}=s_m - \frac{f(s_m)}{f'(s_m)}
\end{equation}
where the index $m$ numbers our attempts at finding the solutions. The exact solution is obtained as $m \rightarrow \infty$ (assuming that it does indeed converge), or numerically at the point when $s_m=s_{m+1}$. This method can be used as stated for the first instant of time, since the wetting front is the initial condition and is provided in this form. For later instants of time the wetting front has been time stepped from the previous one, and will be a sequence of elemental boundaries. On each elemental boundary the coordinates can be obtained as $\tilde{r}(\omega;e,b)\hat{\boldsymbol{r}}+\tilde{z}(\omega;e,b)\hat{\boldsymbol{z}}$ for $-1\leq\omega\leq1$, thus we simply use these coordinates to produce $f(\omega)$ and solve in exactly the same way, except that now the elemental boundary will also have to be stepped onto the adjacent one when $\omega$ exceeds its bounds. It is worth noting that the first and last spines should be included as special cases, since not only is it easy to overstep the end point of an elemental boundary and then have no adjacent element to step into, but the value of $\chi$ is divergent at $C_1$ which cannot be handled numerically.

\subsubsection{Positioning Spines}

To generate the spines we require the values of $r_n$, which control the size of the elements produced. We have two constraints, firstly the spine separation ($r_{n}-r_{n+1}$) should not change suddenly since this will give distorted elements of different sizes next to each other. Thus we shall enforce that
\begin{equation}	\label{eq:spine_max_rate}
	\frac{1}{M_{mh}}\geq\frac{r_{n}-r_{n+1}}{r_{n-1}-r_{n}}\geq M_{mh}
\end{equation}
where $M_{mh}$ is the maximal rate of change of the spine separation. Also there must be spines that intersect $C_1$, $C_2$ and $C_3$, to enable us to have nodes at these points and fill the domain with elements, this shall be reflected in our algorithm. We shall require there to be a minimum spine density of $I_{mh}$ per unit length, thus 
\begin{equation}	\label{eq:spine_max_sep}
	r_{n}-r_{n+1} \leq 1/I_{mh},
\end{equation}
to ensure a decent level of mesh resolution and solution accuracy throughout. We shall denote the smallest separation between spines permitted to be $S_{t}$,
\begin{equation}	\label{eq:spine_min_sep}
	r_{n}-r_{n+1} \geq S_{t}.
\end{equation}
The value of $S_t$ is calculated at each time step to account for the changing shape of the wetted region and get the required resolution. Let $-H$ be the $z$ coordinate of $C_0$, $S_{t1}=(r_f-1)/20$ and $S_{t2}=H/100$. We define $S_t=\min\{S_{t1},S_{t2},S_{mh},1/I_{mh}\}$ where $S_{mh}$ is a parameter dictating the maximal value of $S_t$ allowed.
The spines must not be allowed to separate out so far that they are further apart than they are long, thus we define the number of times longer a spine must be than the separation to the next to be $C_{mh}$, therefore
\begin{equation}	\label{eq:spine_close}
	r_{n}-r_{n+1} \leq \frac{r_n \theta_n}{C_{mh}}.
\end{equation}
A higher level of resolution shall be required at $C_1$ than at any other point, due to the multivalued and singular solutions there, thus we shall start at this point with the smallest elements in the mesh and increase the separation as we move away. There are also these problems at $C_2$, but $C_1$ is on the wetting front which is were we require the highest level of accuracy for the time stepping.

The first spine to be generated shall be that at $C_1$, this is of zero length but is required to generate elements that span from it into the domain (which shall be increasing wedges), thus $r_0=r_f$. The separation between spines $0$ and $1$ should be the smallest in the mesh, so we govern it with the parameter $S_{mh}$, $r_1=r_0-S_{mh}$, which allows us to control how dense the mesh becomes in this region. Note that if $S_{mh}\geq1/I_{mh}$ then $r_1=r_0-(1/I_{mh})$ instead. The separation between spines should now increases at a steady rate, we define $R_{mh}$ to be this rate such that $R_{mh} \leq M_{mh}$ and
\begin{equation}	\label{eq:spine_rapid_sep}
	r_{n+1}=r_n-(r_{n-1}-r_n) R_{mh}.
\end{equation}
If this causes the new spine to break any of the above inequalities, then the value of $r_{n+1}$ should be altered to satisfy the respective equality. These are applied in the order \eqref{eq:spine_max_sep}, \eqref{eq:spine_close}, \eqref{eq:spine_max_rate}, then \eqref{eq:spine_min_sep}.

We next consider how to ensure that the spines align with the point $C_2$, such that one spine passes through this point. Let the distance between the most recently generated spine and $C_2$ be $D$, thus $D=r_n-1$, and the most recent spine separation be $L$, thus $L=r_{n-1}-r_n$. If we are to traverse the distance to $C_2$ in $q$ or $q+1$ equally spaced spines then we require that 
\begin{equation}	\label{eq:spine_next_mode}
	qL \leq D \leq (q+1)L.
\end{equation}
The minimal and maximal distance, $d_-$ and $d_+$, that can be traversed in $q$ spines are, under the constraint \eqref{eq:spine_max_rate}, given by the geometric progression formula
\begin{align}
	d_-&=\frac{L}{M_{mh}} \frac{M_{mh}^{-q}-1}{M_{mh}^{-1}-1},	&	d_+&=L M_{mh} \frac{M_{mh}^{q}-1}{M_{mh}-1}.
\end{align}
Therefore we require that
\begin{gather}
	d_- \leq qL \leq D \leq (q+1)L \leq d_+	\notag\\
	\Rightarrow \qquad
	\frac{M_{mh}^{-q}-1}{1-M_{mh}} \leq q
	\qquad\qquad
	M_{mh} \frac{M_{mh}^{q}-1}{M_{mh}-1}	\geq q+1
	\qquad
\end{gather}
and the value of $q$ can be calculated prior to generating the spines by considering $q=1$ and then increasing its value to the next integer while the inequalities do not hold.

The spines are generated using \eqref{eq:spine_rapid_sep} until \eqref{eq:spine_next_mode} is satisfied. At this point the distance $D$ is divided up into $N$ equal segments that minimise the jump in spine separation, resulting in
\begin{equation}	\label{eq:spine_aim_step}
	r_{n+1}=r_{n} - \frac{D}{N}.
\end{equation}
The value of $N$ is chosen algorithmically by starting with $N=1$ and increasing $N$ to the next integer value while it is true that
\begin{equation}	\label{eq:spine_aim_number}
	\left|\ln\left( \frac{D/(N+1)}{r_{n-1}-r_{n}} \right)\right| < \left|\ln\left( \frac{D/N}{r_{n-1}-r_{n}} \right)\right|.
\end{equation}
such that the change in step size is minimised. If the value of $r_{n+1}$ causes \eqref{eq:spine_max_rate} to be broken, then it is  altered to satisfy the respective equality. Applying this algorithm for each spine generation up to the point $C_2$ produces spines whose separation changes at the maximal rate allowed by \eqref{eq:spine_max_rate} up to a point, and then becomes static. The conditions \eqref{eq:spine_max_sep}-\eqref{eq:spine_close} are not applied.

To generate spines in the region between $C_2$ and $C_3$ the same method is used but with $D=r_n$. I.e. \eqref{eq:spine_rapid_sep} is used \{applying \eqref{eq:spine_max_sep}, \eqref{eq:spine_close}, \eqref{eq:spine_max_rate}, then \eqref{eq:spine_min_sep}\} until \eqref{eq:spine_next_mode} is satisfied, and then \eqref{eq:spine_aim_step} is used with $N$ from \eqref{eq:spine_aim_number} applying \eqref{eq:spine_max_rate}.

\begin{figure}[tbp]
	\centering
	\begin{tabular}{c c}
		\begin{subfigure}[t]{0.9\textwidth}
			\centering
			\includegraphics[width=\textwidth]{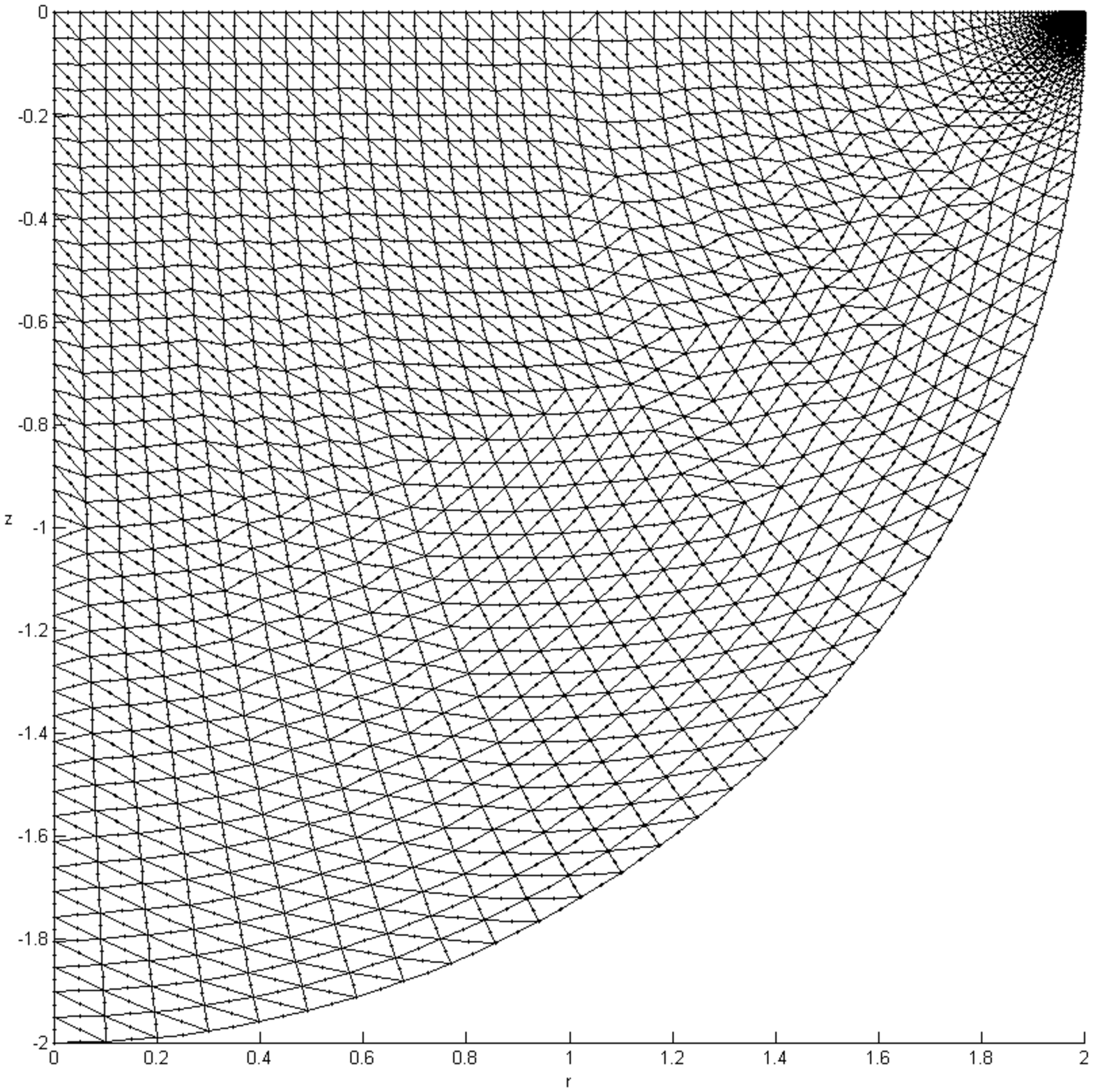}
			\caption{}
			\label{f:example_mesh_all}
		\end{subfigure}
		\\
		\begin{subfigure}[t]{0.9\textwidth}
			\centering
			\includegraphics[width=\textwidth]{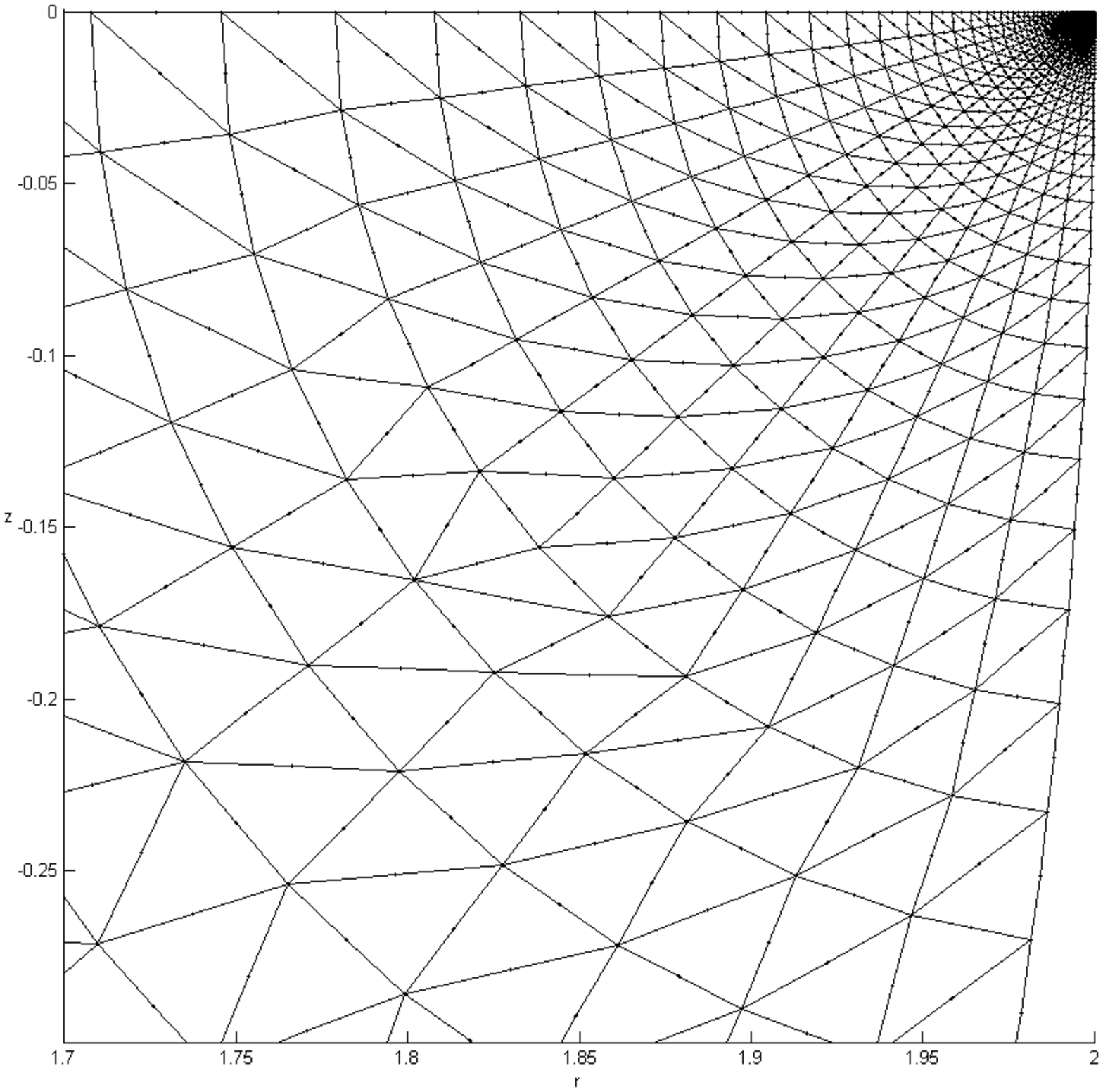}
			\caption{}
			\label{f:example_mesh_focus}
		\end{subfigure}
	\end{tabular}
	\caption{Example mesh generated without refinement at $C_2$. The parameters of the mesh are $R(t)=1$, $\tilde{r}(s)=2\cos(s)$, $\tilde{z}(s)=-2\sin(s)$, $I_{mh}=20$, $S_{mh}=10^{-5}$, $C_{mh}=5$, $M_{mh}=1.4$, and $R_{mh}=1.15$.}
	\label{f:example_mesh}
\end{figure}

An example mesh produced with this method is depicted in figure \ref{f:example_mesh}. It illustrates how we achieve a uniform mesh that has a spine intersecting with $C_2$ and steadily refines around $C_1$.

\subsubsection{Mesh Refinement}

The solution shall not only be singular around $C_1$, but also around $C_2$. The singularity around $C_2$ is less important, since it is not on the wetting front where the solution is required to greatest accuracy, and thus we refine around this point as a secondary consideration. The refinement will have the element sizes changing rapidly, which will decrease the accuracy of the solution, however it will be more accurate than with an unrefined mesh around the singularity which has been found to cause the solution to be poor.

\begin{figure}[t]
	\centering
	\begin{tabular}{c c}
		\begin{subfigure}[t]{0.47\textwidth}
			\centering
			\begin{tikzpicture}
	\begin{scope}
		\clip (-3,-5) rectangle (4,0);
		\draw[dashed] (5,0) circle (8);
		\draw[dashed] (10.5,0) circle (8);
		\draw[dotted] (7.75,0) circle (8);
	\end{scope}
	
	\coordinate (TR) at ($(10.5,0)+(-180:8)$);
	\coordinate (TL) at ($(5,0)+(-180:8)$);
	\coordinate (BR) at ($(10.5,0)+(-150:8)$);
	\coordinate (BL) at ($(5,0)+(-150:8)$);
	
	\coordinate (L)  at ($(TL)!0.5!(BL)$);
	\coordinate (T)  at ($(TL)!0.5!(TR)$);
	\coordinate (R)  at ($(BR)!0.5!(TR)$);
	\coordinate (B)  at ($(BR)!0.5!(BL)$);
	\coordinate (M)  at ($(TL)!0.5!(BR)$);
	
	\draw (R) -- (BR) -- (M) -- (BL) -- (BR) (BL) -- (TL) -- (M) (TL) -- (T) -- (M) -- (R);
	\filldraw
		(TL)	circle (0.05)
		(L)		circle (0.05)
		(BL)	circle (0.05)
		(B)		circle (0.05)
		(BR)	circle (0.05)
		($(TL)!0.5!(T)$)	circle (0.05)
		($(TL)!0.5!(M)$)	circle (0.05)
		($(BL)!0.5!(M)$)	circle (0.05)
		($(BR)!0.5!(M)$)	circle (0.05)
		($(BR)!0.5!(R)$)	circle (0.05);
	\filldraw
		(TR)	circle (0.05);
	\coordinate (TL) at (T);
	\coordinate (BL) at (M);
	\coordinate (BR) at (R);
	\coordinate (L)  at ($(TL)!0.5!(BL)$);
	\coordinate (T)  at ($(TL)!0.5!(TR)$);
	\coordinate (R)  at ($(BR)!0.5!(TR)$);
	\coordinate (B)  at ($(BR)!0.5!(BL)$);
	\coordinate (M)  at ($(TL)!0.5!(BR)$);
	
	\foreach \N in {0,1,...,10}	{
		\draw[dashed] (R) -- (BR) -- (M) -- (BL) -- (BR) (BL) -- (TL) -- (M) (TL) -- (T);
		\filldraw
			(TL)	circle (0.05)
			(L)		circle (0.05)
			(BL)	circle (0.05)
			(B)		circle (0.05)
			(BR)	circle (0.05)
			($(TL)!0.5!(T)$)	circle (0.05)
			($(TL)!0.5!(M)$)	circle (0.05)
			($(BL)!0.5!(M)$)	circle (0.05)
			($(BR)!0.5!(M)$)	circle (0.05)
			($(BR)!0.5!(R)$)	circle (0.05);
		\coordinate (TL) at (T);
		\coordinate (BL) at (M);
		\coordinate (BR) at (R);
		\coordinate (L)  at ($(TL)!0.5!(BL)$);
		\coordinate (T)  at ($(TL)!0.5!(TR)$);
		\coordinate (R)  at ($(BR)!0.5!(TR)$);
		\coordinate (B)  at ($(BR)!0.5!(BL)$);
		\coordinate (M)  at ($(TL)!0.5!(BR)$);
	}
	
	\draw ($(TR)+(0.3,0.3)$) node {$C_2$};
\end{tikzpicture}
			\caption{The method of local refinement of a block.}
			\label{f:spines_zoom}
		\end{subfigure}
		&
		\begin{subfigure}[t]{0.47\textwidth}
			\centering
			\begin{tikzpicture}
	\clip (-2,-3.2) rectangle (4,2);
	
	\coordinate (TR) at ($(10.5,0)+(-180:8)$);
	\coordinate (TL) at ($(5,0)+(-180:8)$);
	\coordinate (BR) at ($(10.5,0)+(-150:8)$);
	\coordinate (BL) at ($(5,0)+(-150:8)$);
	
	\coordinate (L)  at ($(TL)!0.5!(BL)$);
	\coordinate (T)  at ($(TL)!0.5!(TR)$);
	\coordinate (R)  at ($(BR)!0.5!(TR)$);
	\coordinate (B)  at ($(BR)!0.5!(BL)$);
	\coordinate (M)  at ($(TL)!0.5!(BR)$);
	
	\draw[dashed] (R) -- (BR) -- (M) -- (BL) -- (BR) (BL) -- (TL) -- (M) (TL) -- (T);
	\filldraw
		(TL)	circle (0.05)
		(L)		circle (0.05)
		(BL)	circle (0.05)
		(B)		circle (0.05)
		(BR)	circle (0.05)
		($(TL)!0.5!(T)$)	circle (0.05)
		($(TL)!0.5!(M)$)	circle (0.05)
		($(BL)!0.5!(M)$)	circle (0.05)
		($(BR)!0.5!(M)$)	circle (0.05)
		($(BR)!0.5!(R)$)	circle (0.05);
	\coordinate (TL) at (T);
	\coordinate (BL) at (M);
	\coordinate (BR) at (R);
	\coordinate (L)  at ($(TL)!0.5!(BL)$);
	\coordinate (T)  at ($(TL)!0.5!(TR)$);
	\coordinate (R)  at ($(BR)!0.5!(TR)$);
	\coordinate (B)  at ($(BR)!0.5!(BL)$);
	\coordinate (M)  at ($(TL)!0.5!(BR)$);
	
	\draw (TL) -- (TR) -- (BR) -- (BL) -- (TL) -- (BR);
	\filldraw
		(TL)	circle (0.05)
		(T)		circle (0.05)
		(TR)	circle (0.05)
		(R)		circle (0.05)
		(BR)	circle (0.05)
		(B)		circle (0.05)
		(BL)	circle (0.05)
		(L)		circle (0.05)
		(M)		circle (0.05);
		
	\draw ($(TR)+(0.3,0.3)$) node {$C_2$};
\end{tikzpicture}
			\caption{The smallest elements in the refinement.}
			\label{f:spines_zoom_centre}
		\end{subfigure}
	\end{tabular}
	\caption{}
\end{figure}

The refinement is performed by the generation of an alternative block. Instead of generating the block using the method depicted in figure \ref{f:spines_block}, we use the method depicted in figure \ref{f:spines_zoom}. The block depicted is for immediately left of $C_2$, adjacent to the $r$-axis, the block to right of $C_2$ uses a mirrored version of the method discussed.
The three quadrants not containing $C_2$ are filled with four elements as depicted, the extra nodes being midpoints of the sides they are on. This leaves a block remaining that has one quarter the area of the original, which can then be divided up in exactly the same way as the first. This process is repeated until a predefined point has been reached, let us define this to be when the length of the side of the remaining block along the $r$-axis is less than $Z_{mh}$. When this condition is reached the remaining block is split into two elements, see figure \ref{f:spines_zoom_centre}, choosing to have one element containing $C_2$ since the elements containing the singularity induce error, and so we want the total area of such elements to be minimal. It is important that both the blocks are refined the same number of times.

\begin{figure}[tbp]
	\centering
	\includegraphics[width=0.8\textwidth]{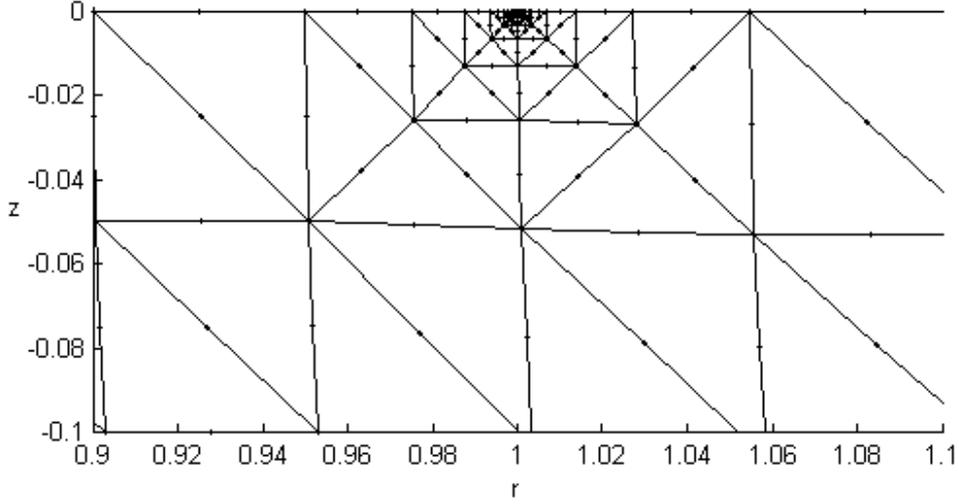}
	\caption{Example mesh generated with refinement at $C_2$, the plot only showing the region around this point. The parameter governing the refinement is $Z_{mh}=10^{-6}$, all other parameters being the same as in figure \ref{f:example_mesh}.}
	\label{f:example_mesh_zoom}
\end{figure}

The refinement of the mesh in figure \ref{f:example_mesh} is depicted in figure \ref{f:example_mesh_zoom}, illustrating the method.

\subsection{Discrete form of the Bulk Equations}

The remaining unknowns are the values of the functions at the nodes, the method of finding these values is explained here. Analytically these are specified by the bulk equations, these bulk equations will be converted into a numerical scheme which is called the Galerkin finite element method. 

We first construct weighted residuals of the bulk equations by volume integrating the equation with weight $\psi_i$. Integration by parts is then used to minimise the level of differentiability required on any function, as well as providing a way to include boundary conditions, preferring to differentiate the interpolation functions over the approximate solutions. Requiring that this form of the equations is satisfied exactly by the approximate solution produces equations that specifies the values of the functions at the $i$th node in terms of the values at the nodes in the elements it is part of. The approximations \eqref{seq:function_discrete} are used to produce this set of linear equations for the unknowns. Since there is one interpolation function and three unknowns for each node, and there are three equations (a vector equation counts as two), the full set of discrete equations will uniquely specify the values of the unknowns (once the boundary conditions are included to remove linearly dependent equations).

\begin{figure}[t]
	\centering
	\tdplotsetmaincoords{50}{10}
\begin{tikzpicture}[tdplot_main_coords]
	\draw[->,loosely dashed] (-0.5,0,0) -- (5,0,0);
	\draw (5.5,0,0) node {$x$};
	\draw[->,loosely dashed] (0,-0.5,0) -- (0,5,0);
	\draw (0,5.5,0) node {$y$};
	\draw[->,loosely dashed] (0,0,-0.5) -- (0,0,2);
	\draw (0,0,2.5) node {$z$};
	
	\def\angleA{20};
	\def\angleB{80};
	\pgfmathsetmacro{\angleM}{(\angleA+\angleB)/2};
	\pgfmathsetmacro{\angleAlpha}{\angleA/2};
	\pgfmathsetmacro{\angleAM}{(\angleA+\angleM)/2};
	\pgfmathsetmacro{\angleBM}{(\angleB+\angleM)/2};
	
	\tdplotsetthetaplanecoords{\angleA}
	\tdplotdrawarc[tdplot_rotated_coords]{(0,0,0)}{4}{90}{180}{}{};
	\tdplotsetthetaplanecoords{\angleB}
	\tdplotdrawarc[tdplot_rotated_coords,densely dashed]{(0,0,0)}{4}{90}{180}{}{};
	
	\tdplotdrawarc{(0,0,0)}{4}{\angleA}{\angleB}{}{}
	
	\draw (0,0) -- (\angleA:4);
	\draw (0,0) -- (\angleB:4);
	\draw (0,0,0) -- (0,0,-4);
	
	\draw[densely dotted] (0,0) -- (\angleM:3);
	
	\tdplotdrawarc[densely dotted,<->]{(0,0,0)}{1}{0}{\angleM}{}{};
	\draw (\angleAlpha:1.3) node {$\alpha$};
	
	\tdplotdrawarc[densely dotted,<->]{(0,0,0)}{2}{\angleA}{\angleM}{}{};
	\draw (\angleAM:2.5) node {$\frac{1}{2}\delta\alpha$};
	
	\tdplotdrawarc[densely dotted,<->]{(0,0,0)}{2}{\angleM}{\angleB}{}{};
	\draw (\angleBM:2.6) node {$\frac{1}{2}\delta\alpha$};
	
	\begin{scope}[canvas is xy plane at z=-3]
		\draw (\angleM:2) node {$\Omega^{wedge}$};
	\end{scope}
 \end{tikzpicture}

 
	\caption{Illustration of the domain $\Omega^{wedge}$ with boundary $\partial\Omega^{wedge}$, this is the part of the wetted region that satisfies $\phi \in [\alpha-\frac{1}{2}\delta\alpha,\alpha+\frac{1}{2}\delta\alpha]$.}
	\label{f:wedge_integrate}
\end{figure}
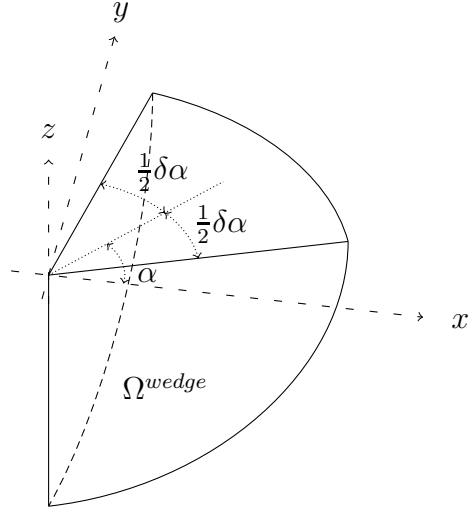

Next we consider the volume that will be integrated over. It must be a three dimensional region, the integrals over which being reducible to integrals over $\Omega_0$. The simplest choice is a wedge of the wetted region, i.e. the part of it that satisfies $\phi\in[\alpha-\frac{1}{2}\delta\alpha,\alpha+\frac{1}{2}\delta\alpha]$ for some $\alpha$, depicted in figure \ref{f:wedge_integrate}. This shall be called $\Omega^{wedge}$ and is considered as $\delta\alpha \rightarrow 0$ to obtain the region $\Omega_0$.

Note that the discrete form produced here is certainly not the only one possible for our system, \eqref{seq:dimless_system}, and not even the only scheme for our choice of interpolation. Stabilized schemes such as that in \cite{stable_darcy} exist but were not found to improve the accuracy of the solution.

\subsubsection{The Continuity Equation}

The dimensionless form of the continuity equation was found to be
\begin{subequations}\begin{alignat}{2}
	\nabla \cdot \boldsymbol{u}&=0 \hspace{1cm} &\forall \: \boldsymbol{r} &\in \Omega_0.	\tag{\ref{eq:continuity}}
\end{alignat}\end{subequations}
The weighted residual form of this is
\begin{align*}
	\int_{\Omega^{wedge}} \psi_i \nabla \cdot \boldsymbol{u} \dif V &=0
	\\
	\Rightarrow \qquad
	\int_{\Omega^{wedge}} (\nabla \psi_i) \cdot \boldsymbol{u} \dif V &=\int_{\partial \Omega^{wedge}} \psi_i \boldsymbol{u} \cdot \hat{\boldsymbol{n}} \dif S
\end{align*}
where $\partial \Omega^{wedge}$ is the surface of $\Omega^{wedge}$ and $\dif S$ is a surface element. As $\delta\alpha \rightarrow 0$, to leading order
\begin{gather*}
	\delta\alpha\int_{\Omega_0} (\nabla \psi_i) \cdot \boldsymbol{u} r \dif r \dif z 
	=\delta\alpha\int_{\partial \Omega_0} \psi_i \boldsymbol{u} \cdot \hat{\boldsymbol{n}} r \dif s 
	+\int_{\Omega_0} \psi_i \boldsymbol{u} \cdot \hat{\boldsymbol{\phi}}\Big|_{\phi=\alpha+\frac{1}{2}\delta\alpha} \dif r \dif z
	-\int_{\Omega_0} \psi_i \boldsymbol{u} \cdot \hat{\boldsymbol{\phi}}\Big|_{\phi=\alpha-\frac{1}{2}\delta\alpha} \dif r \dif z
	\\
	\Rightarrow \qquad
	\int_{\Omega_0} (\nabla \psi_i) \cdot \boldsymbol{u} r \dif r \dif z 
	=\int_{\partial \Omega_0} \psi_i \boldsymbol{u} \cdot \hat{\boldsymbol{n}} r \dif s,
\end{gather*}
where $s$ is the arc length along $\partial \Omega_0$. Let us now define the following
\begin{align}	\label{eq:discrete_ABc}
	A_{ij}&=\int_{\Omega_0} \frac{\partial \psi_i}{\partial r} \psi_j r \dif r \dif z,
	&
	B_{ij}&=\int_{\Omega_0} \frac{\partial \psi_i}{\partial z} \psi_j r \dif r \dif z,
	&
	c_i&=\int_{\partial \Omega_0} \psi_i \boldsymbol{u} \cdot \hat{\boldsymbol{n}} r \dif s. 
\end{align}
Thus, using the approximations in \eqref{seq:function_discrete}, we arrive at the discrete form of the continuity equation
\begin{equation}	\label{eq:discrete_continuity}
	\sum_j\left[A_{ij}u_j + B_{ij}v_j\right] = c_i.
\end{equation}

\subsubsection{Darcy's Equation}

The dimensionless form of Darcy's equation was found to be
\begin{subequations}\begin{alignat}{2}
	\boldsymbol{u}&=-\nabla (p +\gamma z) \hspace{1cm} & \forall \: \boldsymbol{r} &\in \Omega_0.	\tag{\ref{eq:darcy}}
\end{alignat}\end{subequations}
The weighted residual form of this is
\begin{align*}
	\int_{\Omega^{wedge}}\psi_i(\boldsymbol{u}+\nabla p +\gamma \hat{\boldsymbol{z}}) \dif V &=0
	\\
	\Rightarrow \qquad
	\int_{\Omega^{wedge}}\psi_i\boldsymbol{u}\dif V
	-\int_{\Omega^{wedge}}(\nabla\psi_i) p \dif V
	+\gamma \hat{\boldsymbol{z}}\int_{\Omega^{wedge}}\psi_i \dif V
	&=-\int_{\partial\Omega^{wedge}}\psi_i p \hat{\boldsymbol{n}}\dif S.
\end{align*}
As $\delta\alpha \rightarrow 0$, to leading order
\begin{equation*}
	\begin{array}{rr} \displaystyle \vspace{1mm}
		\delta\alpha\int_{\Omega_0}\psi_i\boldsymbol{u} r \dif r \dif z
		-\delta\alpha\int_{\Omega_0}(\nabla\psi_i) p  r \dif r \dif z
		+\delta\alpha\gamma \hat{\boldsymbol{z}}\int_{\Omega_0}\psi_i  r \dif r \dif z
		+ \ldots \\ \displaystyle \ldots
		+\int_{\Omega_0}\psi_i p \hat{\boldsymbol{\phi}}\Big|_{\phi=\alpha+\frac{1}{2}\delta\alpha}\dif r \dif z
		-\int_{\Omega_0}\psi_i p \hat{\boldsymbol{\phi}}\Big|_{\phi=\alpha-\frac{1}{2}\delta\alpha}\dif r \dif z
	\end{array}
	=-\delta\alpha\int_{\partial\Omega_0}\psi_i p \hat{\boldsymbol{n}}r \dif s.
\end{equation*}
Next notice that
\begin{equation*}
	\hat{\boldsymbol{\phi}}\Big|_{\phi=\alpha+\frac{1}{2}\delta\alpha}
	-\hat{\boldsymbol{\phi}}\Big|_{\phi=\alpha-\frac{1}{2}\delta\alpha}
	=
	-2\hat{\boldsymbol{r}}\Big|_{\phi=\alpha}\sin\left(\frac{\delta\alpha}{2}\right),
\end{equation*}
therefore
\begin{equation} \label{eq:integral_discrete_darcy}
	\int_{\Omega_0}\psi_i\boldsymbol{u} r \dif r \dif z
	-\int_{\Omega_0}(\nabla\psi_i) p  r \dif r \dif z
	-\hat{\boldsymbol{r}}\int_{\Omega_0}\psi_i p\dif r \dif z
	+\gamma \hat{\boldsymbol{z}}\int_{\Omega_0}\psi_i  r \dif r \dif z
	=-\int_{\partial\Omega_0}\psi_i p \hat{\boldsymbol{n}}r \dif s.
\end{equation}
Using $\hat{\boldsymbol{n}}=\hat{n}_r \hat{\boldsymbol{r}} + \hat{n}_z \hat{\boldsymbol{z}}$, let
\begin{equation}	\label{eq:discrete_CDabg}
	\begin{array}{c}	\displaystyle
		C_{ij}	= \int_{\Omega_0} \psi_i \psi_j r \dif r \dif z,
		\qquad
		D_{ij}	= \int_{\Omega_0} \psi_i \psi_j \dif r \dif z,
		\\	\displaystyle
		a_{i}	= \int_{\partial\Omega_0} \psi_i p \hat{n}_r r \dif s,
		\qquad
		b_{i}	= \int_{\partial\Omega_0} \psi_i p \hat{n}_z r \dif s,
		\qquad
		g_i		= \int_{\partial\Omega_0} \psi_i r \dif r \dif z.
	\end{array}
\end{equation}
To arrive at a discrete form that a computer can understand, it must be a set of scalar equations. In the bulk it does not matter what direction we choose for these scalar equations, but orthogonal directions are best. Thus we simply choose to scaler product \eqref{eq:integral_discrete_darcy} with $\hat{\boldsymbol{r}}$ and $\hat{\boldsymbol{z}}$, and then use the approximations in \eqref{seq:function_discrete}, to arrive at
\begin{subequations}\label{seq:discrete_darcy}\begin{align}
	\sum_j\left[C_{ij}u_j-(A_{ij}+D_{ij})p_j\right]	&=-a_i,	\\
	\sum_j\left[C_{ij}v_j-B_{ij}p_j\right]	&=-b_i-\gamma g_i.
\end{align}\end{subequations}

\subsection{Discrete form of the Boundary Conditions}

\subsubsection{Essential Boundary Conditions}

The discrete equations \eqref{eq:discrete_continuity} and \eqref{seq:discrete_darcy} are applicable at every node in the bulk. However on the boundary we wish to apply the boundary conditions in \eqref{seq:dimless_system}, and must do so to arrive at the correct number of linearly independent equations. We notice that the continuity equation applies a scalar restriction and thus specifies pressure, whilst Darcy's equation applies a vector restriction and thus specifies velocity (see \cite{fluid_poz} for a fuller justification). In the discrete form the instance of the equations with weight function $\psi_i$ specifies the value of the functions at node $i$. Therefore we can apply the boundary conditions as `essential boundary conditions', replacing the appropriate equation by the specification of the boundary condition. This removes the linearly dependent equations leaving us with the same number of equations as unknowns.

For the conditions
\begin{alignat}{2}
	p&=-1 \hspace{1cm} &\forall \: \boldsymbol{r} &\in \Gamma_0	\tag{\ref{eq:bc_gamma0}}	\\
	p&=0 \hspace{1cm} &\forall \: \boldsymbol{r} &\in \Gamma_2	\tag{\ref{eq:bc_gamma2}}
\end{alignat}
we see that, if node $i$ is on one of these boundaries, we replace \eqref{eq:discrete_continuity} with
\begin{subequations}\label{eq:discrete_bc_pressure}\begin{alignat}{2}
	p_i&=-1 \hspace{1cm}	&\forall \: i \: : \: \boldsymbol{r}_i &\in \Gamma_0		\\
	p_i&=0 \hspace{1cm}		&\forall \: i \: : \: \boldsymbol{r}_i &\in \Gamma_2	
\end{alignat}\end{subequations}

For the condition
\begin{alignat}{2}
	\boldsymbol{u}\cdot\hat{\boldsymbol{n}}&=0 \hspace{1cm} &\forall \: \boldsymbol{r} &\in \Gamma_1 \cup \Gamma_3,	\tag{\ref{eq:bc_gamma13}}
\end{alignat}
neither of \eqref{seq:discrete_darcy} are for $\boldsymbol{u}\cdot\hat{\boldsymbol{n}}$, we chose to have one for $\boldsymbol{u}\cdot\hat{\boldsymbol{r}}$ and the other for $\boldsymbol{u}\cdot\hat{\boldsymbol{z}}$. Thus we must use a new rotated form of the discrete equations. Let us define orthogonal constant unit vectors in the $r$-$z$ plane, $\hat{\boldsymbol{N}}$ and $\hat{\boldsymbol{T}}$, such that if node $i$ is on $\Gamma_1$ or $\Gamma_3$ then $\hat{\boldsymbol{N}}=\hat{\boldsymbol{n}}$ at $\boldsymbol{r}=\boldsymbol{r}_i$ and $\hat{\boldsymbol{T}}$ points in the anticlockwise direction around the boundary. Writing $\hat{\boldsymbol{N}}=\hat{N}_r\hat{\boldsymbol{r}}+\hat{N}_z\hat{\boldsymbol{z}}$ we see that $\hat{\boldsymbol{T}}=-\hat{N}_z\hat{\boldsymbol{r}}+\hat{N}_r\hat{\boldsymbol{z}}$. Thus the condition \eqref{eq:bc_gamma13} and tangential component of \eqref{eq:integral_discrete_darcy} are, respectively,
\begin{subequations}	\label{eq:discrete_bc_velocity}	\begin{align}
	\hat{N}_r u_i + \hat{N}_z v_i &= 0,	\\
	\sum_j\left[-\hat{N}_z C_{ij} u_j + \hat{N}_r C_{ij} v_j + (\hat{N}_z A_{ij} - \hat{N}_r B_{ij} + \hat{N}_z D_{ij}) p_j\right] &= \hat{N}_z a_i - \hat{N}_r b_i - \hat{N}_r \gamma g_i.
\end{align}\end{subequations}
and are used in place of \eqref{seq:discrete_darcy} for $i \: : \: \boldsymbol{r}_i \in \Gamma_1 \cup \Gamma_3$.

\subsubsection{Natural Boundary Conditions}

The objects $a_i$, $b_i$ and $c_i$ are boundary integrals of the unknowns $p$ and $\boldsymbol{u}\cdot\hat{\boldsymbol{n}}$. If the required variable is specified on the domain of integration as a boundary condition then this is a ``natural boundary condition'' and the integral is taken directly from the condition. If the value is not known then it can be obtained from the approximations in \eqref{seq:function_discrete}. Let
\begin{equation}	\label{eq:discrete_EF}
	E_{ij}=\int_{\partial\Omega_0} \psi_i \psi_j \hat{n}_r r \dif s,
	\qquad
	F_{ij}=\int_{\partial\Omega_0} \psi_i \psi_j \hat{n}_z r \dif s,
\end{equation}
therefore
\begin{equation}	\label{eq:discrete_abc}
	a_i=\sum_j E_{ij} p_j,
	\qquad
	b_i=\sum_j F_{ij} p_j,
	\qquad
	c_i=\sum_j \left[ E_{ij} u_j + F_{ij} v_j \right].
\end{equation}
Note that when using \eqref{eq:1d_master_integration} each term in the sum can be chosen to be of the natural or approximate form individually.

\subsubsection{Boundary Conditions at the Corners}

In the mesh there are nodes at each of the corners $C_0$, $C_1$, $C_2$ and $C_3$, and we must choose which of the boundary conditions to apply at each corner. However, in all tests the solutions produced with each boundary condition were indistinguishable. We have arbitrarily chosen to use pressure boundary conditions at all corners except for $C_0$ at which the normal velocity condition is applied.

\subsection{Summary of the Spatial Method}

First the spines are generated from the position of the wetting front, either from the initial condition $r=r(s)$, $z=z(s)$ or the set of elemental boundaries obtained from time stepping the wetting front. The spines are constructed from the values of $r_n$, where $r_0=r_f$, $r_1=r_0-S_{mh}$ and then the algorithm in \eqref{eq:spine_rapid_sep} is used, applying the constraints \eqref{eq:spine_max_sep}, \eqref{eq:spine_close}, \eqref{eq:spine_max_rate}, then \eqref{eq:spine_min_sep}. This proceeds until the condition \eqref{eq:spine_next_mode} is reached with $D=r_n-1$, at which point \eqref{eq:spine_aim_step} is used with $N$ from \eqref{eq:spine_aim_number} applying the constraint \eqref{eq:spine_max_rate}. The constant parameter for each spine ($\chi_n$) can then be found from \eqref{eq:spines_chi} with $(r,z)=(r_n,0)$, and from this all other parameters of the spine using \eqref{eq:spines_Rrho}, \eqref{eq:spines_theta}, \eqref{eq:spines_h} and \eqref{eq:spines_J}. The elements are then produced algorithmically between the spines using the method discussed in \S\ref{sss:spines_to_elements}.

From the above we have the mesh of nodes over which to calculate the solution, this is done by each node having three equations for its values. For the bulk nodes these equations are \eqref{eq:discrete_continuity} and \eqref{seq:discrete_darcy}, where the value of $i$ is the global node number of the considered node. For a node at a boundary that has the pressure condition \eqref{eq:bc_gamma0} and \eqref{eq:bc_gamma2} we use \eqref{eq:discrete_bc_pressure} and \eqref{seq:discrete_darcy}. For a node at a boundary that has the velocity condition \eqref{eq:bc_gamma13} we use \eqref{eq:discrete_continuity} and \eqref{eq:discrete_bc_velocity}. The variables involved in these equations are defined as integrals in \eqref{eq:discrete_ABc} and \eqref{eq:discrete_CDabg}. The terms in the integrands are defined in \eqref{eq:normal}, \eqref{eq:local_interp_master_def} and \eqref{eq:boundary_interp_master_def}, the integrals being performed over master coordinates using \eqref{eq:2d_master_integration} and \eqref{eq:1d_master_integration}, with the coordinate transformations having Jacobian \eqref{eq:jacobian} and derivative \eqref{eq:dsdomega}. The coordinate transformations these describe are defined in \eqref{eq:elem_iso_coord_trans} and \eqref{eq:bound_iso_coord_trans}. In cases where $a_i$, $b_i$ or $c_i$ are required on regions of the boundary where the integrated variable is not supplied as a boundary condition, \eqref{eq:discrete_abc} is used to find the value, where the variables are defined in \eqref{eq:discrete_EF}.

These equations are constructed as a matrix and then solved using standard methods.

\subsection{Numerical Testing}

\begin{figure}[ptb]
	\centering
	\begin{tabular}{c c}
		\begin{subfigure}[t]{0.47\textwidth}
			\centering
			\includegraphics[width=\textwidth]{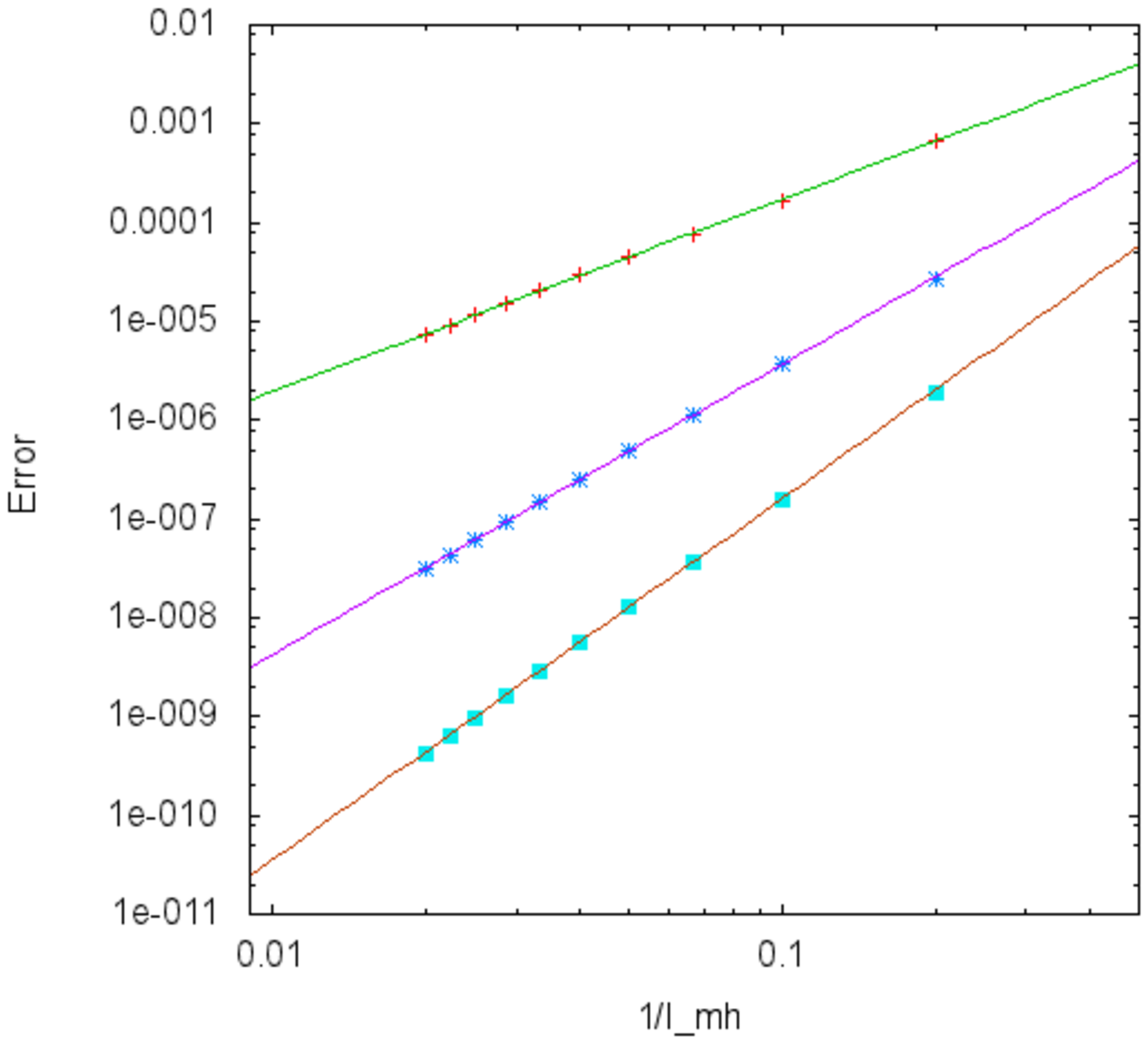}
			\caption{Convergence test for the mesh on a linear polynomial $P=1-z$. Plotted is: top, relative error on $v$ with line of best fit $1.54\cdot10^{-2}\cdot I_{mh}^{-1.95}$; middle, relative error on $p$ with line of best fit $3.21\cdot10^{-3}\cdot I_{mh}^{-2.94}$; bottom, integrated error on $p$ with line of best fit $7.40\cdot10^{-4}\cdot I_{mh}^{-3.66}$.}
			\label{f:test_linear}
		\end{subfigure}
		&
		\begin{subfigure}[t]{0.47\textwidth}
			\centering
			\includegraphics[width=\textwidth]{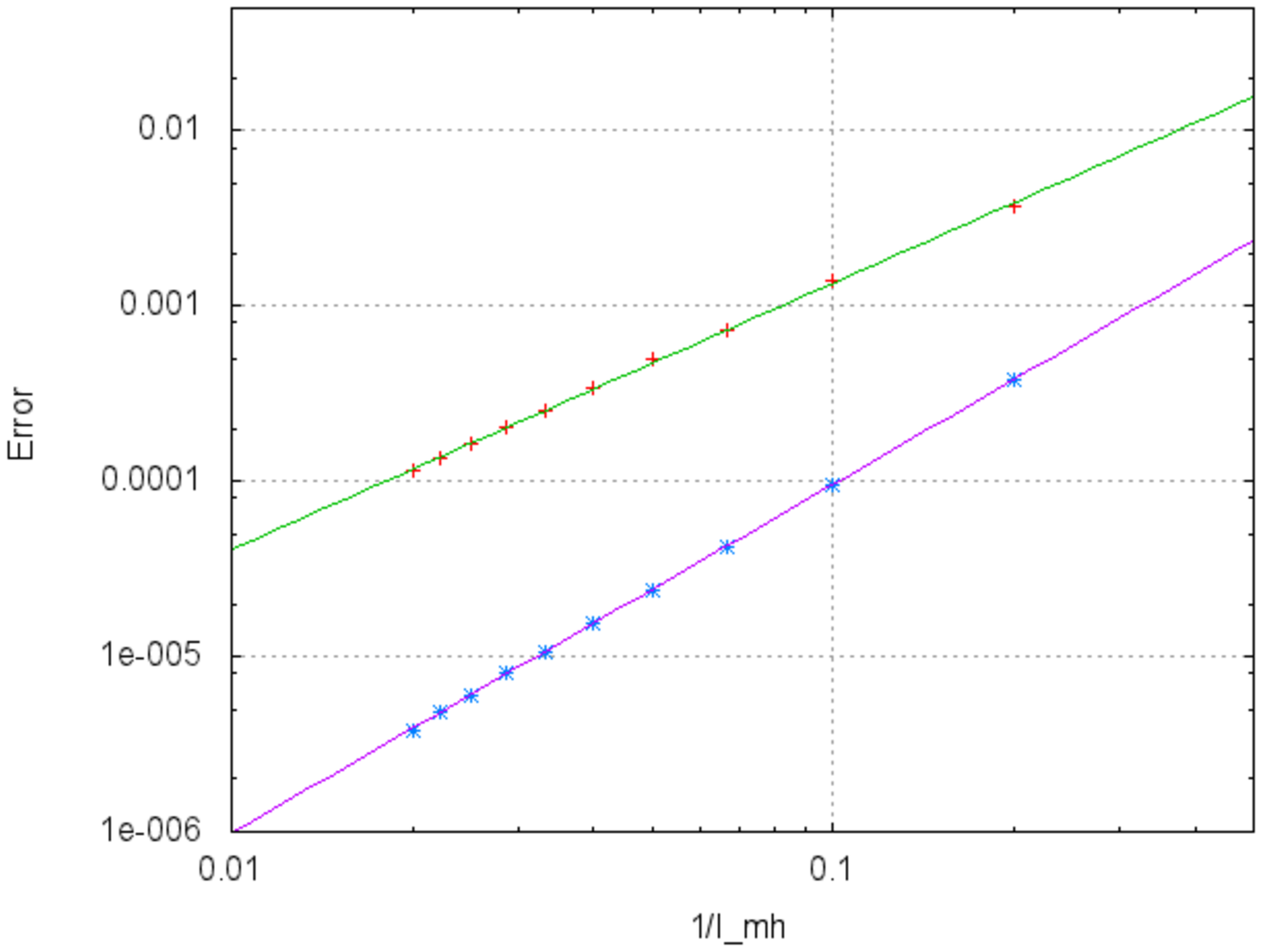}
			\caption{Convergence test for the mesh on a linear polynomial $P=10-z+r^2-2z^2+3r^2z-2z^3$. Plotted is: top, relative error on $p$ with line of best fit $4.48\cdot10^{-2}\cdot I_{mh}^{-1.52}$; bottom, integrated error on $p$ with line of best fit $9.38\cdot10^{-3}\cdot I_{mh}^{-1.99}$.}
			\label{f:text_cubic}
		\end{subfigure}
	\end{tabular}
	\caption{Plotted is the maximal or integrated errors for several runs of numerical solver. The parameters of the mesh are $R(t)=1$, $r(s)=2\cos(s)$, $z(s)=-2\sin(s)$, $S_{mh}=10^{10}$, $Z_{mh}=10^{10}$, $C_{mh}=1$, $M_{mh}=1.4$, and $R_{mh}=1.15$. Also, $\gamma=0$ and boundary conditions on $\hat{\boldsymbol{n}}\cdot\boldsymbol{u}$ are applied on $\Gamma_1$ and $\Gamma_3$.}
\end{figure}

\begin{figure}[ptb]
	\centering
	\begin{tabular}{c c}
		\begin{subfigure}[t]{0.47\textwidth}
			\centering
			\includegraphics[width=\textwidth]{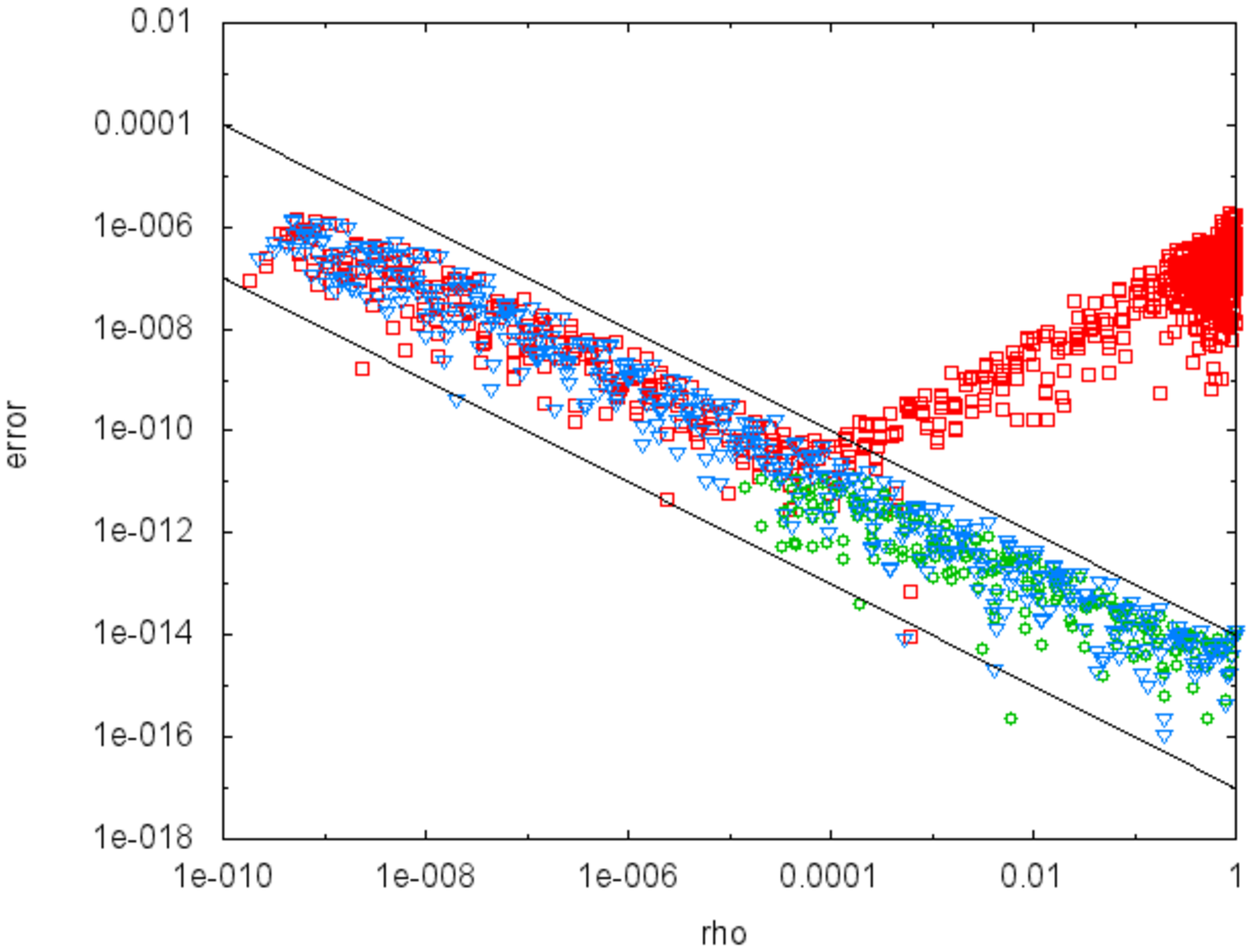}
			\caption{Absolute error on pressure. Lines are $10^{-14}/\rho$ and $10^{-17}/\rho$.}
			\label{f:test_zoom_p}
		\end{subfigure}
		&
		\begin{subfigure}[t]{0.47\textwidth}
			\centering
			\includegraphics[width=\textwidth]{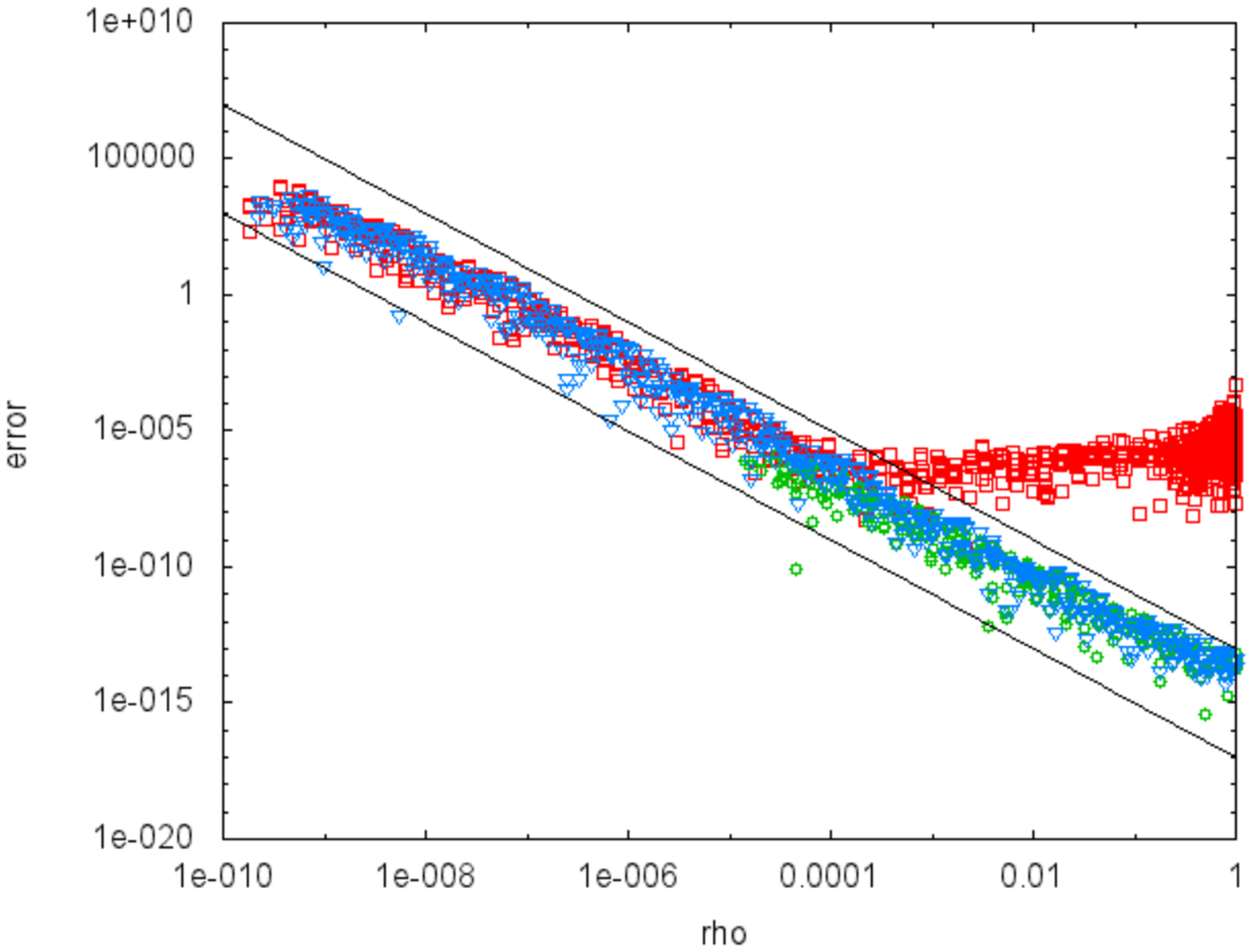}
			\caption{Absolute error on radial velocity. Lines are $10^{-13}/\rho^2$ and $10^{-17}/\rho^2$.}
			\label{f:text_zoom_u}
		\end{subfigure}
	\end{tabular}
	\caption{Errors on three runs of the solver, the error on every node plotted as a function of the distance from $C_2$, $\rho$. The data are for runs with $P=1$ and $Z_{mh}=10^{-5}$ ($\fullmoon$), $P=1$ and $Z_{mh}=10^{-10}$ ($\triangledown$) and $P=1-z$ and $Z_{mh}=10^{-10}$ ($\square$). $\gamma=0$ for all runs.}
\end{figure}

To perform error analysis on the code we consider exact analytic solutions to the bulk equations. From these analytic solutions boundary conditions can be deduced and the numerical solver run with these conditions. This should reproduce the analytic solution, and any difference between the analytic solution and the numerical solution is numerical error.
Combining \eqref{eq:continuity} and \eqref{eq:darcy} we obtain the equation for pressure $\nabla^2 p=0$. Considering a cubic polynomial solution in axisymmetric cylindrical coordinates the general form is, denoting the analytic solution by $P$, $U$ and $V$,
\begin{align}
	P(r,z)&=P_1+P_2z+P_3r^2-2P_3z^2-3P_4r^2z+2P_4z^3,	\\
	U(r,z)&=-2P_3r+6P_4rz,	\\
	V(r,z)&=-P_2-\gamma+4P_3z+3P_4r^2-6P_4z^2.
\end{align}
We consider three types of error: absolute, relative and integrated, which are for pressure
\begin{equation*}
	E_a=p-P, \qquad E_r=\frac{p-P}{P}, \qquad E_i=\frac{\int_{\Omega_0} (p-P)^2 \dif r \dif z}{\int_{\Omega_0} P^2 \dif r \dif z}
\end{equation*}
respectively.

First we examine the convergence properties as the mesh is refined. We do this by setting $S_{mh}$ to be very large such that all spines are constructed in a uniform distribution approximately $1/I_{mh}$ apart. By changing the value of $I_{mh}$ the convergence properties can be seen. Figure \ref{f:test_linear} shows how, for a linear polynomial, the convergence is very rapid. For higher order polynomials, as in figure \ref{f:text_cubic}, the convergence is slower, but for $I_{mh}>10$ the solution is acceptable. In regions where we are not having to refine the mesh the solution is well behaved and so this level of resolution should be sufficient.

The refinement around $C_1$ is steady and so will not produce errors until $S_{mh}\lesssim10^{-7}$, at which point the fact that the value of the Jacobian is less than $10^{-14}$ may start to produce errors from machine precision. The refinement around $C_2$ is much more rapid and the error from machine precision will become a problem much more rapidly. This is clearly shown in figure \ref{f:test_zoom_p}, where the absolute errors for one linear and two constant solutions are plotted at every node against $\rho=\sqrt{(r-1)^2+z^2}$. As the nodes get closer to $C_2$ the error on pressure grows as $\rho^{-1}$, where $\rho$ characterises the size of the elements. For the constant solutions this error is the only error and so it is shown across the range of values. For the linear solution there is a region in which the error due to the other inaccuracies dominates, but as the elements get smaller there comes a point when the error caused by the rapidly changing element size dominates. From figure \ref{f:text_zoom_u} we see that, for the linear polynomial, the convergence of velocity caused by the mesh refinement is zero in the region where pressure is converging. This is worrying since this is for a linear polynomial, which have the highest rate of convergence. For other solutions the error in velocity will likely increase throughout the refinement. However, for the singularity at $C_2$ the refinement is required to stabilise the solution, and the solution is not required at this point, only at the wetting front to perform the time-stepping. This aspect of our mesh is the least desirable and in any future work should be improved upon.

\subsection{Time Stepping the Wetting Front}

The time stepping of the front will be discussed in several parts. First we shall discuss the stepping of a front with a set of known velocities, then the process by which velocities are extracted from a solution, and finally the scheme of time-stepping that is to be used. To number the nodes on the wetting front we shall use the subscript $i$, this should not cause confusion with the global node numbers since we will not be using them in this subsection. The numbering scheme will number the node at $C_1$ as $0$ and use consecutive natural numbers as we move towards the node at $C_0$ up to a highest value of $N$.

Firstly, time stepping once the velocities are known. Let the velocity of the surface at node $i$ be $v_{s,i}$, the coordinate of the node be $\boldsymbol{r}_i (t)$, the unit normal at this node be $\hat{\boldsymbol{n}}_i$ and the amount to time step be $\Delta t$. The position of the nodes after time-stepping is
\begin{subequations}\begin{equation}
	\boldsymbol{r}_i(t+\Delta t)=\boldsymbol{r}_i(t) + v_{s,i} \hat{\boldsymbol{n}}_i \Delta t \hspace{1cm} \forall \: i \in \{1,2,\ldots,N-1\}.
\end{equation}
At either end the stepping is performed using the assumption that the velocities are locally constant, which means that they are stepped by
\begin{align}
	r_0(t+\Delta t)&=r_0(t) + \frac{v_{s,0}}{\hat{n}_{r,i}} \Delta t	\\
	z_N(t+\Delta t)&=z_N(t) + \frac{v_{s,0}}{\hat{n}_{z,i}} \Delta t.
\end{align}\end{subequations}

The velocities can be found from the solution at the time $t$ either by taking the values of the solution at the nodes that the problem is solved over or, if the node to step is not part of the solution mesh, by simple interpolation using \eqref{seq:function_discrete_boundary}. However, this will cause problems since the error on the node fluctuates from one node to the next, i.e. if the error on the normal velocity is $\delta$ at node $i$ then it will be $-\delta$ at nodes $i-1$ and $i+1$. This error would cause the wetting front at the next time step to have fluctuations in it, which has been found to cause situations where the fluctuations build and build. To solve this problem a simple smoothing algorithm is employed. The use of a standard splines smoother may also be suitable, but that is not what has been used. We smooth not only the velocities, but also the normals to aid the stepping if fluctuations do start to build, to produce the smoothed variables $\bar{v}_{s,i}$ and $\hat{\bar{\boldsymbol{n}}}$. The smoothing algorithm to remove the fluctuating errors is presented below for velocity, and is the same for the normals.
\begin{align}
	\bar{v}_{s,i}&=\frac{2 v_{s,i}+v_{s,i+1}+v_{s,i-1}}{4}	\hspace{1cm} \forall \: i \in \{2,3,\ldots,N-1\}	\\
	\bar{v}_{s,1}&=\frac{1}{2}\left( v_{s,1} + \sum_{j=0}^{3} \bar{v}_{s,4-j} \psi_j^B(\omega(\chi(\boldsymbol{r}_1)))  \right)	\\
	\bar{v}_{s,0}&=\frac{1}{2}\left( v_{s,0} + \sum_{j=0}^{3} \bar{v}_{s,4-j} \psi_j^B(\omega))  \right)	\\
	\bar{v}_{s,N}&=\frac{1}{2}\left( v_{s,N} + \sum_{j=0}^{3} \bar{v}_{s,N-1-j} \psi_j^B(\omega(\chi(\boldsymbol{r}_N)))  \right)
\end{align}
In the equation for $\bar{v}_{s,1}$, $\omega(\chi(\boldsymbol{r}_1))$ denotes the process by which the value of $\chi(\boldsymbol{r}_1)$ is found, and then the Newton-Raphson method is used on the the boundary made up of nodes $4,3,2$ to find the value of $\omega$ that has the correct value of $\chi$. This process is described in subsection \ref{ss:Spines}. Similar notation is used in the equation for $\bar{v}_{s,N}$, except that the boundary is made up of the nodes $N-1,N-2,N-3$. In the equation for $\bar{v}_{s,0}$, the value of $\omega$ is found by solving for $z=0$ in the boundary made up of nodes $4,3,2$. This process removes the main contribution of the error along the bulk of the wetting front during time stepping in the simplest manner whilst reducing the spatial accuracy of the solution slightly. At the corners we interpolate along to perform the averaging. In the current implementation we use this stabilisation twice on the velocities before performing the time-stepping.

We time-step using Heun's method (also known as the improved Euler's method) where first the velocities at one instant of time are found, then a trial time step is performed and the velocities are found at this time. The actual time-step is performed by using the average of the velocity at time $t$ and at the trial step. The velocity at the trial time step must be found at the node that was projected from the front at time $t$, and not at the nodes that now form the mesh, which is done by interpolation. Note that here when we say velocity we mean the two component vector, i.e. the normal velocity \textit{and} the normal direction. This method is of second order convergence, which is deemed to be sufficient for our problem. 

The size of the time-step that is used is not fixed, but is adjusted to restrict the rate of change of the contact angle CA2 and the rate of change of the local curvature of the surface at both the trial step and a secondary trial step taken from the trial step forward $\Delta t$.

\subsection{Measurements}

In our results we will discuss the volume flux into the wetted region, the total volume of the wetted region and the contact angle variation. These are calculated as follows. The volume influx is
\begin{equation}
	F=\int_{0}^{2\pi} \dif\phi \int_{\Gamma_2} r \dif l \left(\boldsymbol{u} \cdot -\hat{\boldsymbol{n}}\right)=-2\pi \int_{0}^{1} v  r \dif r.
\end{equation}
The total volume of the wetted region is 
\begin{equation}
	V=\int_{0}^{2\pi} \dif \phi \int_{\Omega_0} r \dif r \dif z = 2\pi \int_{\Omega_0} r \dif r \dif z.
\end{equation}
The contact angle of interest is CA2 (from the introduction). This is the angle subtended at $C_1$ between $\Gamma_0$ and $\Gamma_1$, and will be denoted $\theta_1$. Using the notation from the previous section, where $i$ represents the node number along the wetting front, this is calculated by
\begin{equation}
	\theta_1=\frac{1}{5}\sum_{i=1}^{5} \arctan\left(\frac{z_{i}-z_{i-1}}{r_{i}-r_{i-1}}\right).
\end{equation}
Due to the curvature of the wetting front, this will always produce a slight underestimate, but this can be taken into consideration when evaluating the results. Also, the variation of the mesh at each time step will cause the approximation to fluctuate. We can ignore this since it it is an artefact of our method of extracting data from our numerical scheme and not the scheme itself.

\section{Numerical and Asymptotic Analysis}\label{s:simple}

\subsection{Initial Conditions}

We shall first examine numerical solutions for a single instant of time, for which the initial condition for the wetting front shall be the only wetting front geometry. Then we will look at asymptotic analysis that justifies the behaviour that we see. Finally we will look at some time evolutions of the wetting front. We make the simplification that the initial $\Gamma_0$ is a segment of an ellipse and subtends a contact angle $\theta_1$ to the boundary $\Gamma_1$, this angle is CA2 from the introduction. Let the radial coordinate of $C_1$ be $r_f$ and the intersection of the wetting front with the axis of symmetry be at $z=-H$. The equation for the wetting front is
\begin{subequations}\begin{align}
	r(s)&=b\cos(s+s_0),	\\
	z(s)&=a-(a+H)\sin(s+s_0),	\\
\intertext{where}
	a&=\frac{H^2}{r_f\tan(\theta_1)-2H},	\\
	b&=(a+H)\sqrt{\frac{r_f}{a\tan(\theta_1)}},	\\
	\sin(s_0)&=\frac{a}{a+H}.
\end{align}\end{subequations}

\subsection{Pressure and Velocity Distributions} \label{ss:pv_dist}

\begin{figure}[ptb]
	\centering
	\begin{tabular}{c}
		\begin{subfigure}[t]{0.8\textwidth}
			\centering
			\includegraphics[width=\textwidth]{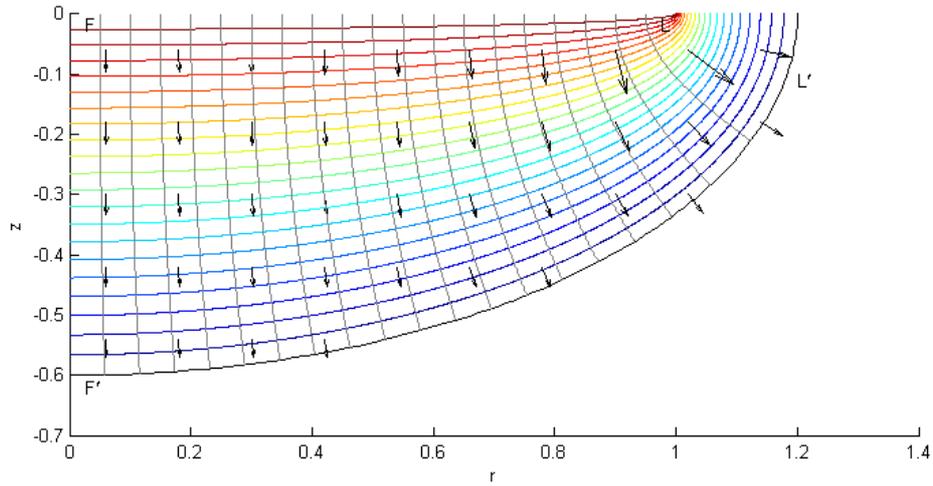}
			\caption{$\gamma=0$}
			\label{f:uvp_small_ng}
		\end{subfigure}
		\\
		\begin{subfigure}[t]{0.8\textwidth}
			\centering
			\includegraphics[width=\textwidth]{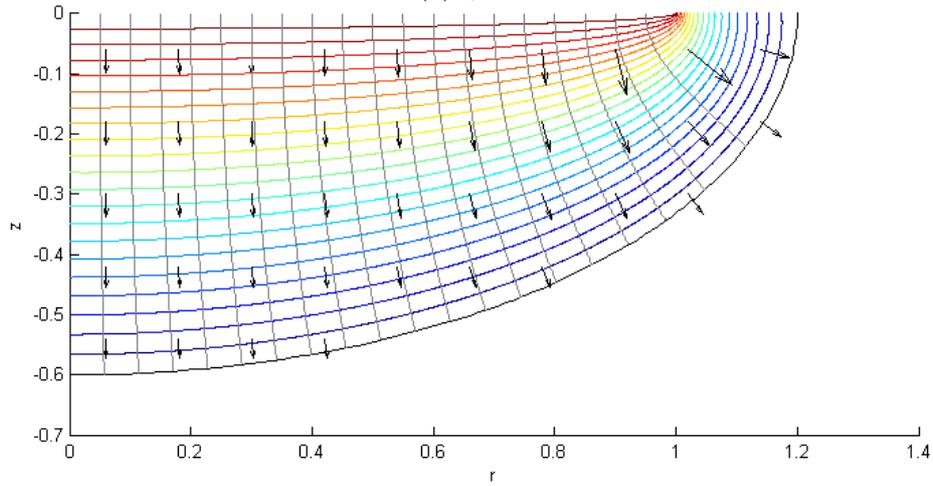}
			\caption{$\gamma=0.2$}
			\label{f:uvp_small_g}
		\end{subfigure}
	\end{tabular}
	\caption{Two plots of velocity and pressure for the region $\theta_1=\pi/2$, $r_f=1.2$ and $H=0.6$, (a) for without gravity and (b) for with gravity. In this domain the pressure gradient is sufficiently high that the gravitational effect is negligible and the pressure and velocity distributions are almost identical.}
	\label{sf:uvp_small}
\end{figure}

\begin{figure}[ptb]
	\centering
	\includegraphics[width=0.8\textwidth]{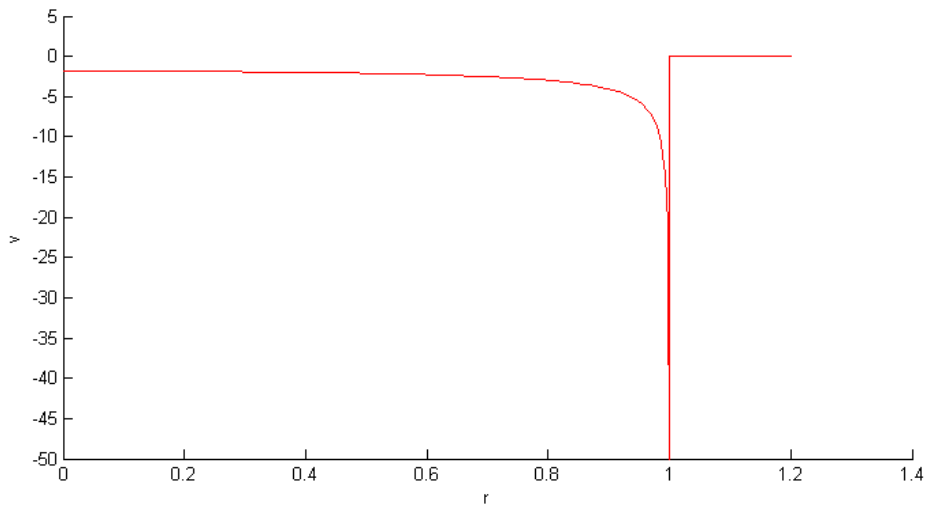}
	\caption{The distribution of axial velocity along the surface of the porous substrate, $z=0$, in the case plotted in figure \ref{f:uvp_small_ng}.}
	\label{f:v_small}
\end{figure}

\begin{figure}[ptb]
	\centering
	\begin{tabular}{c}
		\begin{subfigure}[t]{0.95\textwidth}
			\centering
			\includegraphics[height=0.43\textheight]{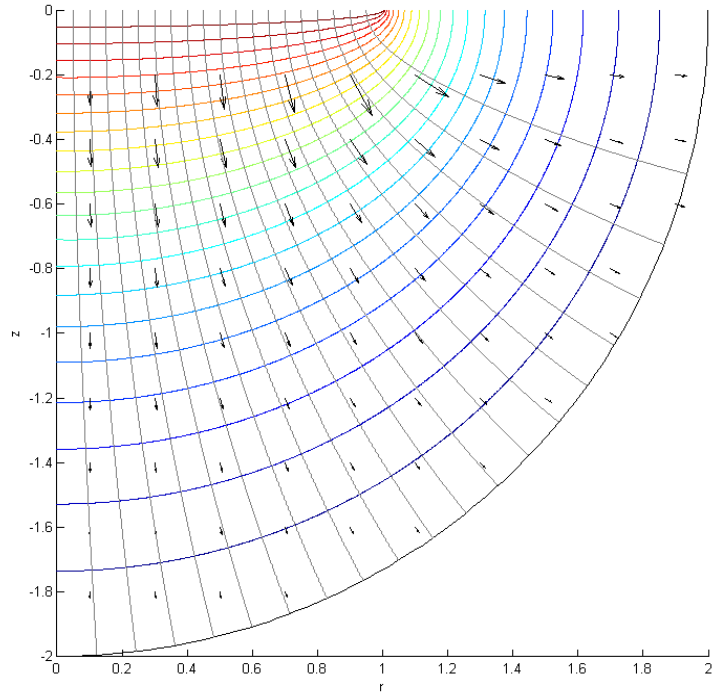}
			\caption{$\gamma=0$}
			\label{f:uvp_medium_ng}
		\end{subfigure}
		\\
		\begin{subfigure}[t]{0.95\textwidth}
			\centering
			\includegraphics[height=0.43\textheight]{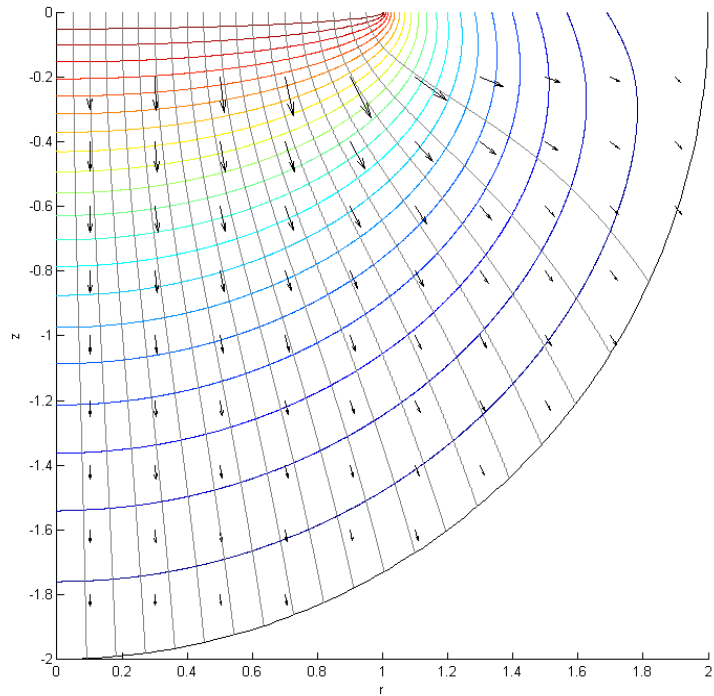}
			\caption{$\gamma=0.2$}
			\label{f:uvp_medium_g}
		\end{subfigure}
	\end{tabular}
	\caption{Two plots of velocity and pressure for the region $\theta_1=\pi/2$, $r_f=2$ and $H=2$, (a) for without gravity and (b) for with gravity. In this domain we see that gravity causes the fluid to flow downward as can be seen from the streamlines, especially the streamline at largest $r$. }
	\label{sf:uvp_medium}
\end{figure}

\begin{figure}[ptb]
	\centering
	\begin{tabular}{c}
		\begin{subfigure}[t]{0.95\textwidth}
			\centering
			\includegraphics[height=0.43\textheight]{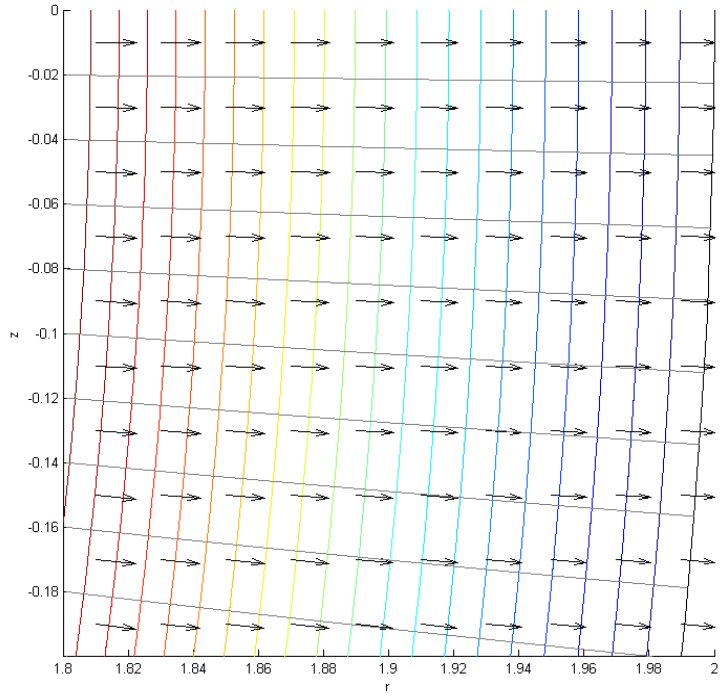}
			\caption{$\gamma=0$}
			\label{f:uvp_z_medium_ng}
		\end{subfigure}
		\\
		\begin{subfigure}[t]{0.95\textwidth}
			\centering
			\includegraphics[height=0.43\textheight]{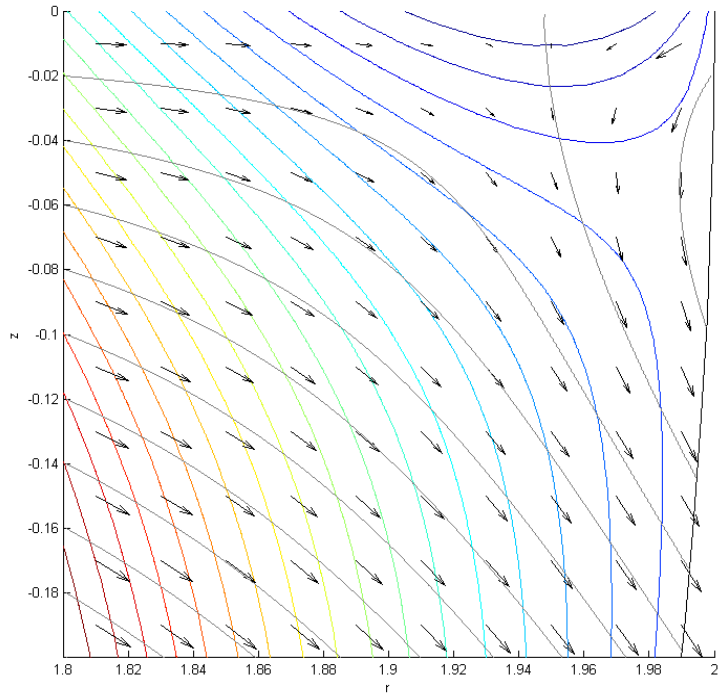}
			\caption{$\gamma=0.2$}
			\label{f:uvp_z_medium_g}
		\end{subfigure}
	\end{tabular}
	\caption{An enlargement around $C_1$ for the plots in figure \ref{sf:uvp_medium}. For the case without gravity the streamlines all intersect with the free surface approximately at the perpendicular, meaning that the free surface will propagate approximately uniformly. With gravity there is a region of the free surface that is not fed by the drawing area, the region near the contact line receding and that below advancing. All of the fluid in this region is noticeably affected by gravity.}
	\label{sf:uvp_z_medium}
\end{figure}

\begin{figure}[ptb]
	\centering
	\begin{tabular}{c}
		\begin{subfigure}[t]{0.95\textwidth}
			\centering
			\includegraphics[height=0.43\textheight]{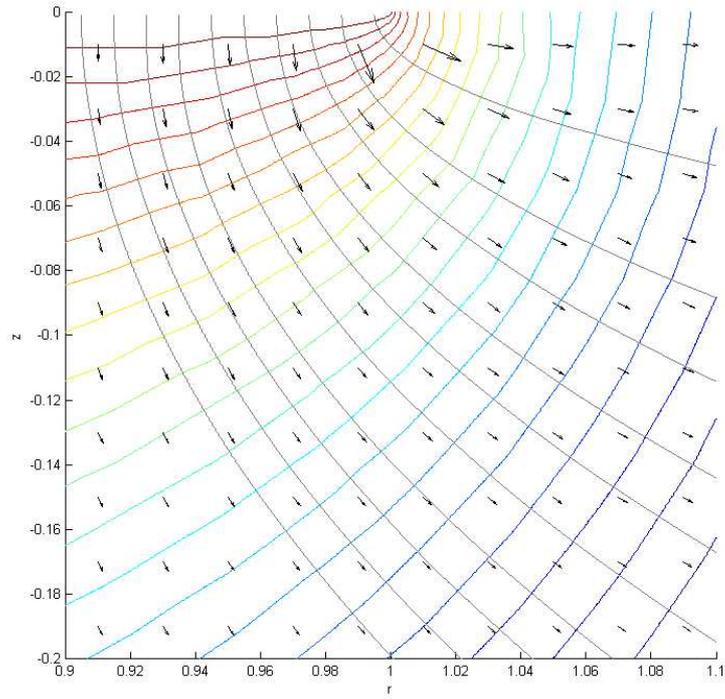}
			\caption{$\gamma=0$}
			\label{f:uvp_zc2_medium_ng}
		\end{subfigure}
		\\
		\begin{subfigure}[t]{0.95\textwidth}
			\centering
			\includegraphics[height=0.43\textheight]{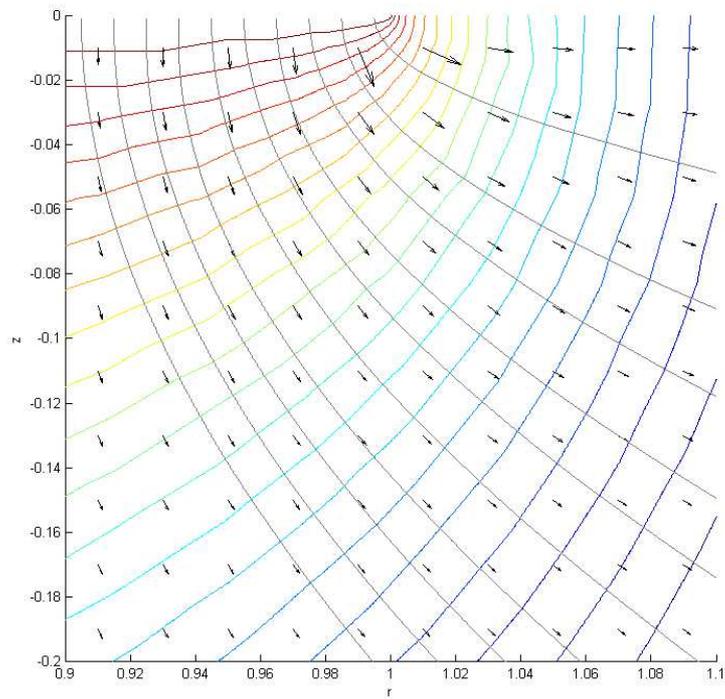}
			\caption{$\gamma=0.2$}
			\label{f:uvp_zc2_medium_g}
		\end{subfigure}
	\end{tabular}
	\caption{An enlargement around $C_2$ for the plots in figure \ref{sf:uvp_medium}. The plots appear similar, the high pressure gradient means that the effect of gravity is negligible. It is clear that there is a high volume flux through $\Gamma_2$ local to $C_2$.}
	\label{sf:uvp_zc2_medium}
\end{figure}

\begin{figure}[ptb]
	\centering
	\begin{tabular}{c}
		\begin{subfigure}[t]{0.95\textwidth}
			\centering
			\includegraphics[height=0.43\textheight]{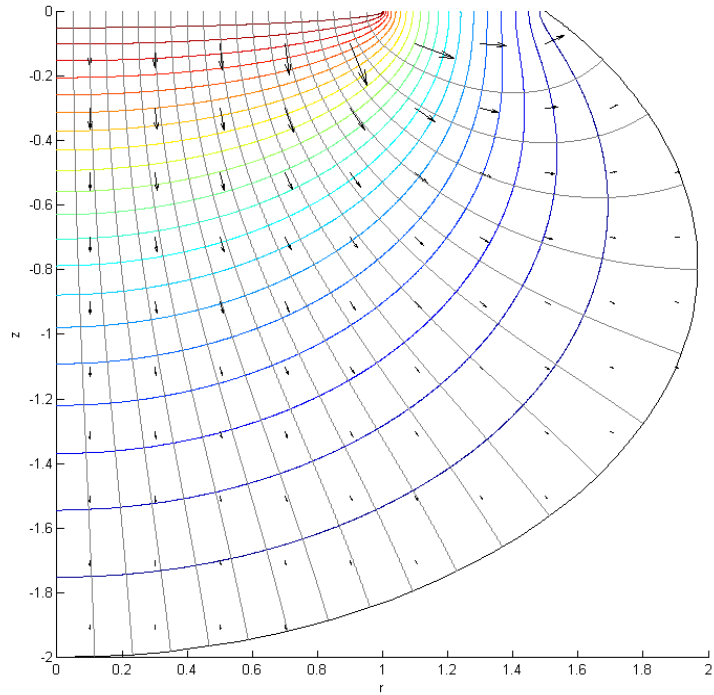}
			\caption{$\gamma=0$}
			\label{f:uvp_obtuse_ng}
		\end{subfigure}
		\\
		\begin{subfigure}[t]{0.95\textwidth}
			\centering
			\includegraphics[height=0.43\textheight]{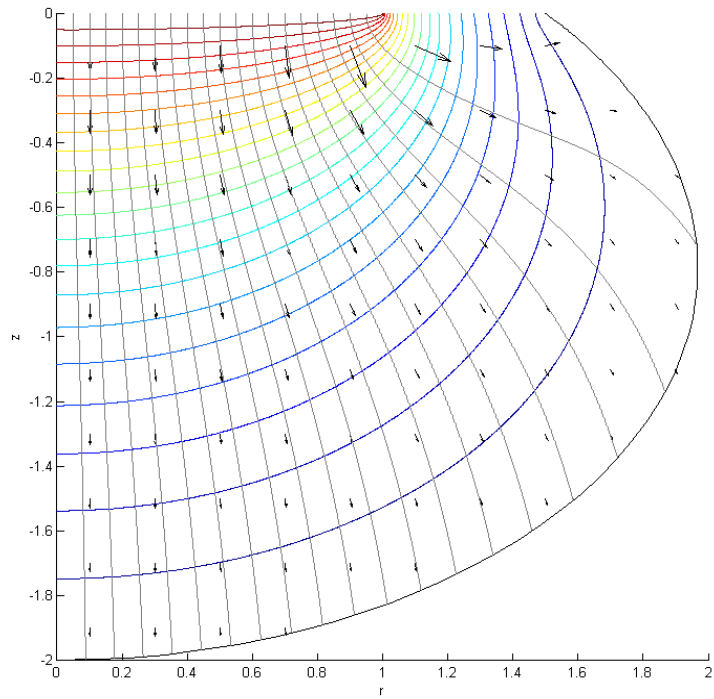}
			\caption{$\gamma=0.2$}
			\label{f:uvp_obtuse_g}
		\end{subfigure}
	\end{tabular}
	\caption{Two plots of velocity and pressure for the region $\theta_1=0.8\pi$, $r_f=1.5$ and $H=2$, (a) for without gravity and (b) for with gravity. This plot has the same qualitative features as figure \ref{sf:uvp_medium}.}
	\label{sf:uvp_obtuse}
\end{figure}

\begin{figure}[ptb]
	\centering
	\begin{tabular}{c}
		\begin{subfigure}[t]{0.95\textwidth}
			\centering
			\includegraphics[height=0.43\textheight]{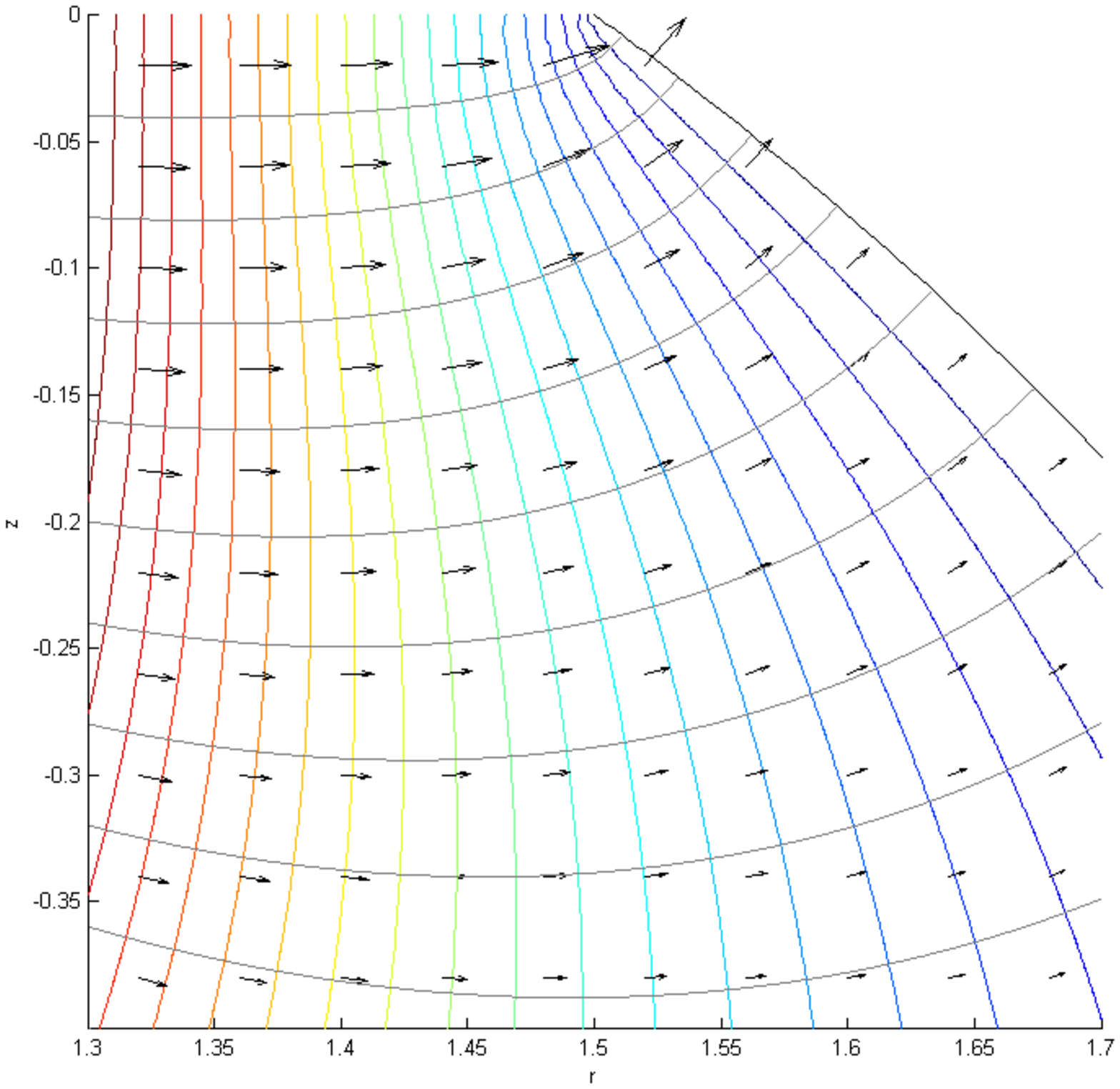}
			\caption{$\gamma=0$}
			\label{f:uvp_z_obtuse_ng}
		\end{subfigure}
		\\
		\begin{subfigure}[t]{0.95\textwidth}
			\centering
			\includegraphics[height=0.43\textheight]{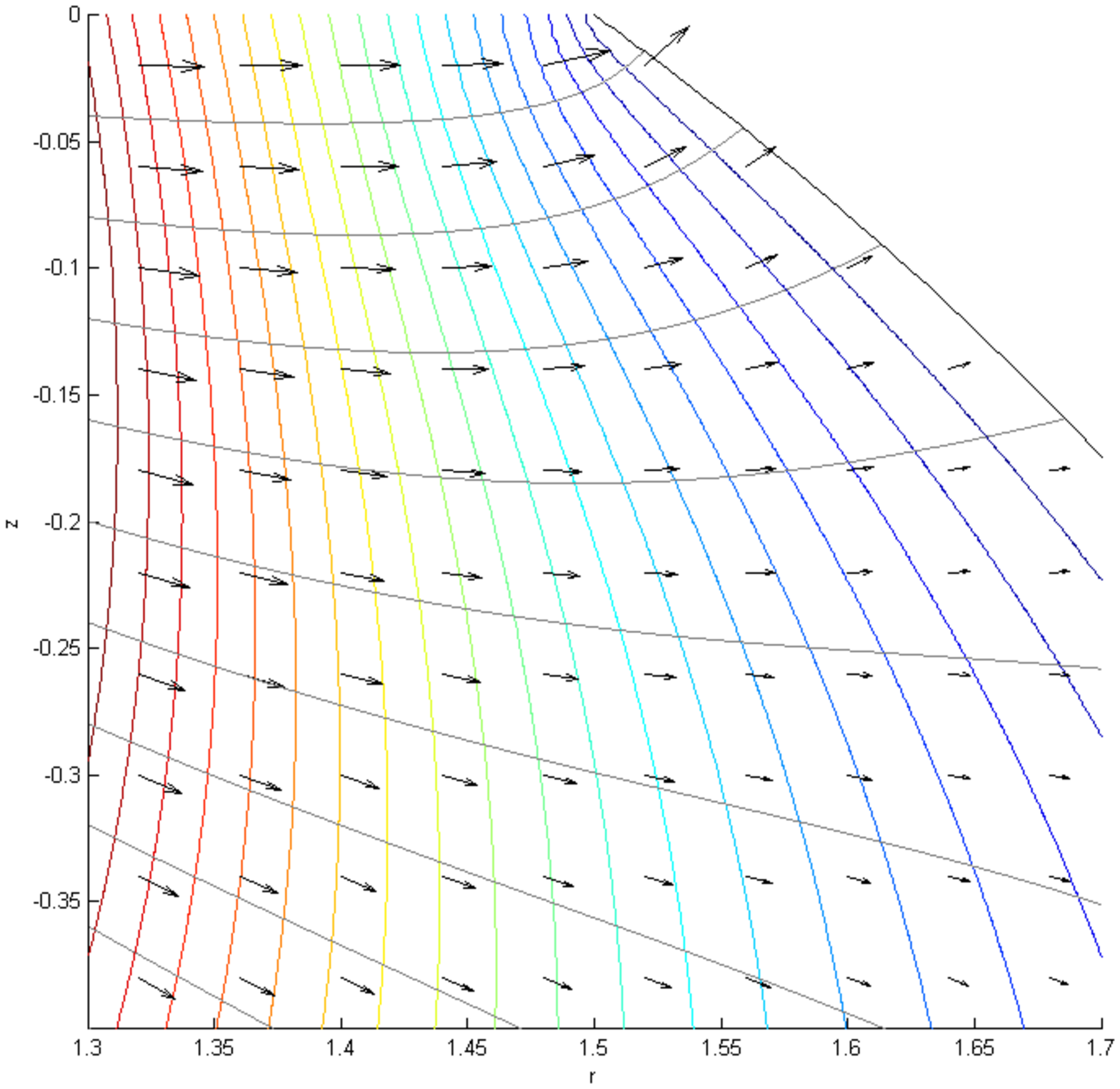}
			\caption{$\gamma=0.2$}
			\label{f:uvp_z_obtuse_g}
		\end{subfigure}
	\end{tabular}
	\caption{An enlargement around $C_1$ for the plots in figure \ref{sf:uvp_obtuse}. We see that the plots are qualitatively the same local to $C_1$, the dominant effect being that the velocity of the wetting front is highest near $C_1$ and reduces along.}
	\label{sf:uvp_z_obtuse}
\end{figure}

In this section we plot the velocity and pressure distributions within the wetted region for various wetting fronts, to give the reader a qualitative understanding of the solution before we perform the asymptotic analysis. We do this for solutions that do not include gravity ($\gamma=0$) and for a small, but certainly not negligible, gravitational effect ($\gamma=0.2$). See figure \ref{f:uvp_small_ng} as an example of such a plot. The plot is in the $r$-$z$ plane, with the wetting front plotted in black. Pressure contours are plotted in colours that represent the value of pressure, red for high pressures and blue for low pressures. Example streamlines are plotted in grey, and a small number of velocity vectors are plotted in black. In this plot we also label some intervals of the boundary which will be used for other cases but not labelled on their plots. The intervals $F$ and $F'$ extend from the axis of symmetry to the first streamline plotted on $\Gamma_2$ and $\Gamma_0$ respectively. $L$ and $L'$ are the parts of $\Gamma_2$ and $\Gamma_0$ between the last streamline plotted and the contact line $C_2$ and $C_1$ respectively.

This first pair of plots, figure \ref{sf:uvp_small}, reveal that, for a small domain, the pressure gradient dominates the effect of gravity such that the plots appear almost identical. Looking more closely, the separation of the pressure contours close to the wetting front is approximately the same along the length of the wetting front. Due the the velocity being proportional to the pressure gradient the wetting front should propagate approximately uniformly along its length, at least at first. The pressure contours close to the point $C_2$ at $(1,0)$ are very closely packed, revealing enormous velocities close to this point. Finally, the streamlines that enter the wetted region at large $r$ spread out much more than those that enter at small $r$. As the wetting front advances, the volume increase due to the advancement of a segment of the wetting front between to streamlines must come from the influx of volume through the drawing area between these same streamlines. Therefore, the volume flux through the section of the drawing area $L$ must be sufficient to supply the segment of the wetting front $L'$. The area it has to supply is enormous in comparison to the area that is supplied by the section $F$, which is $F'$, especially when axisymmetry is taken into account. The volume flux though the drawing area is vastly greater near $r=1$ than it is near $r=0$. This is seen clearly in figure \ref{f:v_small}, the axial velocity is singular at $C_2$, which is why $L$ can supply enough fluid to feed $L'$.

Figure \ref{sf:uvp_medium} show how, for a larger domain, gravity has an effect. The pressure contours are spread out close to the wetting front, revealing the smaller pressure gradient which is now of the same order as the gravitational effect. We also see that, for the plot with gravity, the pressure gradient close to $\Gamma_1$ is angled upward to counter gravity, which is the result of enforcing that the normal velocity on this surface is zero. The streamlines are angled downwards in the case with gravity in comparison to the case without, showing how the fluid is falling under its action. In the plot \ref{f:uvp_medium_g} we see even more starkly how much greater the segment of the front fed by the section of the drawing area $L$ is than the segment fed by $F$. In fact, in this case, it is too large. Figure \ref{f:uvp_z_medium_g} shows an enlargement around $C_1$. We see that there is a region of the wetting front around $C_1$ that is not fed by the drawing area, and is cut off by a streamline that starts at around $(1.96,0)$. In this cut off region the fluid at the top is receding and at the bottom advances, as the fluid 'slumps' under the action of gravity. The plot without gravity , figure \ref{f:uvp_z_medium_ng}, does not show this behaviour, instead the pressure gradient is very uniform and the velocities at the wetting front are approximately perpendicular to it. In this case the front will advance uniformly.

Examining the behaviour local to $C_2$, figure \ref{sf:uvp_zc2_medium} again reveals that the velocities close to contact line are very large. In addition, the streamlines that start at a larger value of $r$ spread out more from their neighbours more than those at smaller $r$. It is also of note that the pressure and velocity distribution around this point is not affected by gravity, due to the huge pressure gradients.

The next case that we consider is that of an obtuse contact angle, $\theta_1>\pi/2$. The large scale pressure and velocity distribution is qualitatively the same as for the previous case, with gravity causing the velocities far from the drawing area to fall rather than rise. However, the pressure distribution near $\Gamma_1$ appears the similar in the two cases. Figure \ref{sf:uvp_z_obtuse} is an enlargement around $C_1$, and it is seen that in this region the pressure and velocity fields are indeed very similar, appearing identical very close to $C_1$. The pressure contours are very closely spaced around $C_1$, and spread out as we move along the wetting front, from this we deduce that the velocity is very large at the contact line and is smaller further from it, causing the contact angle to reduce as the front propagates. 

In this section we have found that there is some interesting behaviour close to the points $C_1$ and $C_2$. We shall next look at the results in these regions and investigate the leading order terms that dominate the behaviour.

\subsection{Local Behaviour in Numerical Results}	\label{ss:num_asymp}

\begin{figure}[ptb]
	\centering
	\begin{tabular}{c c}
		\begin{subfigure}[t]{0.47\textwidth}
			\centering
			\includegraphics[width=\textwidth]{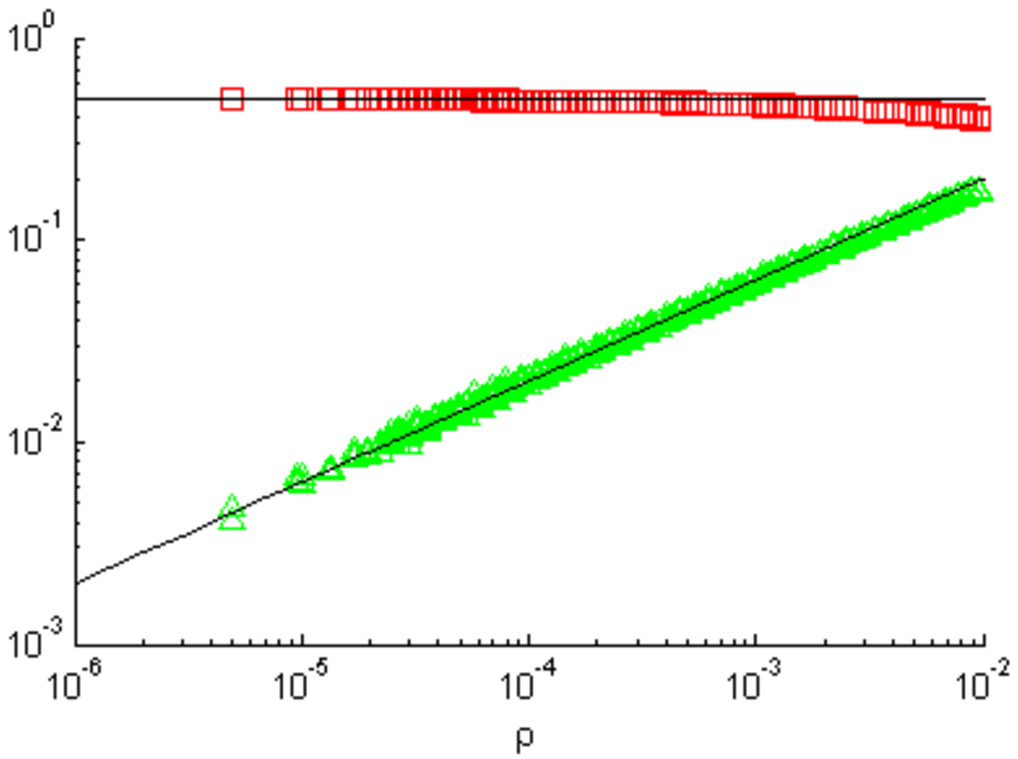}
			\caption{$\square$ is $-u/\sqrt{3}$ with line $0.5$ and $\triangle$ is $-v/(\sin(3\theta/2)\sin(\theta-\pi/3)+\cos(3\theta/2)\cos(\theta-\pi/3))$ with curve $2\sqrt{\rho}$.}
			\label{f:simple_c1_acute_rh}
		\end{subfigure}
		&
		\begin{subfigure}[t]{0.47\textwidth}
			\centering
			\includegraphics[width=\textwidth]{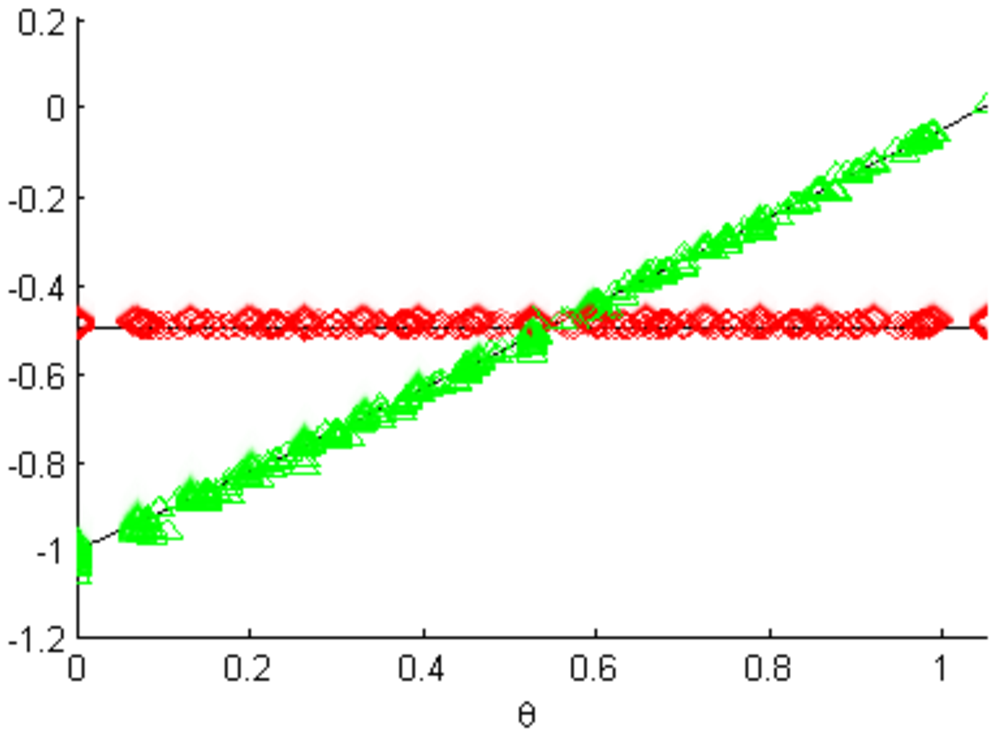}
			\caption{$\Diamond$ is $u/\sqrt{3}$ with line at $-0.5$ and $\triangle$ is $v/\sqrt{\rho}$ with curve $-2[\sin(3\theta/2)\sin(\theta-\pi/3)+\cos(3\theta/2)\cos(\theta-\pi/3)]$.}
			\label{f:simple_c1_acute_th}
		\end{subfigure}
	\end{tabular}
	\caption{Plots for $\theta_1=\pi/3$, $\gamma=0.5$ around $C_1$.}
	\label{sf:simple_c1_acute}
\end{figure}

\begin{figure}[ptb]
	\centering
	\begin{tabular}{c c}
		\begin{subfigure}[t]{0.47\textwidth}
			\centering
			\includegraphics[width=\textwidth]{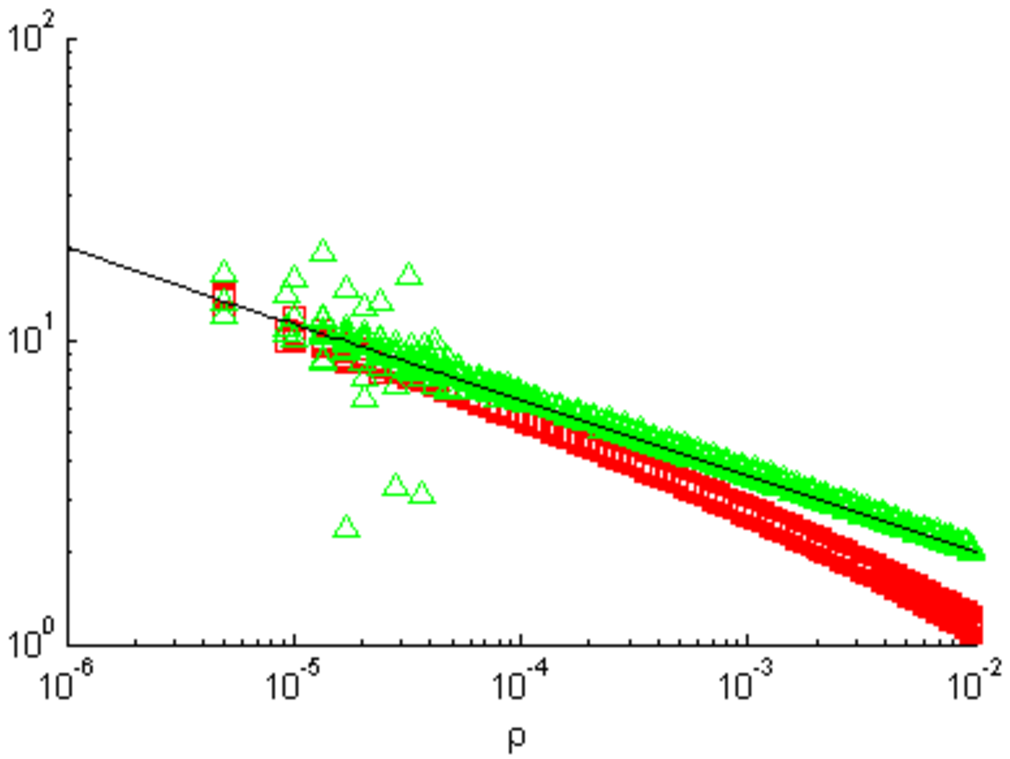}
			\caption{$\square$ is $u/(-\sin(3\theta/4)\cos(\theta-2\pi/3)+\cos(3\theta/4)\sin(\theta-2\pi/3))$ and $\triangle$ is $v/(\sin(3\theta/4)\sin(\theta-2\pi/3)+\cos(3\theta/4)\cos(\theta-2\pi/3))$ with curve $0.64/\rho^{1/4}$.}
			\label{f:simple_c1_obtuse_rh}
		\end{subfigure}
		&
		\begin{subfigure}[t]{0.47\textwidth}
			\centering
			\includegraphics[width=\textwidth]{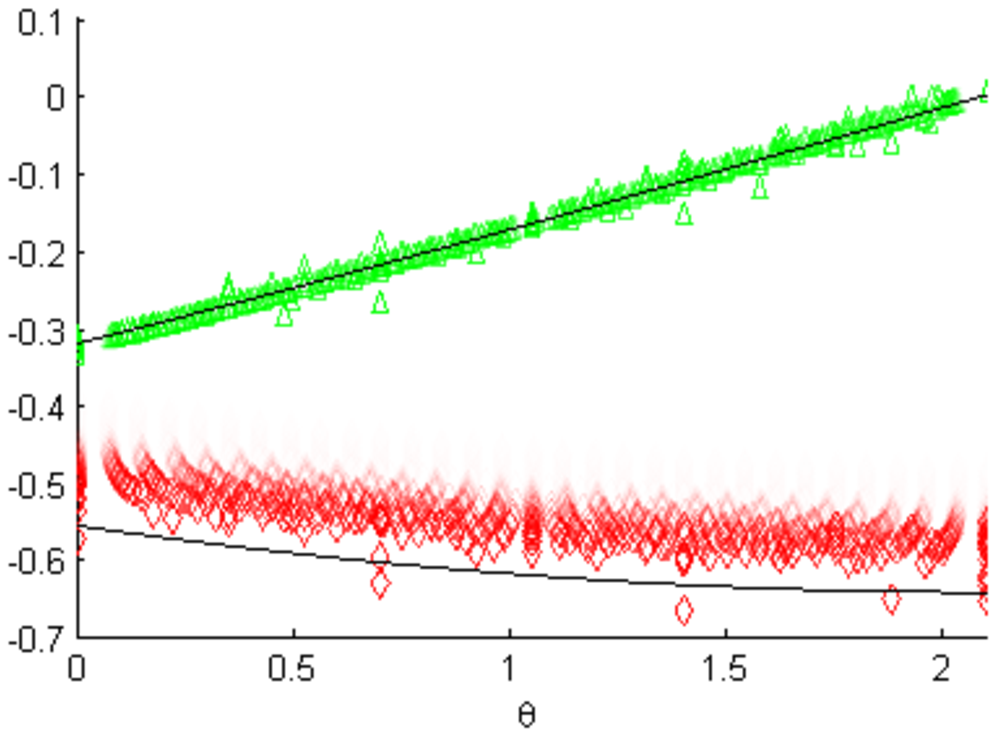}
			\caption{$\Diamond$ is $u \rho^{1/4}$ with curve $0.64[-\sin(3\theta/4)\cos(\theta-2\pi/3)+\cos(3\theta/4)\sin(\theta-2\pi/3)]$ and $\triangle$ is $v \rho^{1/4}$ with curve $0.64[\sin(3\theta/4)\sin(\theta-2\pi/3)+\cos(3\theta/4)\cos(\theta-2\pi/3)]$.}
			\label{f:simple_c1_obtuse_th}
		\end{subfigure}
	\end{tabular}
	\caption{Plots for $\theta_1=2\pi/3$, $\gamma=0.5$ around $C_1$.}
	\label{sf:simple_c1_obtuse}
\end{figure}

\begin{figure}[ptb]
	\centering
	\begin{tabular}{c c}
		\begin{subfigure}[t]{0.47\textwidth}
			\centering
			\includegraphics[width=\textwidth]{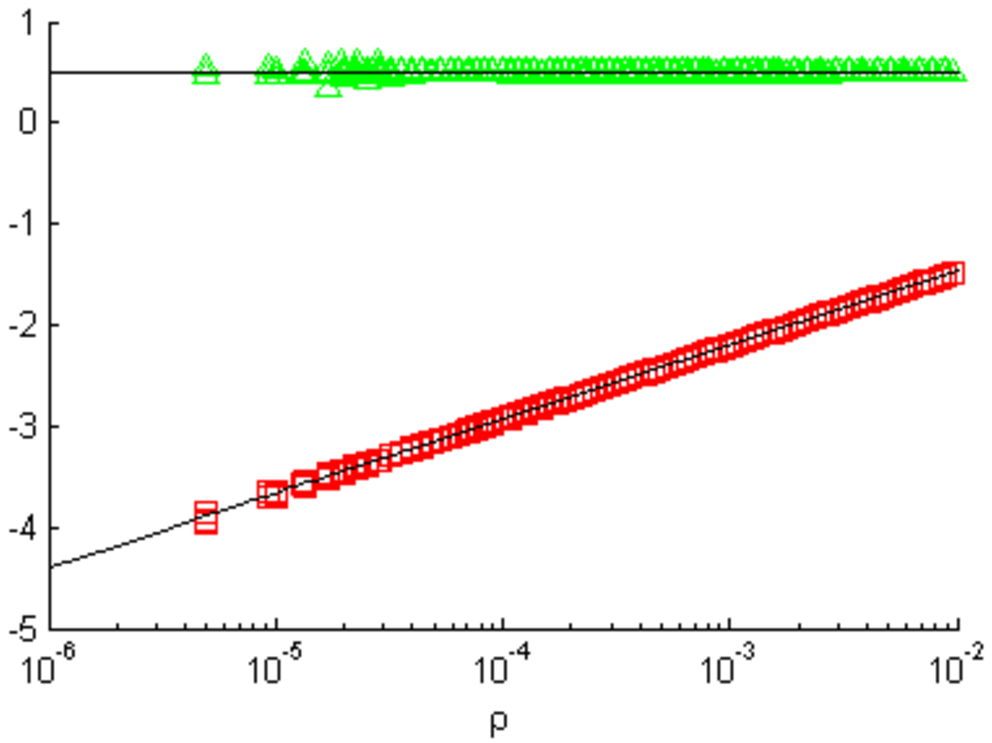}
			\caption{$\square$ is $u-0.5$ with line $\ln(\rho)/\pi$ and $\triangle$ is $v/(2\theta/\pi-1)$ with line $0.5$.}
			\label{f:simple_c1_right_rh}
		\end{subfigure}
		&
		\begin{subfigure}[t]{0.47\textwidth}
			\centering
			\includegraphics[width=\textwidth]{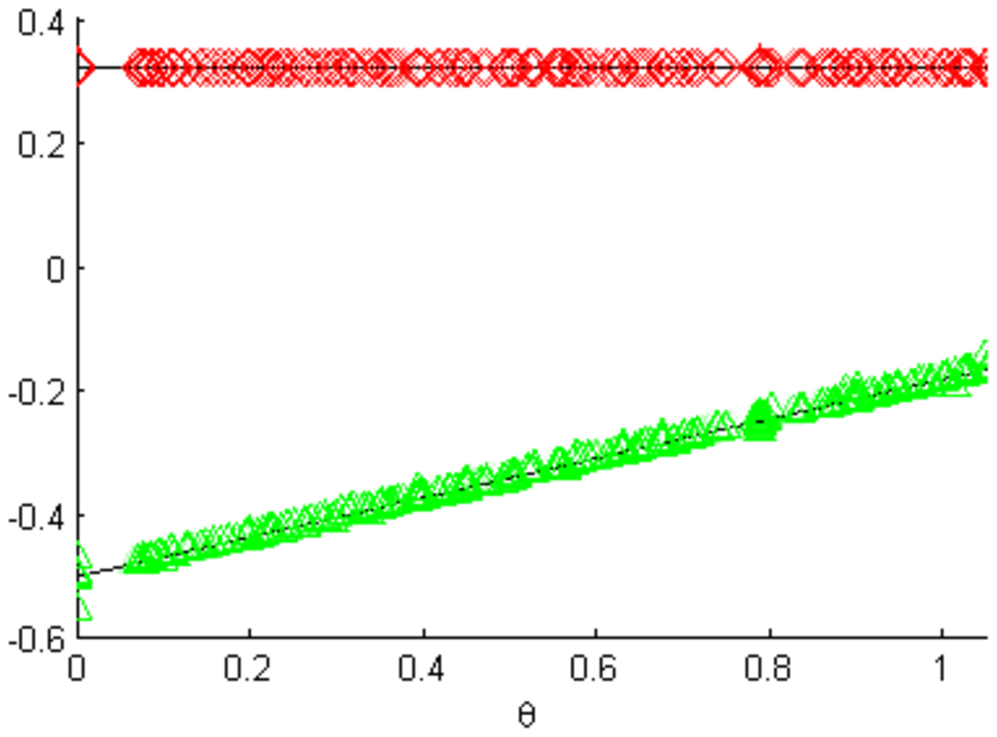}
			\caption{$\Diamond$ is $(u-0.5)/\ln(\rho)$ with line $1/\pi$ and $\triangle$ is $v$ with line $(\theta/\pi)-0.5$.}
			\label{f:simple_c1_right_th}
		\end{subfigure}
	\end{tabular}
	\caption{Plots for $\theta_1=\pi/2$, $\gamma=0.5$ around $C_1$.}
	\label{sf:simple_c1_right}
\end{figure}

\begin{figure}[ptb]
	\centering
	\begin{tabular}{c c}
		\begin{subfigure}[t]{0.47\textwidth}
			\centering
			\includegraphics[width=\textwidth]{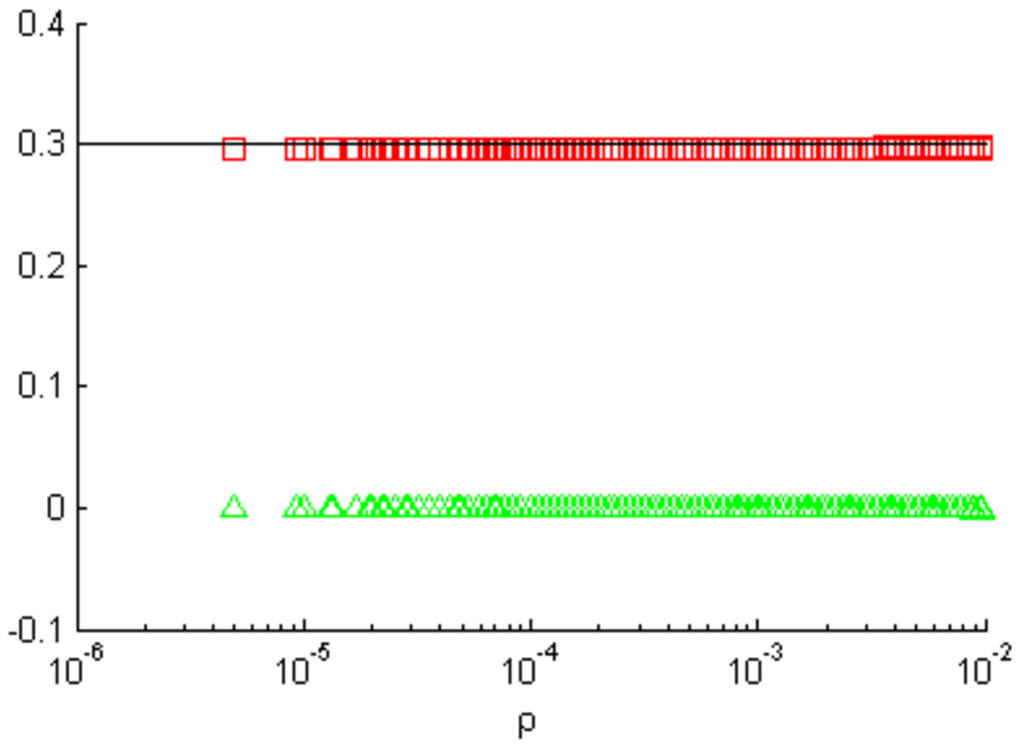}
			\caption{$\square$ is $u$ with line at $0.3$ and $\triangle$ is $v$.}
			\label{f:simple_c1_right_ng_rh}
		\end{subfigure}
		&
		\begin{subfigure}[t]{0.47\textwidth}
			\centering
			\includegraphics[width=\textwidth]{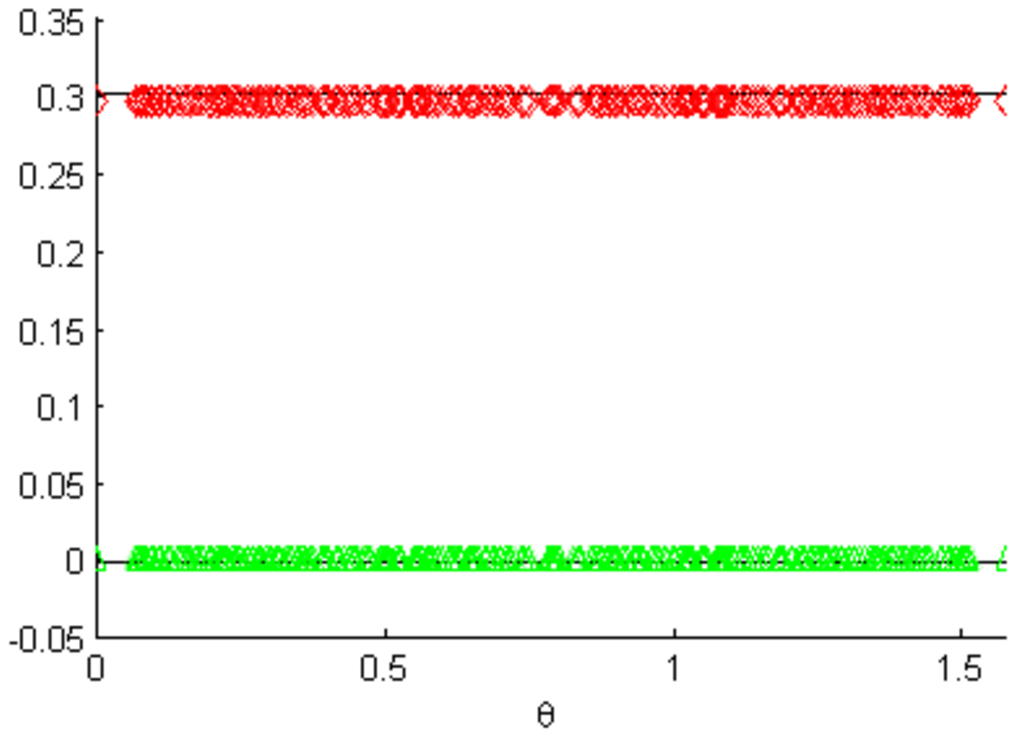}
			\caption{$\Diamond$ is $u$ with line at $0.3$ and $\triangle$ is $v$.}
			\label{f:simple_c1_right_ng_th}
		\end{subfigure}
	\end{tabular}
	\caption{Plots for $\theta_1=\pi/2$, $\gamma=0$ around $C_1$}
	\label{sf:simple_c1_right_ng}
\end{figure}

\begin{figure}[ptb]
	\centering
	\begin{tabular}{c c}
		\begin{subfigure}[t]{0.47\textwidth}
			\centering
			\includegraphics[width=\textwidth]{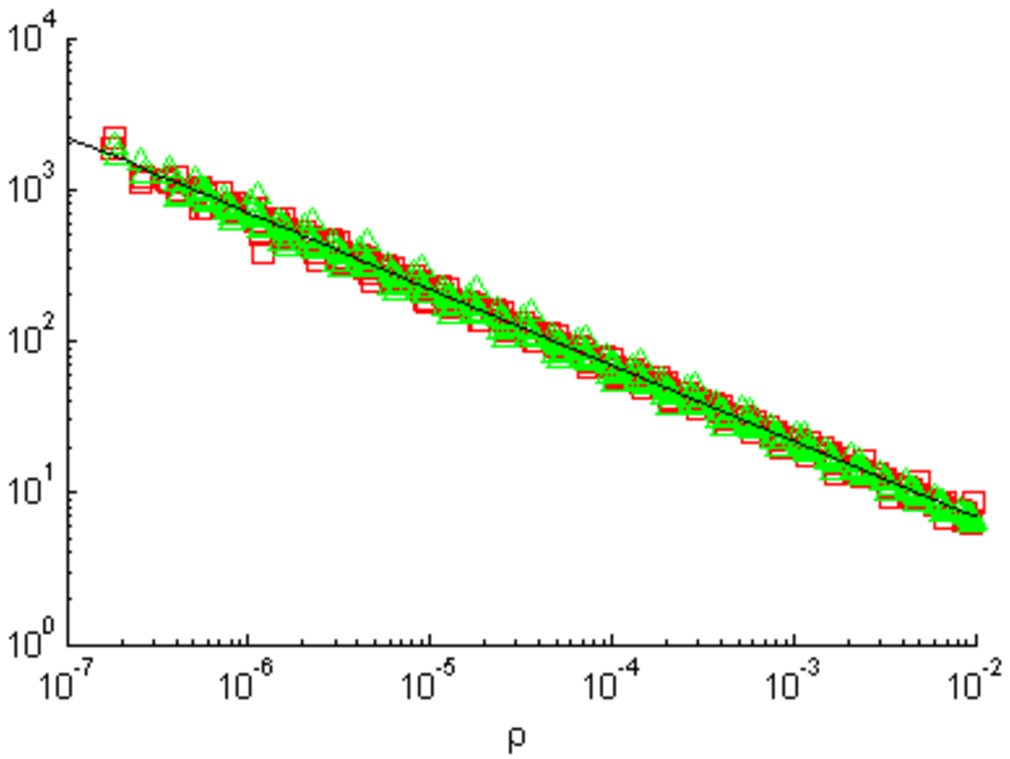}
			\caption{$\square$ is $u/(-\sin(\theta/2)\cos(\theta)+\cos(\theta/2)\sin(\theta))$ and $\triangle$ is $v/(-\sin(\theta/2)\sin(\theta)-\cos(\theta/2)\cos(\theta))$, curve is $0.7/\sqrt{\rho}$.}
			\label{f:simple_c2_rh}
		\end{subfigure}
		&
		\begin{subfigure}[t]{0.47\textwidth}
			\centering
			\includegraphics[width=\textwidth]{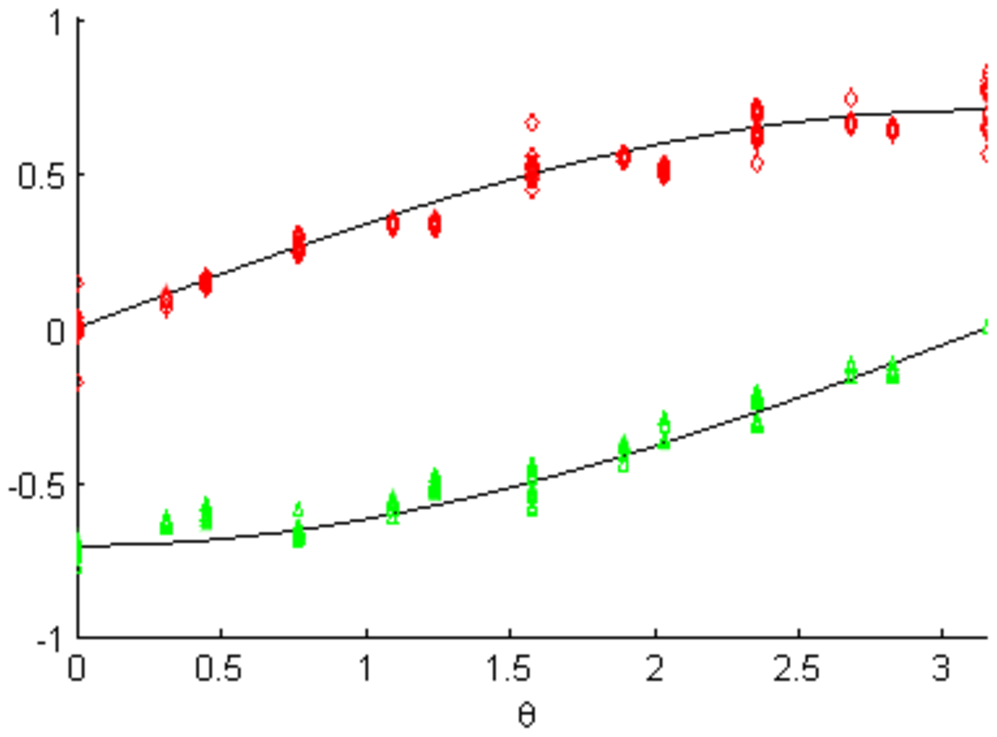}
			\caption{$\Diamond$ is $u\sqrt{\rho}$ with curve $0.7[-\sin(\theta/2)\cos(\theta)+\cos(\theta/2)\sin(\theta)]$ and $\triangle$ is $v\sqrt{\rho}$ with curve $0.7[-\sin(\theta/2)\sin(\theta)-\cos(\theta/2)\cos(\theta)]$.}
			\label{f:simple_c2_th}
		\end{subfigure}
	\end{tabular}
	\caption{Plots for $\gamma=0$ around $C_2$.}
	\label{sf:simple_c2}
\end{figure}

We produce numerical solutions for different $\theta_1$ in the regions around $C_1$ and $C_2$ to examine the locally dominant behaviour. Our purpose is to investigate observed multivalued points and singularities in the solutions for velocity, which shall reveal some fundamental issues in the current formulation of this phenomenon. Around each of $C_1$ and $C_2$ we use a local polar coordinate systems with distance from the point of interest $\rho$ and angle $\theta$, as defined by figure \ref{f:asymptotics}. The curves and lines plotted on the graphs are the leading order terms from the asymptotic analysis that is performed in the next section, and are included for later comparison. Also note that the scattering of points at small $\rho$ is due to numerical error when evaluating singularities with the current scheme.

We now consider the solutions around $C_1$ for different values of $\theta_1$ and $\gamma$ for very small values of $\rho$. 

In figure \ref{sf:simple_c1_acute} we plot the values of the velocities $u$ and $v$ for $\theta_1=\pi/3$ and $\gamma=0.5$. From it we see that $u$ tends to a constant as $\rho$ becomes small, whilst the $\rho$ dependence of $v$ is $v=O(\sqrt{\rho})$. Therefore the solution is single valued and bounded, and can easily used for simulating the propagation of the wetting front. In figure \ref{sf:simple_c1_obtuse} we consider the case $\theta_1=2\pi/3$ and $\gamma=0.5$. Here the velocities diverge as $u,v=O(\rho^{-1/4})$. For the case $\theta_1=\pi/2$ and $\gamma=0.5$, figure \ref{sf:simple_c1_right}, we see that $u$ diverges as $u=O(\ln(\rho))$, whilst $v$ is multivalued at $C_1$. For $\theta_1=\pi/2$ and $\gamma=0$, figure \ref{sf:simple_c1_right_ng}, these issues do not occur, $u$ being constant and $v=0$.

Considering the solution around $C_2$ we have singularities in both components of velocity as $u,v=O(1/\sqrt{\rho})$, as seen in figure \ref{sf:simple_c2}. The angular dependence is also plotted, although there are issues with our mesh resolution around this point so the quality of the angular dependence is not high. The implications of the divergent velocities around $C_2$ are important and shall be discussed later.

Next we will verify the results that we have obtained numerically using local asymptotic solutions. This will give a full picture of the range of behaviours that exist and allow us to physically interpret them.

\subsection{Asymptotic Analysis}	\label{ss:asymp}

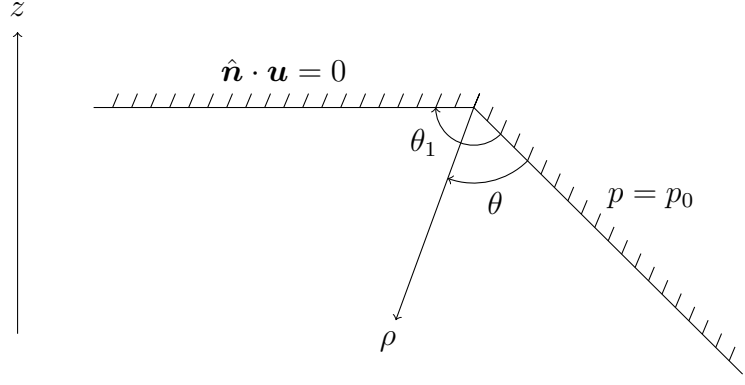
\begin{figure}[tbp]
	\centering
	\begin{tikzpicture}
	\draw (-180:5) -- (0,0) --(-45:5);
	\foreach \n in {0,0.25,...,4.9}	{
		\draw (-180:\n) -- +(67.5:0.2) (-45:\n) -- +(67.5:0.2);
	}
	
	\draw[->]	(-6,-3) -- (-6,1);
	\draw		(-6,1.3) node {$z$};
	
	\draw[->]	(0,0) -- (-110:3);
	\draw	(-110:3.3) node {$\rho$};
	
	\draw[<-]	(-110:1) arc (-110:-45:1);
	\draw	(-77.5:1.3) node {$\theta$};
	
	\draw[->]	(-45:0.5) arc (-45:-180:0.5);
	\draw	(-145:0.8) node {$\theta_1$};
	
	\draw	(-180:2.5)+(90:0.5) node {$\hat{\boldsymbol{n}}\cdot\boldsymbol{u}=0$}
			(-45:2.5)+(45:0.8) node {$p=p_0$};
\end{tikzpicture}
	\caption{The wedge that the regions around $C_1$ and $C_2$ tend to asymptotically, with subtended angle $\theta_1$.}
	\label{f:asymptotics}
\end{figure}

Let us consider the domain asymptotically as we tend towards the contact lines $C_1$ and $C_2$. As we do this the curvature on the length scale we are observing tends to zero, thus the wetting front tends to a plane, the curvature of the contact line (due to it being a circle) tends to zero, and the domain of the flow tends towards a two dimensional wedge. In both cases the boundary with $\hat{\boldsymbol{n}}\cdot\boldsymbol{u}=0$ (which is $\Gamma_1$) is horizontal, so we choose to consider the wedge depicted in \ref{f:asymptotics}, with contact angle $\theta_1$ and local polar coordinates $\rho$ and $\theta$ such that $z=\rho \sin(\theta-\theta_1)$. The local components of velocity are $u_\rho=\boldsymbol{u}\cdot\hat{\boldsymbol{\rho}}$ and $u_\theta=\boldsymbol{u}\cdot\hat{\boldsymbol{\theta}}$, where $\hat{\boldsymbol{\rho}}$ and $\hat{\boldsymbol{\theta}}$ are the basis vectors of the local polar coordinate system. These are related to the components $u$ and $v$ by
\begin{subequations}\label{seq:local_u_c1}\begin{align}
	u&=-u_\rho \cos(\theta-\theta_1) + u_\theta \sin(\theta-\theta_1),	\\
	v&=u_\rho \sin(\theta-\theta_1)+u_\theta \cos(\theta-\theta_1),
\end{align}\end{subequations}
for $C_1$, and for $C_2$
\begin{subequations}\label{seq:local_u_c2}\begin{align}
	u&=-u_\rho \cos(\theta) + u_\theta \sin(\theta),	\\
	v&=-u_\rho \sin(\theta)-u_\theta \cos(\theta).
\end{align}\end{subequations}

The equations in the wedge region are, using \eqref{eq:darcy} to eliminate velocity,
\begin{subequations}\begin{alignat}{2}
	\nabla^2 p&=0	\hspace{1cm} &\forall \: \theta &\in [0,\theta_1],	\\
	p&=p_0	\hspace{1cm} & \mathrm{on} \: \theta&=0,	\\
	\frac{\partial (p+\gamma \rho \sin(\theta-\theta_1))}{\partial \theta}&=0	\hspace{1cm} & \mathrm{on} \: \theta&=\theta_1.
\end{alignat}\end{subequations}
We make the change of variables $\tilde{p}=p-p_0+\gamma \rho \sin(\theta-\theta_1)$ to obtain
\begin{subequations}\begin{alignat}{2}
	\nabla^2\tilde{p}&=0	\hspace{1cm} &\forall \: \theta &\in [0,\theta_1],	\\
	\tilde{p}&=-\gamma \rho \sin(\theta_1)	\hspace{1cm} & \mathrm{on} \: \theta&=0,	\\
	\frac{\partial\tilde{p}}{\partial \theta}&=0	\hspace{1cm} & \mathrm{on} \: \theta&=\theta_1.
\end{alignat}\end{subequations}
It is observed that, for $\theta_1\neq\pi/2$, this set of equations has a solution
\begin{equation}
	\tilde{p}_1=-\gamma \rho \sin(\theta_1) \left[\cos(\theta)+\tan(\theta_1)\sin(\theta)\right]
\end{equation}
and for $\theta_1=\pi/2$ it has a solution
\begin{equation}
	\tilde{p}_2=\frac{2\gamma}{\pi} \sin(\theta) \rho \ln (\rho)-\gamma \rho \cos(\theta) \left[ 1 - \frac{2}{\pi}\theta \right].
\end{equation}
Defining $\hat{p}=\tilde{p}-\tilde{p}_1$ for $\theta_1\neq\pi/2$ and $\hat{p}=\tilde{p}-\tilde{p}_2$ for $\theta_1=\pi/2$, the equations become
\begin{subequations}\begin{alignat}{2}
	\nabla^2\hat{p}&=0	\hspace{1cm} &\forall \: \theta &\in [0,\theta_1],	\\
	\hat{p}&=0	\hspace{1cm} & \mathrm{on} \: \theta&=0,	\\
	\frac{\partial\hat{p}}{\partial \theta}&=0	\hspace{1cm} & \mathrm{on} \: \theta&=\theta_1.
\end{alignat}\end{subequations}
This is now soluble using separation of variables, the solution is
\begin{equation}
	\hat{p}=\sum_{n\in\mathbb{Z}} \left[ c_n \rho^{(n+\frac{1}{2})\frac{\pi}{\theta_1}} \sin\left(\left[n+\frac{1}{2}\right]\frac{\pi}{\theta_1}\theta\right)\right].
\end{equation}
where the values $c_n$ are arbitrary constants. Observing that in our numerical solution the pressure is bounded, the sum is truncated to $n\geq0$, this is the solution obtained in \cite[(3.11)]{dynam_angle_shk} except that there the velocity was restricted to be bounded also, and only the case $\theta_1=\pi$ was considered. For $\theta_1\neq\pi/2$ we obtain the solution
\begin{subequations}\begin{align}
	p&=\sum_{n=0}^{\infty} \left[ c_n \rho^{(n+\frac{1}{2})\frac{\pi}{\theta_1}} \sin\left(\left[n+\frac{1}{2}\right]\frac{\pi}{\theta_1}\theta\right)\right]
		-\gamma \rho \frac{\sin(\theta)}{\cos(\theta_1)}+p_0
	,\\
	u_\rho&=-\sum_{n=0}^{\infty} \left[ c_n \left(n+\frac{1}{2}\right)\frac{\pi}{\theta_1} \rho^{(n+\frac{1}{2})\frac{\pi}{\theta_1}-1} \sin\left(\left[n+\frac{1}{2}\right]\frac{\pi}{\theta_1}\theta\right)\right]
		+ \gamma \sin(\theta_1) \left[\cos(\theta)+\tan(\theta_1)\sin(\theta)\right]
	,\\
	u_\theta&=-\sum_{n=0}^{\infty} \left[ c_n \left(n+\frac{1}{2}\right)\frac{\pi}{\theta_1} \rho^{(n+\frac{1}{2})\frac{\pi}{\theta_1}-1} \cos\left(\left[n+\frac{1}{2}\right]\frac{\pi}{\theta_1}\theta\right)\right]
		+ \gamma \sin(\theta_1) \left[\tan(\theta_1)\cos(\theta)-\sin(\theta)\right]	\label{eq:asymp_uth_1}
	,
\end{align}\end{subequations}
and for $\theta_1=\pi/2$
\begin{subequations}\begin{align}
	p&=\sum_{n=0}^{\infty} \left[ c_n \rho^{2n+1} \sin\left(\left[2n+1\right]\theta\right)\right]
		+ \frac{2\gamma}{\pi} \sin(\theta) \rho \ln (\rho)+\gamma \rho \cos(\theta) \frac{2}{\pi}\theta + p_0
	,\\
	u_\rho&=-\sum_{n=0}^{\infty} \left[ c_n \left(2n+1\right) \rho^{2n} \sin\left(\left[2n+1\right]\theta\right)\right]
		-\frac{2\gamma}{\pi} \sin(\theta) \left[\ln (\rho)+1\right]-\gamma \cos(\theta) \left[\frac{2}{\pi}\theta-1\right]
	,\\
	u_\theta&=-\sum_{n=0}^{\infty} \left[ c_n \left(2n+1\right) \rho^{2n} \cos\left(\left[2n+1\right]\theta\right)\right]
		-\frac{2\gamma}{\pi} \cos(\theta) \left[\ln (\rho)+1\right]+\gamma \sin(\theta) \left[\frac{2}{\pi}\theta-1\right]	\label{eq:asymp_uth_2}
	.
\end{align}\end{subequations}

Let us now consider the leading order solutions as $\rho \rightarrow 0$ in the cases relevant to our model. We shall deduce the components of velocity $u$ and $v$ using equations \eqref{seq:local_u_c1} and \eqref{seq:local_u_c2}, for these components the leading order terms sometimes cancel and in these cases the second order terms shall be stated for this function only. In all cases only sufficient terms to understand the numerical results in the previous section are presented.

For the region around $C_2$ the wedge subtends an angle $\theta_1=\pi$ and $p_0=0$, to leading order
\begin{subequations}\begin{align}
	p& \sim c_0 \rho^{\frac{1}{2}} \sin\left(\frac{1}{2}\theta\right)	,\\
	u_\rho& \sim -c_0 \frac{1}{2}\rho^{-\frac{1}{2}} \sin\left(\frac{1}{2}\theta\right)	,\\
	u_\theta& \sim -c_0 \frac{1}{2} \rho^{-\frac{1}{2}} \cos\left(\frac{1}{2}\theta\right)	,\\
	u &\sim -c_0 \frac{1}{2} \rho^{-\frac{1}{2}} \left[-\sin\left(\frac{1}{2}\theta\right)\cos(\theta)+\cos\left(\frac{1}{2}\theta\right)\sin(\theta)\right],	\\
	v &\sim -c_0 \frac{1}{2} \rho^{-\frac{1}{2}} \left[-\sin\left(\frac{1}{2}\theta\right)\sin(\theta)-\cos\left(\frac{1}{2}\theta\right)\cos(\theta)\right].
\end{align}\end{subequations}

For $C_1$ we have $p_0=-1$. We consider four cases, firstly for $\theta_1>\pi/2$, or $\gamma=0$ and $\theta_1\neq\pi/2$, to leading order
\begin{subequations}\begin{align}
	p+1& \sim c_0 \rho^{\frac{\pi}{2\theta_1}} \sin\left(\frac{\pi}{2\theta_1}\theta\right)	,\label{eq:asymp_c1_1_p}\\
	u_\rho& \sim -c_0 \frac{\pi}{2\theta_1} \rho^{\frac{\pi}{2\theta_1}-1} \sin\left(\frac{\pi}{2\theta_1}\theta\right)	,\\
	u_\theta& \sim -c_0 \frac{\pi}{2\theta_1} \rho^{\frac{\pi}{2\theta_1}-1} \cos\left(\frac{\pi}{2\theta_1}\theta\right),	\label{eq:asymp_c1_1_theta}\\
	u &\sim -c_0 \frac{\pi}{2\theta_1} \rho^{\frac{\pi}{2\theta_1}-1}\left[ -\sin\left(\frac{\pi}{2\theta_1}\theta\right) \cos(\theta-\theta_1) + \cos\left(\frac{\pi}{2\theta_1}\theta\right) \sin(\theta-\theta_1)\right],	\\
	v &\sim -c_0 \frac{\pi}{2\theta_1} \rho^{\frac{\pi}{2\theta_1}-1}\left[ \sin\left(\frac{\pi}{2\theta_1}\theta\right) \sin(\theta-\theta_1) + \cos\left(\frac{\pi}{2\theta_1}\theta\right) \cos(\theta-\theta_1)\right].
\end{align}\end{subequations}
For the components of velocity the power of $\rho$ is less than zero, so all are singular. Secondly for $\theta_1<\pi/2$ and $\gamma\neq0$,
\begin{subequations}\begin{align}
	p+1& \sim -\gamma \rho \frac{\sin(\theta)}{\cos(\theta_1)}	,\\
	u_\rho& \sim \gamma \sin(\theta_1) \left[\cos(\theta)+\tan(\theta_1)\sin(\theta)\right]	,\\
	u_\theta& \sim \gamma \sin(\theta_1) \left[\tan(\theta_1)\cos(\theta)-\sin(\theta)\right]	,\\
	u &\sim -\gamma \sin^2(\theta_1) [\tan(\theta_1)+\cot(\theta_1)],	\\
	v &\sim -c_0 \frac{\pi}{2\theta_1} \rho^{\frac{\pi}{2\theta_1}-1}\left[ \sin\left(\frac{\pi}{2\theta_1}\theta\right) \sin(\theta-\theta_1) + \cos\left(\frac{\pi}{2\theta_1}\theta\right) \cos(\theta-\theta_1)\right].
\end{align}\end{subequations}
The radial component of velocity is constant, and the axial component has power of $\rho$ greater than zero, so is finite. The final two cases are for $\theta_1=\pi/2$, for $\gamma=0$
\begin{subequations}\begin{align}
	p+1 & \sim \rho\left[ c_0 \sin(\theta) \right]	,\\
	u_\rho & \sim -c_0 \sin(\theta)	,\\
	u_\theta & \sim -c_0 \cos(\theta)	,\\
	u & \sim c_0	,\\
	v & \sim 3 c_1 \rho^2 \sin(2\theta)	,
\end{align}\end{subequations}
so both components are finite. For $\gamma\neq0$
\begin{subequations}\begin{align}
	p+1 & \sim \rho \ln (\rho) \left[ \frac{2\gamma}{\pi} \sin(\theta) \right] + \rho \left[c_0 \sin(\theta) + \gamma \cos(\theta) \frac{2}{\pi} \theta \right]	,\\
	u_\rho & \sim \ln (\rho) \left[-\frac{2\gamma}{\pi} \sin(\theta) \right] +\left[-c_0 \sin(\theta) - \frac{2 \gamma}{\pi} \sin(\theta) - \gamma \cos(\theta) \left(\frac{2}{\pi}\theta-1\right)\right] 	,\\
	u_\theta & \sim \ln (\rho) \left[-\frac{2\gamma}{\pi} \cos(\theta) \right] +\left[-c_0 \cos(\theta) - \frac{2 \gamma}{\pi} \cos(\theta) + \gamma \sin(\theta) \left(\frac{2}{\pi}\theta-1\right)\right]	,\\
	u & \sim \ln (\rho)\left[\frac{2\gamma}{\pi}\right]  + \left[ c_0 + \frac{2\gamma}{\pi} \right]	,\\
	v & \sim \gamma\left[\frac{2}{\pi}\theta-1\right]	,
\end{align}\end{subequations}
so the radial component is singular and the axial component is multivalued at $C_1$.

Curves of the forms obtained above are plotted in \cref{sf:simple_c1_acute,sf:simple_c1_obtuse,sf:simple_c1_right,sf:simple_c1_right_ng}, and fit the data plotted very well. We shall next discuss the physical meaning of these equations.

\subsection{Interpretation of the Asymptotic Analysis} \label{ss:asym_phys}

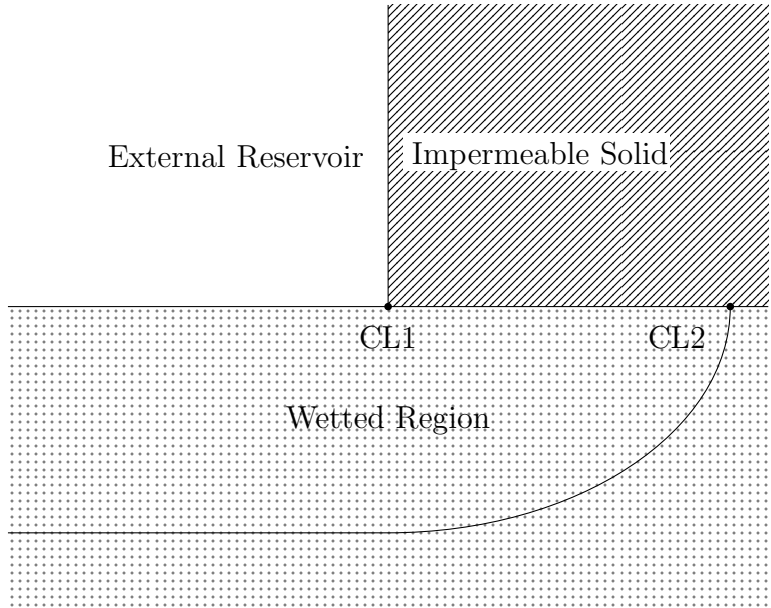
\begin{figure}[ptb]
	\centering
	\begin{tikzpicture}
	\draw (-5,0) -- (5,0);
	\fill[pattern=dots,pattern color=gray] (-5,0) rectangle (5,-4);
	
	\draw (0,0) -- (0,4);
	\fill[pattern=north east lines] (0,0) rectangle (5,4);
	
	\begin{scope}
		\clip (0,0) rectangle (5,-4);
		\draw (0,0) ellipse (4.5 and 3);
	\end{scope}
	\draw (-5,-3) -- (0,-3);
	
	\draw (-2,2) node {External Reservoir};
	\fill[color=white] (0.2,1.8) rectangle (3.7,2.3);
	\draw (2,2) node {Impermeable Solid};
	\draw (0,-1.5) node {Wetted Region};
	
	\fill (0,0) circle (0.05);
	\fill (4.5,0) circle (0.05);
	
	\draw
		(3.8,-0.4) node {CL2}
		(0,-0.4) node {CL1};
\end{tikzpicture}
	\caption{Example configuration of two dimension flow for which we can examine the validity of the equations in the problem formulation.}
	\label{f:darcy_problem}
\end{figure}

\begin{figure}[ptb]
	\centering
	\begin{tabular}{c c}
		\begin{subfigure}[t]{0.47\textwidth}
			\centering
			\begin{tikzpicture}
	\def\L{2};
	\coordinate (L) at (0,0);
	\coordinate (R) at (\L,0);
	
	
	\draw (L)+(-2,0) -- ($(R)+(1,0)$) (L) -- +(-45:3);
	\draw[dotted] ($(R)+(1,0)$) -- +(0.5,0);
	\foreach \n in {-1.75,-1.5,...,3}	{
		\draw ($(L)+(\n,0)$) -- +(112.5:0.2);
	}
	
	\foreach \n in {0.25,0.5,...,2.9}	{
		\draw[->] ($(L)+(-45:\n)$) -- +(45:0.3/\n);
	}
	
	\draw[scale=1,domain=-1.3:0,smooth,variable=\x,black] plot ({\L-0.3*0.3-1+cosh(\x)},{-0.3+sinh(\x)});
	\draw[scale=1,domain=0:0.3,smooth,variable=\x,black] plot ({\L-0.3*0.3+\x*\x},{-0.3+\x});
\end{tikzpicture}
			\caption{$v_s = c\rho^{-n}$ where $-1<n<0$, or $v_s = -c\ln(\rho)$}
			\label{f:asymptotic_advance_sing}
		\end{subfigure}
		\vspace{0.5cm}
		&
		\begin{subfigure}[t]{0.47\textwidth}
			\centering
			\begin{tikzpicture}
	\def\L{1};
	\coordinate (L) at (0,0);
	\coordinate (R) at (\L,0);
	
	
	\draw (L)+(-2,0) -- ($(R)+(1,0)$) (L) -- +(-45:3);
	\draw[dotted] ($(R)+(1,0)$) -- +(0.5,0);
	\foreach \n in {-1.75,-1.5,...,2}	{
		\draw ($(L)+(\n,0)$) -- +(112.5:0.2);
	}
	
	\foreach \n in {0.25,0.5,...,2.9}	{
		\draw[->] ($(L)+(-45:\n)$) -- +(45:0.3);
	}
	
	\draw (R) -- +(-45:3);
\end{tikzpicture}
			\caption{$v_s = c$}
			\label{f:asymptotic_advance_const}
		\end{subfigure}
		\vspace{0.5cm}
		\\
		\begin{subfigure}[t]{0.47\textwidth}
			\centering
			\begin{tikzpicture}
	\def\L{0};
	\coordinate (L) at (0,0);
	\coordinate (R) at (\L,0);
	
	
	\draw (L)+(-2,0) -- ($(R)+(1,0)$) (L) -- +(-135:3);
	\draw[dotted] ($(R)+(1,0)$) -- +(0.5,0);
	\foreach \n in {-1.75,-1.5,...,1}	{
		\draw ($(L)+(\n,0)$) -- +(112.5:0.2);
	}
	
	\foreach \n in {0.25,0.5,...,2.9}	{
		\draw[->] ($(L)+(-135:\n)$) -- +(-45:{0.3*sqrt(\n)} );
	}
	
	\draw[scale=1,domain=0:2.5,smooth,variable=\x,black] plot ({0.5*sqrt(\x)-(\x/sqrt(2))},{-0.5*sqrt(\x)-\x/sqrt(2)});
\end{tikzpicture}
			\caption{$v_s = c\rho^n$ where $0<n<1$}
			\label{f:asymptotic_advance_root}
		\end{subfigure}
		\vspace{0.5cm}
		&
		\begin{subfigure}[t]{0.47\textwidth}
			\centering
			\begin{tikzpicture}
	\def\L{2};
	\coordinate (L) at (0,0);
	\coordinate (R) at (\L,0);
	
	
	\draw (L)+(-2,0) -- ($(R)+(1,0)$) (L) -- +(-135:3);
	\draw[dotted] ($(R)+(1,0)$) -- +(0.5,0);
	\foreach \n in {-1.75,-1.5,...,3}	{
		\draw ($(L)+(\n,0)$) -- +(112.5:0.2);
	}
	
	\foreach \n in {0.25,0.5,...,2.9}	{
		\draw[->] ($(L)+(-135:\n)$) -- +(-45:0.3*\n);
	}
	
	\draw ($(L)$) -- +(-100:3);
\end{tikzpicture}
			\caption{$v_s = c\rho$}
			\label{f:asymptotic_advance_lin}
		\end{subfigure}
		\vspace{0.5cm}
		\\
		\vspace{1cm}
		\begin{subfigure}[t]{0.47\textwidth}
			\centering
			\begin{tikzpicture}
	\def\L{2};
	\coordinate (L) at (0,0);
	\coordinate (R) at (\L,0);
	
	
	\draw (L)+(-2,0) -- ($(R)+(1,0)$) (L) -- +(-135:3);
	\draw[dotted] ($(R)+(1,0)$) -- +(0.5,0);
	\foreach \n in {-1.75,-1.5,...,3}	{
		\draw ($(L)+(\n,0)$) -- +(112.5:0.2);
	}
	
	\foreach \n in {0.25,0.5,...,2.9}	{
		\draw[->] ($(L)+(-135:\n)$) -- +(-45:{0.3*\n*sqrt(\n)} );
	}
	
	\draw[scale=1,domain=0:2,smooth,variable=\x,black] plot ({0.5*\x*sqrt(\x)-(\x/sqrt(2))},{-0.5*\x*sqrt(\x)-(\x/sqrt(2))});
\end{tikzpicture}
			\caption{$v_s = c\rho^n$ where $n>1$}
			\label{f:asymptotic_advance_poly}
		\end{subfigure}
	\end{tabular}
	\caption{Illustrations of the different dynamics caused by the various powers of $\rho$ in the expansions of $v_s$. The left curve in each figure is the wetting front at some time, and the right curve is the front that it evolves into. The value of $c$ is a constant value, that is irrelevant for the dynamics (so long as it is non-zero), the only thing that matters is the power. For values of $c$ that are negative, the front moves in the opposite direction.}
\end{figure}
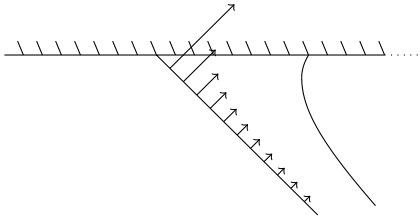
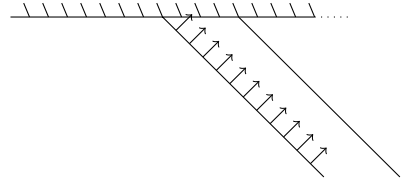
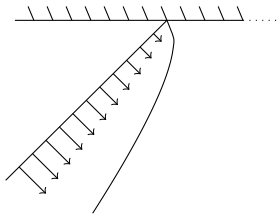
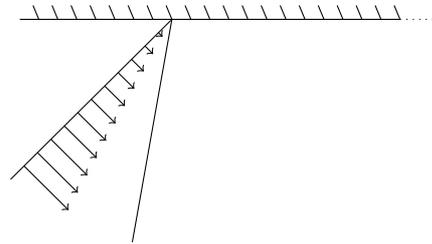
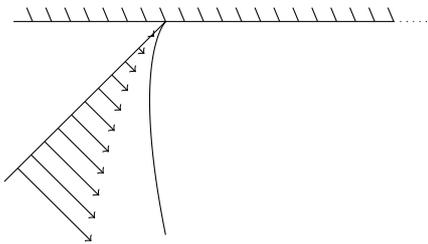

In our analysis we obtained that the velocities are singular at $C_2$, and at $C_1$ for the case $\theta_1>\pi/2$ and for $\theta_1=\pi/2$ when $\gamma\neq0$. These singularities are all integrable, i.e. they diverge as $\rho^n$ where $n>-1$ or as $\ln(\rho)$. They are called integrable because the integral of the velocity over any finite surface will be finite, which means that flux of volume through any finite surface will be finite. The physical interpretation of the singularities it that a finite volume of fluid is moving through a point or line per unit time. 

The singularities are the symptom of a fundamental problem in our problem formulation. Darcy's equation is believed to describe slow creping flows in porous materials where the effect of inertia is negligible. The singular velocities the we observe are inconstant with this. For a particle that passes through one of these singular points its velocity will start out finite, become divergent and then become finite again. The velocity and acceleration of such a particle are certainly not small. Therefore, one of our equations must be un-physical. To examine which equation this is let us temporarily examine the two dimensional flow depicted in figure \ref{f:darcy_problem}. We can be sure that the boundary condition $\boldsymbol{u}\cdot\hat{\boldsymbol{n}}=0$ is correct between the two contact lines because the surface of the porous medium is covered by an impermeable solid. The wetted region cannot penetrate the impermeable solid, nor can it retreat away from it because that would create a vacuum. The most that can happen is that CL2 recedes causing the wetted region to `peal off,' but this still leaves a finite amount of time with the boundary condition valid. The boundary condition $p=0$ on the drawing area was established using analysis of the scales of the pressures. For this to be wrong there would have to be a boundary layer in the external reservoir just above the drawing area, but this cannot be the case due to the very low volume flux into the wetted region. Of course, if the singular velocity also existed in the external reservoir then this would cause there to be very high pressures and velocity gradients which may change the solution, but this would not solve the fundamental problem. The slow imbibition of a highly viscous fluid into a low porosity solid should not cause a boundary layer due to high stresses in the external reservoir. Therefore, the singularities \emph{must} arise due to inadequacies in Darcy's equation, and not in the boundary conditions. Even \emph{if} the boundary conditions in the asymptotic analysis are not physically correct for \emph{this} phenomenon, they are physically correct for \textit{a} phenomenon, and so cannot be what is fundamentally wrong with the problem formulation. From this we identify the point $C_2$, the contact line CL1 at the edge of the drawing area, to be a place at which improvements to Darcy's equation could be tested. Such an improvement would almost certainly need to include inertial effects, and perhaps long range viscous diffusion effects also. One of the improvements that is discussed in the introduction may be what is required, although none of these were developed to rectify an issue like the one we face and so this is unlikely.

However, the volume of fluid that passes through $\Gamma_2$ into the porous solid is likely to be almost the same for any improvement (since the fluid is drawn in to feed the advancement of the wetting front which dictates the volume of fluid required) and will simply be distributed more evenly along the portion of $\Gamma_2$ that is close to $C_2$. It is also possible that the imbibition will be slower because the volume flux though the drawing area is suppressed. This requires further investigation.

We shall now discuss the behaviour local to $C_1$ in the various cases in the previous section, that is the local distribution of the normal velocity of the front, which is $v_s=\boldsymbol{u}\cdot\hat{\boldsymbol{n}}=-u_\theta$ on $\theta=0$. We must assume that the behaviour occurring with Darcy's equation will be qualitatively the same as for an equation that suppresses the velocities that we see, and also for a formulation where a dynamic contact angle is used. Whether this is a reasonable assumption should be verified.

First the case when $\theta_1<\pi/2$, from \eqref{eq:asymp_uth_1} the leading order terms in the expansion of the surface velocity are
\begin{equation}
	v_s\sim  - \gamma \sin(\theta_1)\tan(\theta_1) + c_0 \frac{\pi}{2\theta_1} \rho^{\frac{\pi}{2\theta_1}-1} + c_1 \frac{3\pi}{2\theta_1} \rho^{\frac{3\pi}{2\theta_1}-1}.
\end{equation}
These first three terms have been included because they reveal three of the five behaviours that the wetting front can undertake, the three that exist for this case. The first term is constant across the wetting front, so moves all of the wetting front equally as illustrated in figure \ref{f:asymptotic_advance_const}. The value of the term is negative and so it is causing the wetting front to recede, although other terms will balance this in a wetting process causing the front to advance. Physically this can be understood as gravity attempting to reshape the wetted region such that it extends further downwards and has less of its mass at its top. The second term is a power of $\rho$ that is between zero and one, as illustrated in figure \ref{f:asymptotic_advance_root}. This causes the contact angle $\theta_1$ to change rapidly and does not cause the contact line to advance. The third and all subsequent terms are of a higher power than one, illustrated in \ref{f:asymptotic_advance_poly}, so they do not affect the contact angle or move the contact line, and only have an influence further along the wetting front.

Next the case when $\theta_1=\pi/2$, this time we extract the leading order terms from \eqref{eq:asymp_uth_2} to arrive at
\begin{equation}	\label{eq:asym_vs_right}
	v_s\sim  \frac{2\gamma}{\pi} \ln(\rho) + \left[ \frac{2\gamma}{\pi} + c_0 \right] + 3 c_1 \rho^2 .
\end{equation}
The first term is singular, as illustrated by figure \ref{f:asymptotic_advance_sing}. By the sign of the coefficient we see that the contact line is receding at a singular velocity, gravity is rapidly increasing the contact angle as it causes the fluid to fall. From the second term  we see that gravity is also causing the fluid to advance, so that the fluid is indeed receding near the surface of the solid substrate, and advancing below as in figure \ref{f:uvp_z_medium_g}. The constant term also includes an unspecified constant, which could cause the front to either advance or recede. The third term and all subsequent terms are, as before, of the type depicted in \ref{f:asymptotic_advance_poly}, affecting neither the contact angle nor the contact lines position.

Finally the case $\theta_1>\pi/2$ is very similar to the first case, except that the terms are of different orders and so have different effects. Ordering the terms by their dominance we see that
\begin{equation}	\label{eq:asym_vs_obtuse}
	v_s\sim  c_0 \frac{\pi}{2\theta_1} \rho^{\frac{\pi}{2\theta_1}-1} - \gamma \sin(\theta_1)\tan(\theta_1) + c_1 \frac{3\pi}{2\theta_1} \rho^{\frac{3\pi}{2\theta_1}-1}.
\end{equation}
The term that is now first is singular, as illustrated by figure \ref{f:asymptotic_advance_sing}. If $\gamma\neq0$ then we would anticipate that for $\theta_1\approx\pi/2$ that the contact line would be receding and the contact angle increasing, because this is the behaviour seen at $\pi/2$. For the contact angle to be physical it must be that eventually $c_0=0$ at some $\theta_1\in(\pi/2,\pi)$, otherwise the contact angle will increase to infinity. However, the behaviour may not be so trivial as there being a particular value of $\theta_1$ for each $\gamma$ at which $c_0=0$, it may be that the contact angle varies in a manner that depends on the geometry of the entire wetting front, increasing and decreasing until the entire wetting front has reached a suitable geometry. The second term has the same meaning as it did in the first case (where it was the first term). The third term causes different behaviour depending on $\theta_1$. For $\theta_1\in(\pi/2,3\pi/4)$ the power of $\rho$ is greater than unity, so does not affect the contact angle or move the contact line. For $\theta_1=3\pi/4$ power is one and affects the contact angle as illustrated in figure \ref{f:asymptotic_advance_lin}. For $\theta_1\in(3\pi/4,\pi)$ the power is between zero and one, so affects the contact angle as illustrated in figure \ref{f:asymptotic_advance_root}. All subsequent terms have power greater than one, and so do not affect the contact angle or move the contact angle.

It is important to realise that the terms that we discuss do add together, and so one term affecting the contact angle and another moving the wetting front in the far field will cause both the angle to change and the wetting front to move. In all cases the wetting front has the ability to advance, since they either have a constant term, or a singular term and a high power term. That is all cases except $\theta_1<\pi/2$ and $\gamma=0$ where the contact angle must change up to $\pi/2$ before the contact line can advance, and $\theta_1=\pi/2$ and $\gamma=0$ where the contact angle cannot change.

For the cases where the wetting front does actually recede, we have the additional issue that our problem formulation is only valid for wetting processes. We must assume that the de-wetting and re-wetting processes have the same physics as the wetting process. This should be verified.

Numerically speaking, any simulations that are run will not be able to simulate the singular behaviour with the accuracy that is desired for prediction. The numerical scheme would need to be specially designed to cope with this behaviour, and ours was not because we did not anticipate such an un-physical solution. However, we can produce some qualitative predictions which may be useful in guiding future developments in this area.

\subsection{Numerical Simulations} \label{ss:simulate}

\subsubsection{Large Initial Wetted Regions}

\begin{figure}[ptb]
	\centering
	\begin{tabular}{c c}
		\begin{subfigure}[t]{0.45\textwidth}
			\centering
			\includegraphics[width=\textwidth]{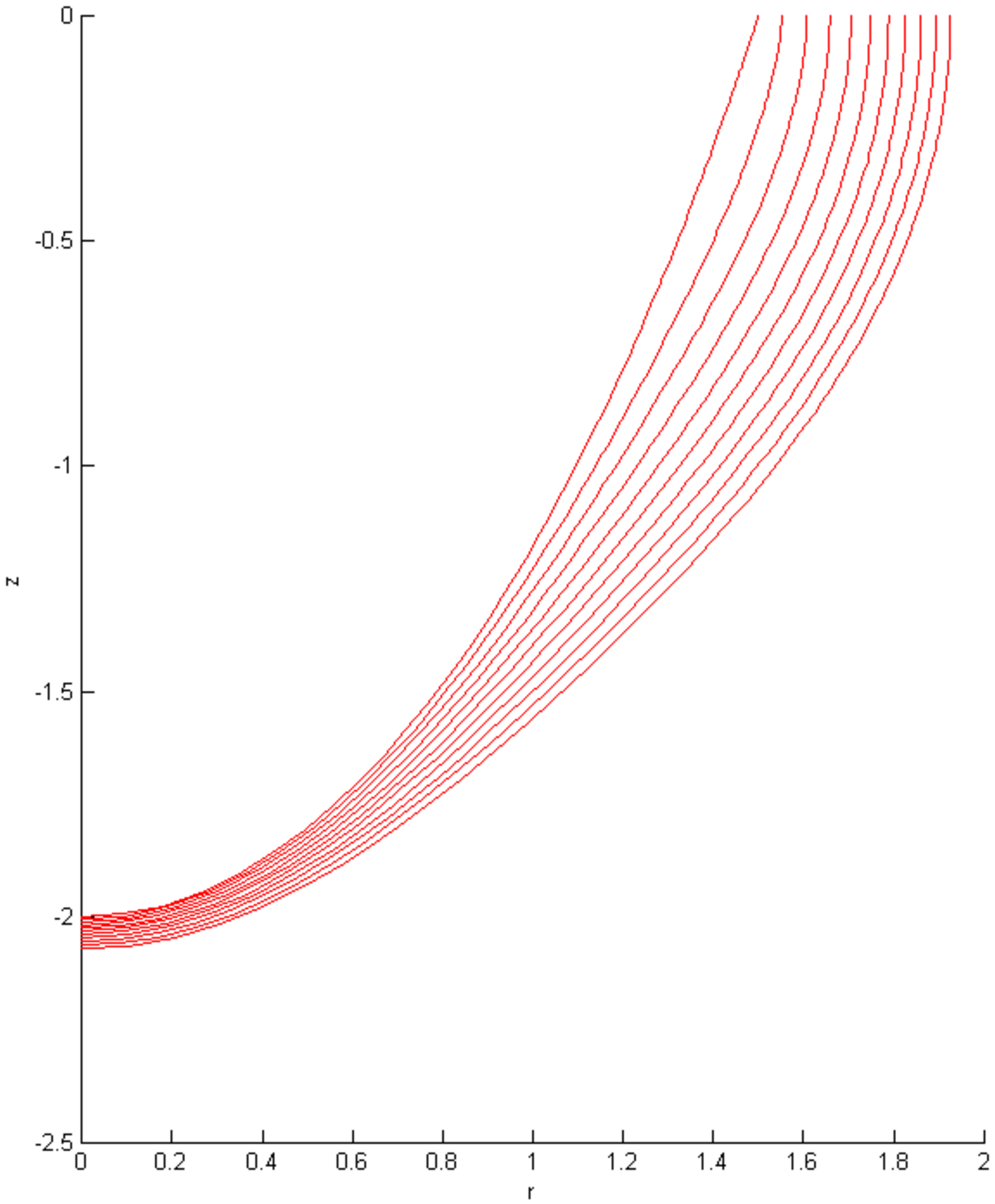}
			\caption{$\gamma=0$}
			\label{f:front_acute_ng}
		\end{subfigure}
		&
		\begin{subfigure}[t]{0.45\textwidth}
			\centering
			\includegraphics[width=\textwidth]{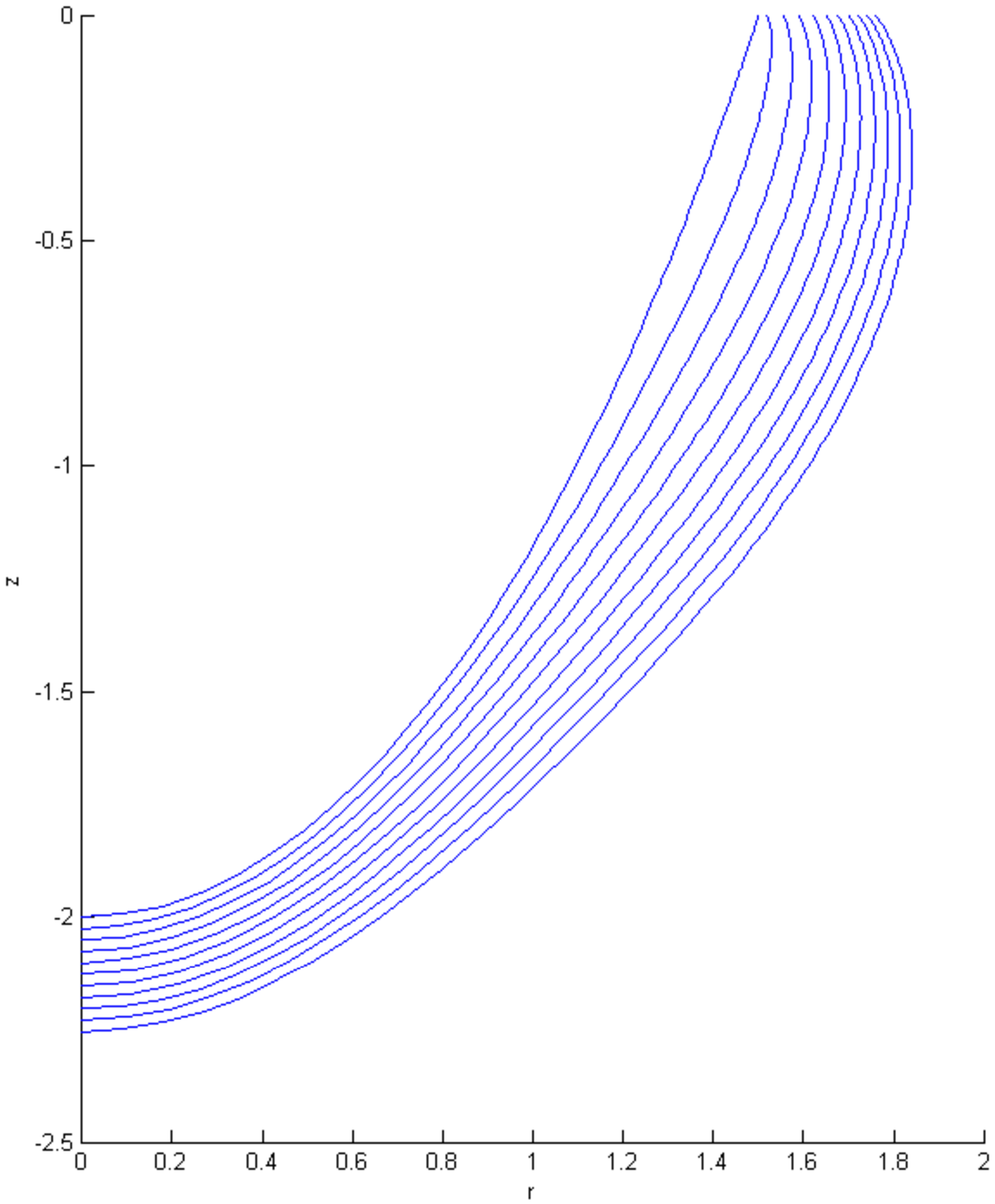}
			\caption{$\gamma=0.2$}
			\label{f:front_acute_g}
		\end{subfigure}
	\end{tabular}
	\\
	\begin{subfigure}[t]{\textwidth}
		\centering
		\includegraphics[width=0.45\textwidth]{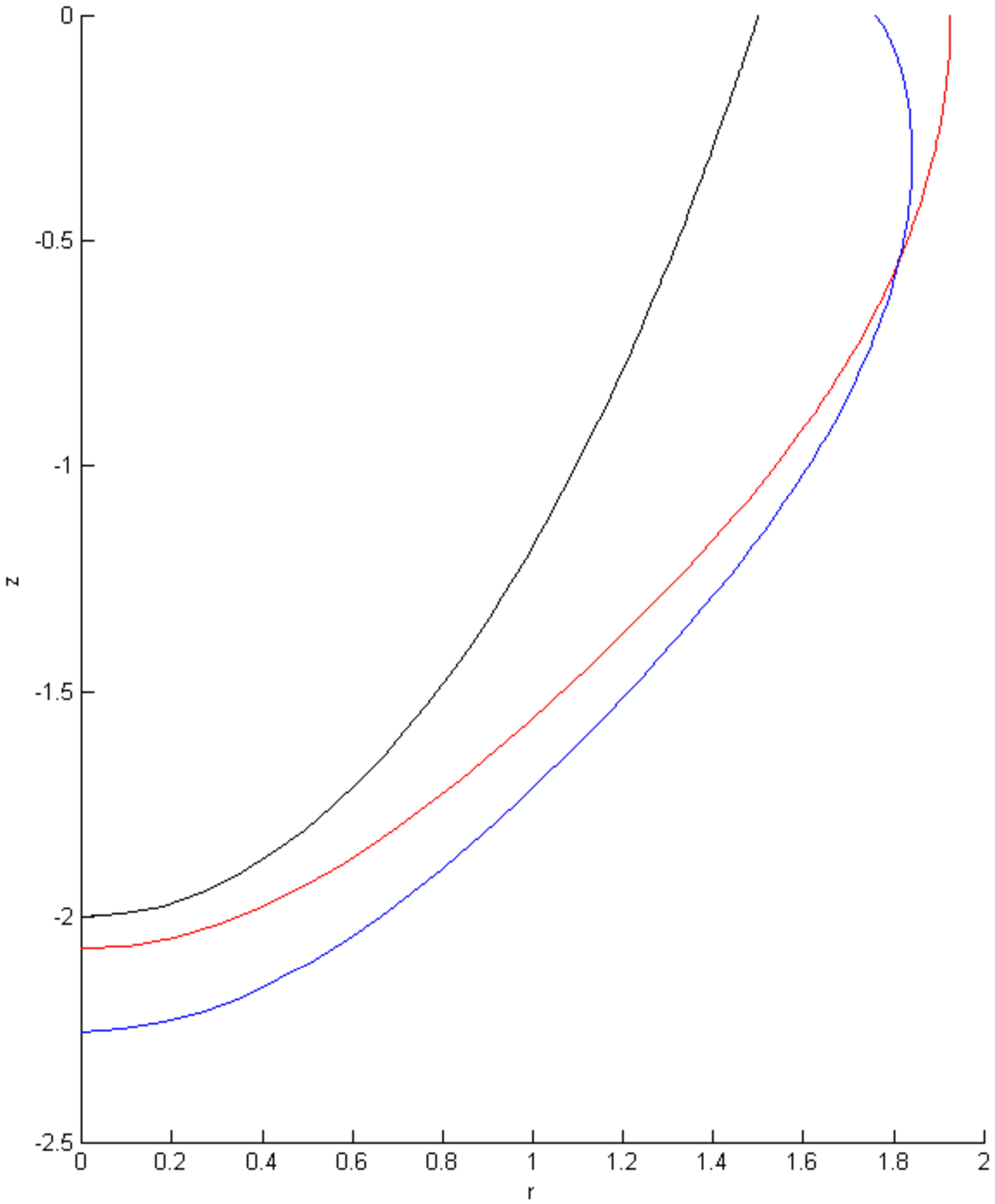}
		\caption{Comparison of the wetting fronts at time $t=1$, with the original front in black.}
		\label{f:front_acute_comp}
	\end{subfigure}
	\\
	\begin{subfigure}[t]{0.9\textwidth}
		\centering
		\includegraphics[width=\textwidth]{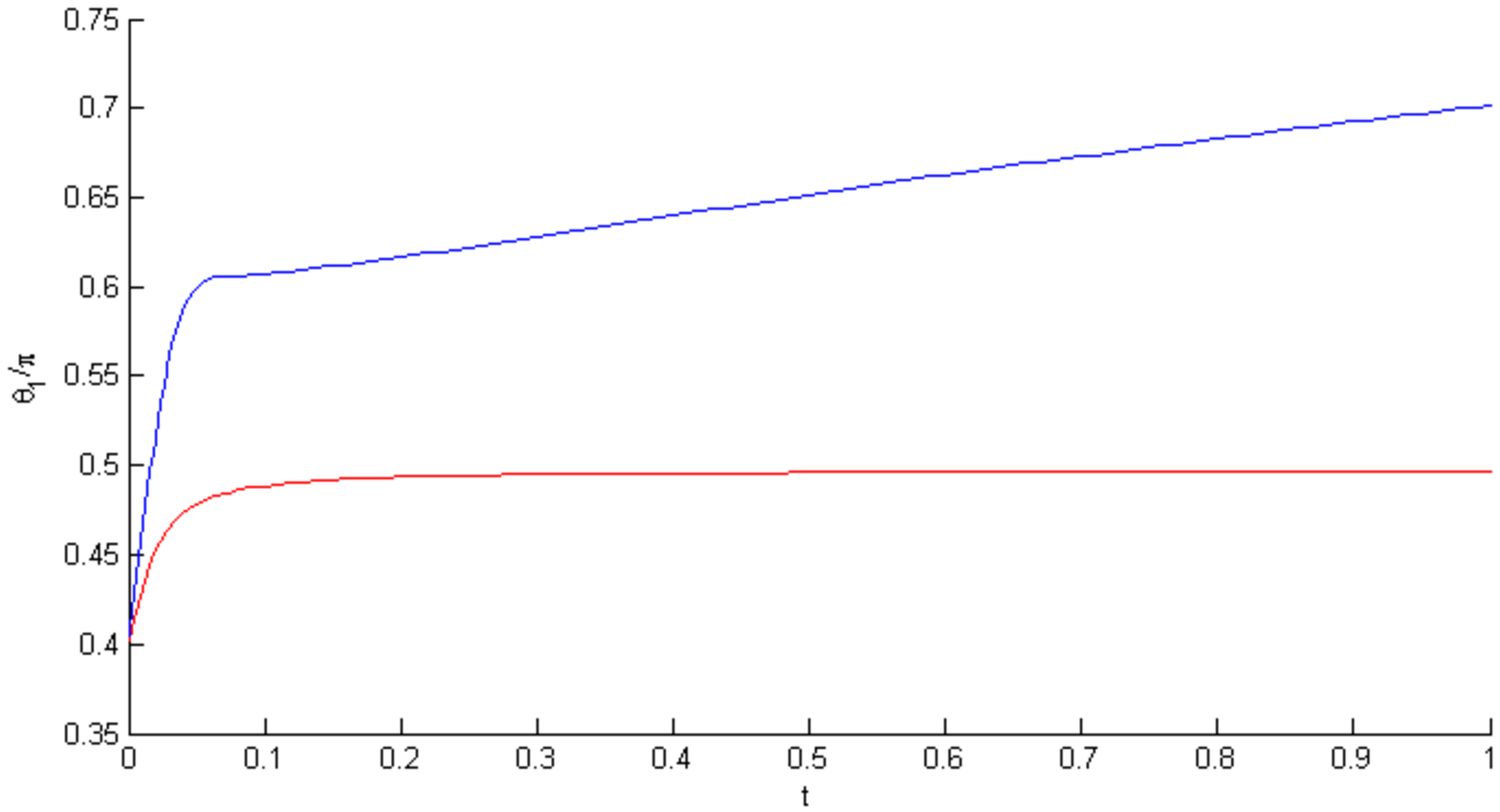}
		\caption{The variation of $\theta_1$ with time.}
		\label{f:front_acute_ang}
	\end{subfigure}
	\caption{Plots depicting the dynamics of the wetting front for initial conditions $\theta_1=0.4\pi$, $r_f=1.5$ and $H=2$. Red curves are for without gravity and blue are for with gravity. (a) and (b) include the wetting front at times $t \in \{0,0.1,\ldots,1\}$.}
	\label{sf:acute_set}
\end{figure}

\begin{figure}[ptb]
	\centering
	\begin{tabular}{c c}
		\begin{subfigure}[t]{0.45\textwidth}
			\centering
			\includegraphics[width=\textwidth]{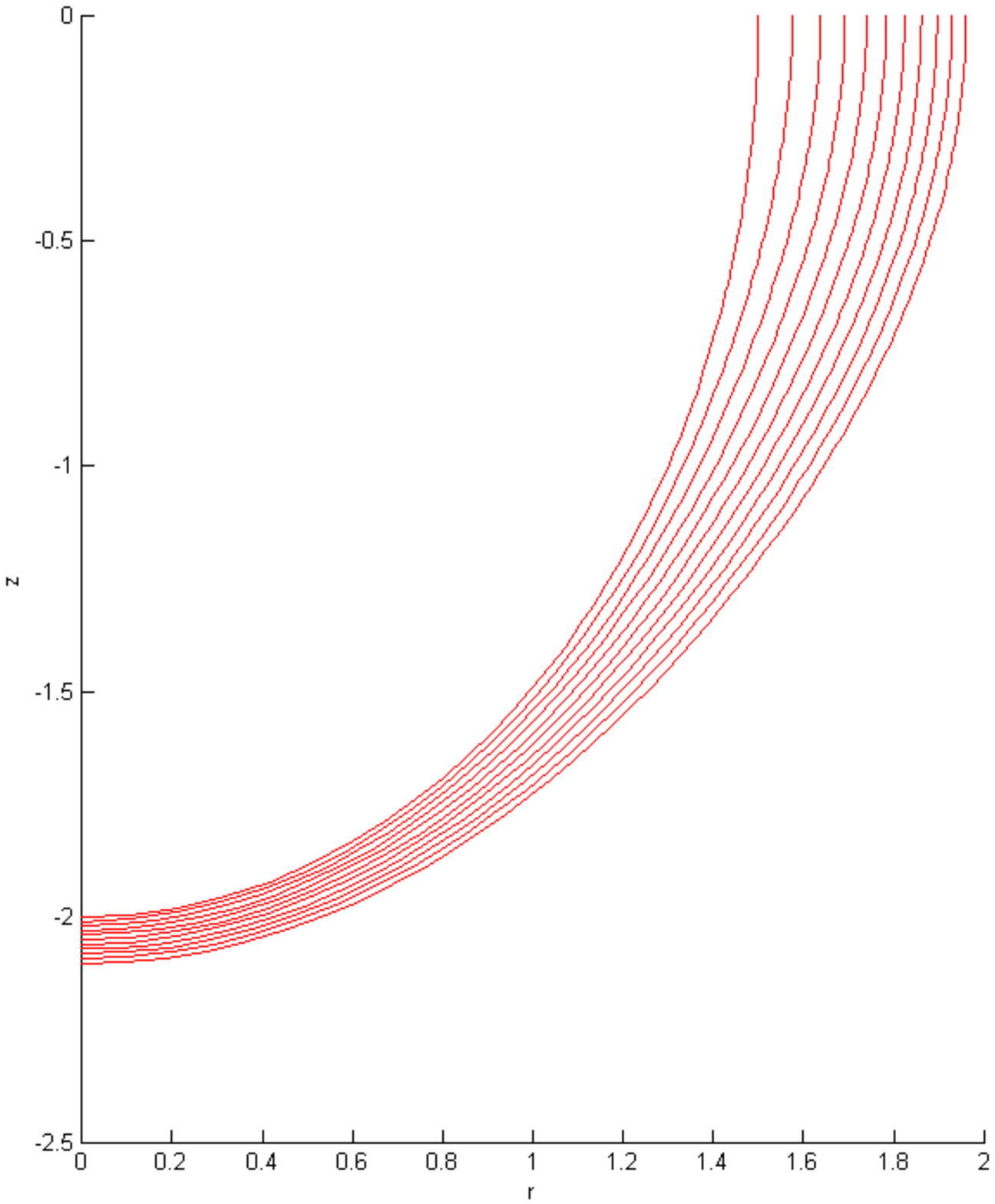}
			\caption{$\gamma=0$}
			\label{f:front_right_ng}
		\end{subfigure}
		&
		\begin{subfigure}[t]{0.45\textwidth}
			\centering
			\includegraphics[width=\textwidth]{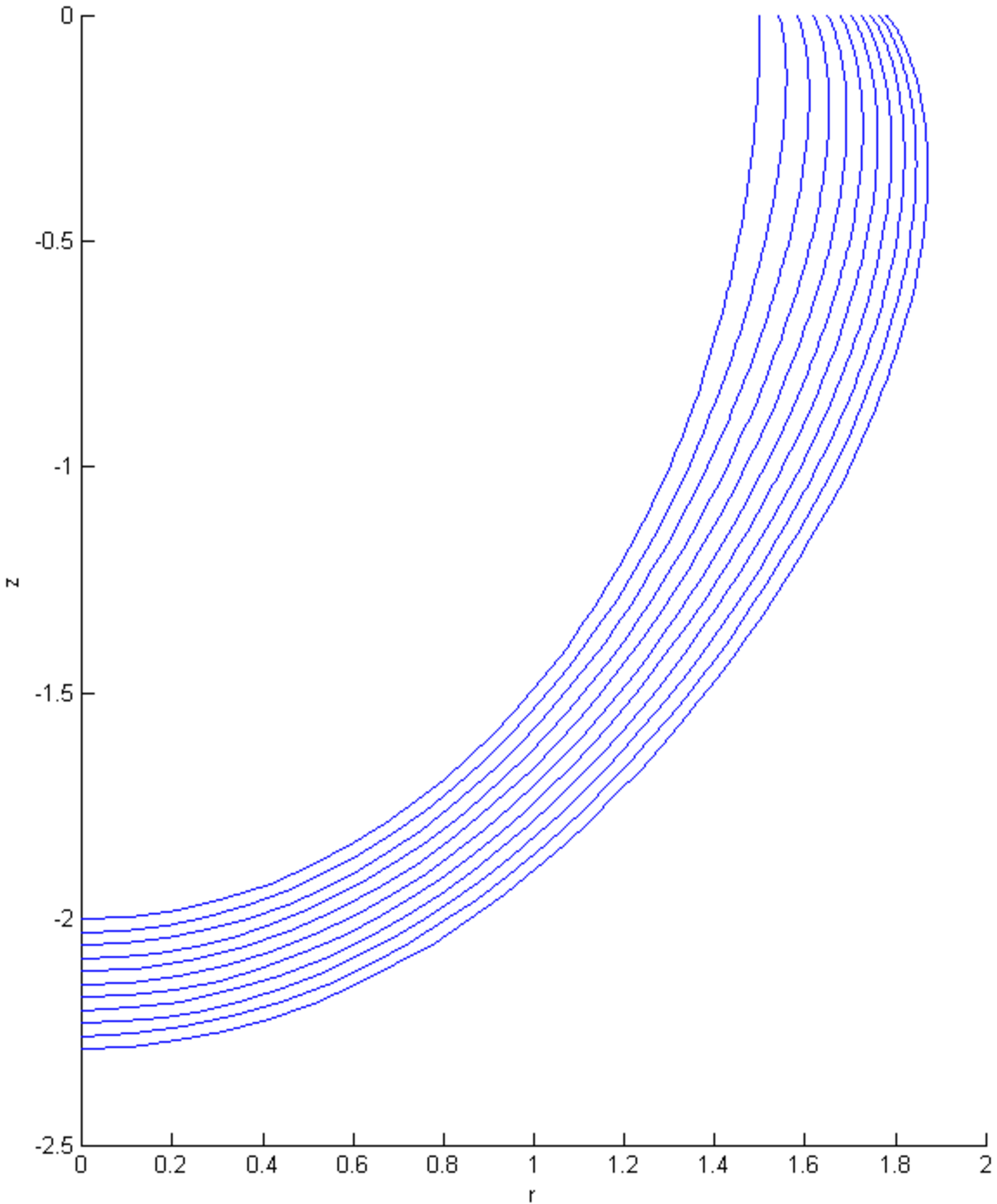}
			\caption{$\gamma=0.2$}
			\label{f:front_right_g}
		\end{subfigure}
	\end{tabular}
	\\
	\begin{subfigure}[t]{\textwidth}
		\centering
		\includegraphics[width=0.45\textwidth]{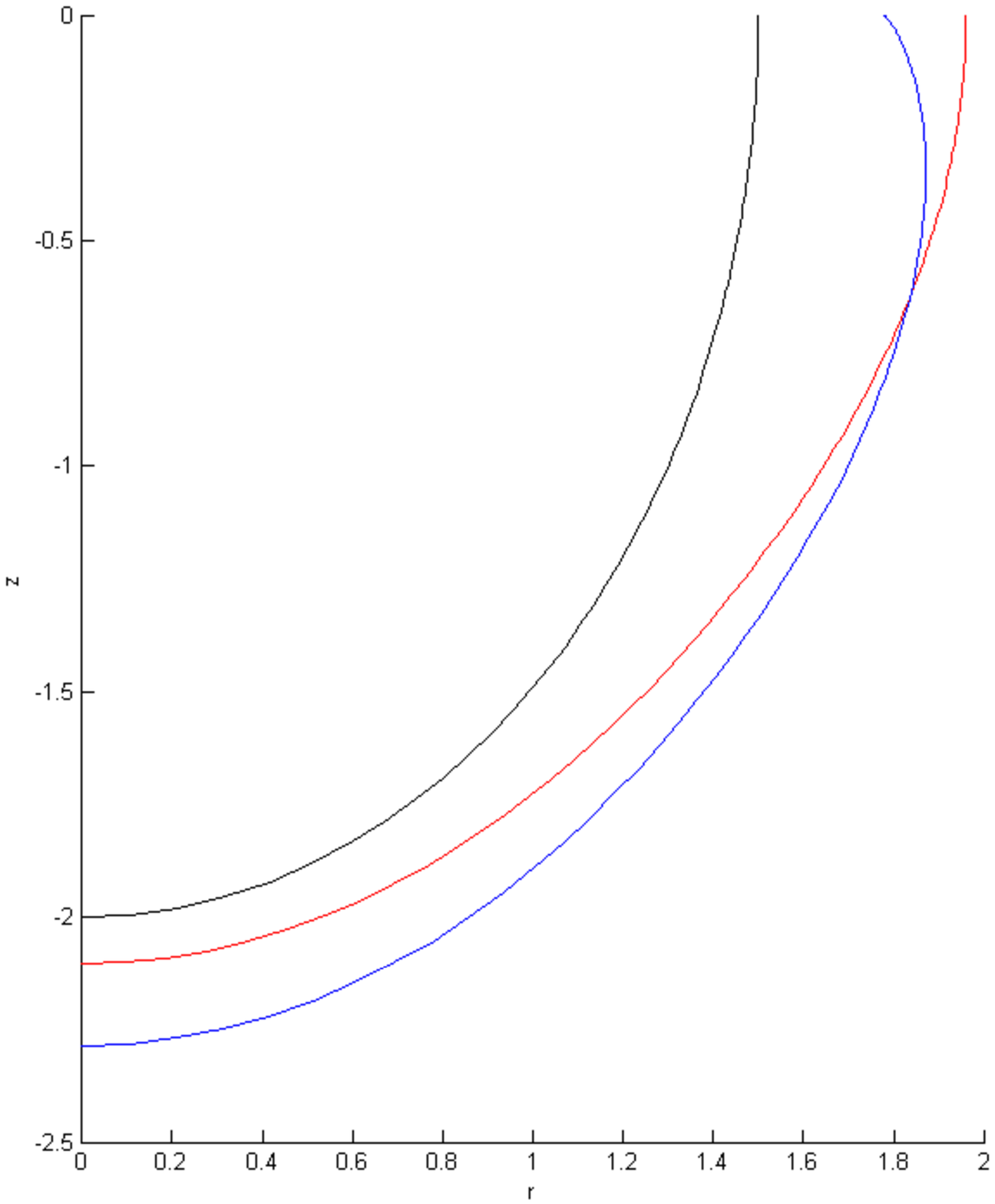}
		\caption{Comparison of the wetting fronts at time $t=1$, with the original front in black.}
		\label{f:front_right_comp}
	\end{subfigure}
	\\
	\begin{subfigure}[t]{0.9\textwidth}
		\centering
		\includegraphics[width=\textwidth]{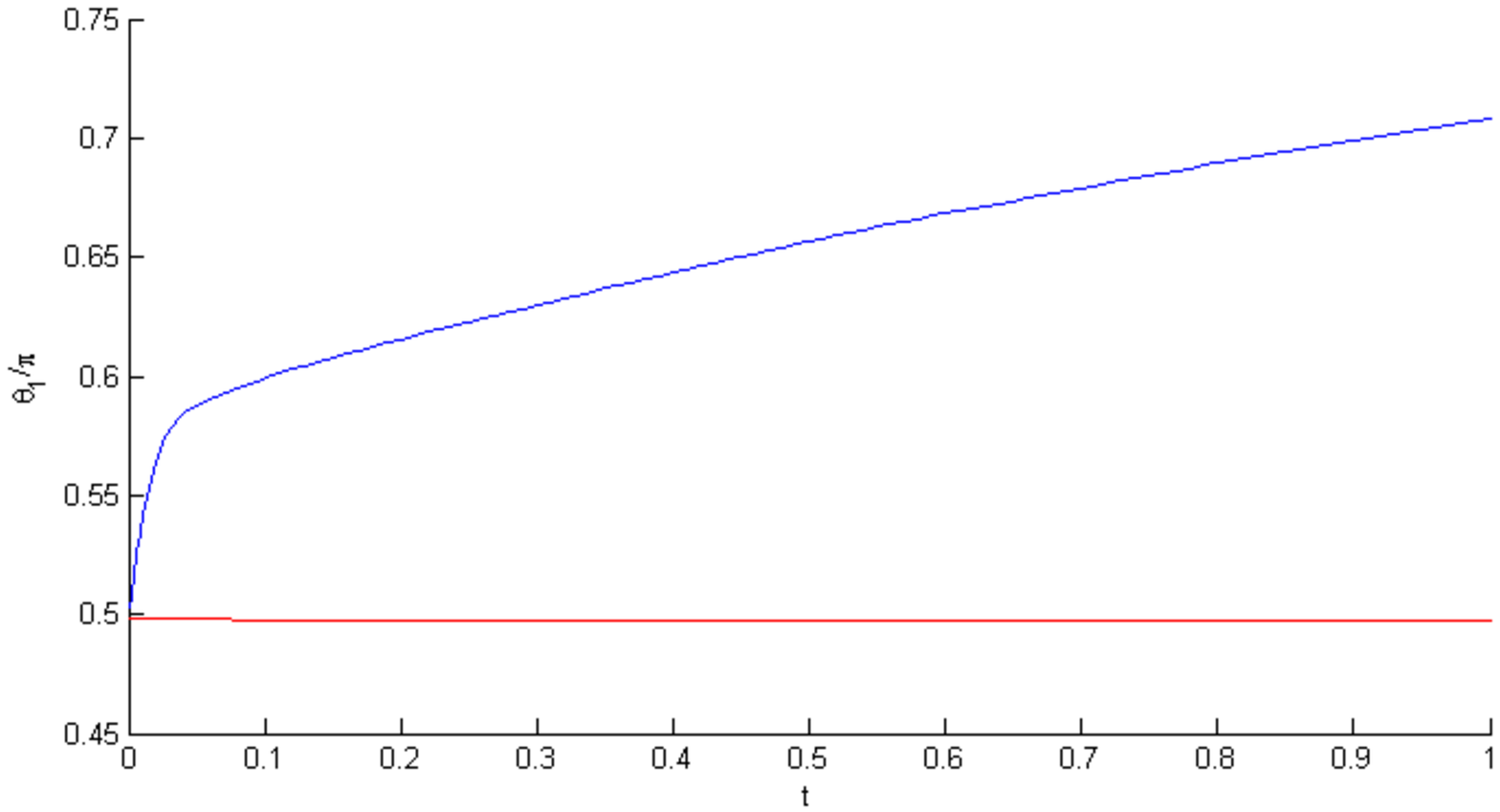}
		\caption{The variation of $\theta_1$ with time.}
		\label{f:front_right_ang}
	\end{subfigure}
	\caption{Plots depicting the dynamics of the wetting front for initial conditions $\theta_1=0.5\pi$, $r_f=1.5$ and $H=2$. Red curves are for without gravity and blue are for with gravity. (a) and (b) include the wetting front at times $t \in \{0,0.1,\ldots,1\}$.}
	\label{sf:right_set}
\end{figure}

\begin{figure}[ptb]
	\centering
	\begin{tabular}{c c}
		\begin{subfigure}[t]{0.45\textwidth}
			\centering
			\includegraphics[width=\textwidth]{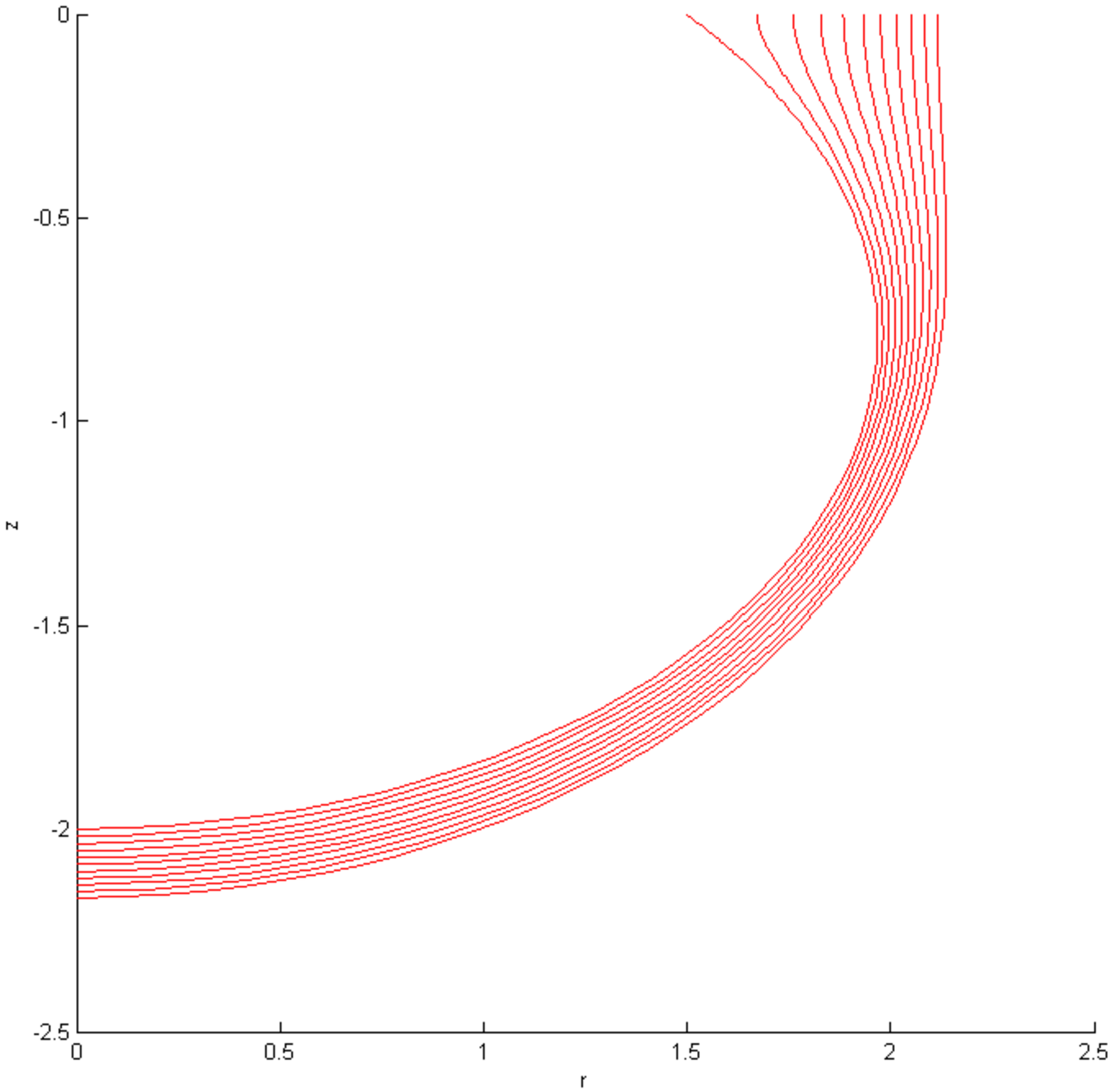}
			\caption{$\gamma=0$}
			\label{f:front_obtuse_ng}
		\end{subfigure}
		&
		\begin{subfigure}[t]{0.45\textwidth}
			\centering
			\includegraphics[width=\textwidth]{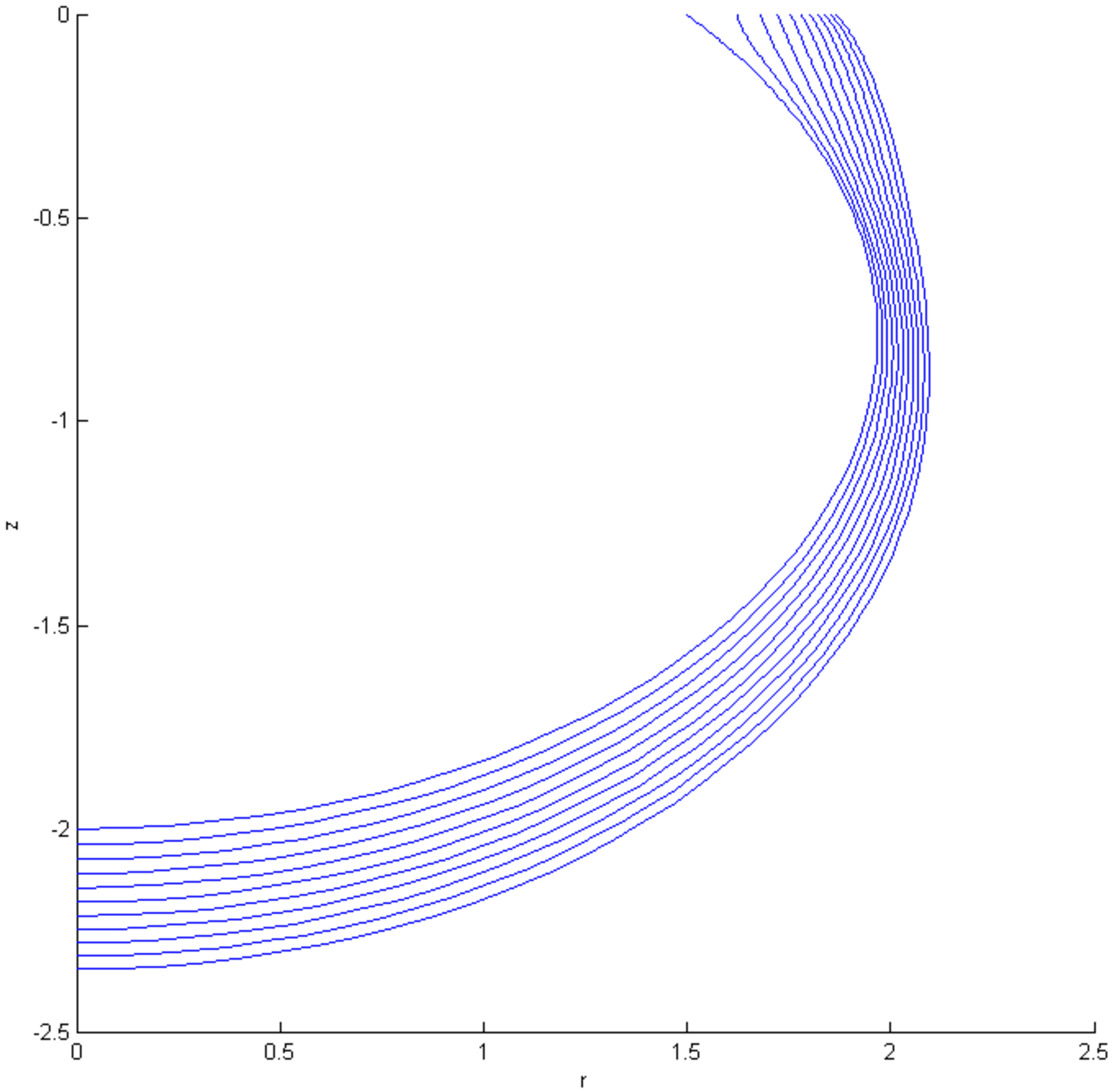}
			\caption{$\gamma=0.2$}
			\label{f:front_obtuse_g}
		\end{subfigure}
	\end{tabular}
	\\
	\begin{subfigure}[t]{\textwidth}
		\centering
		\includegraphics[width=0.45\textwidth]{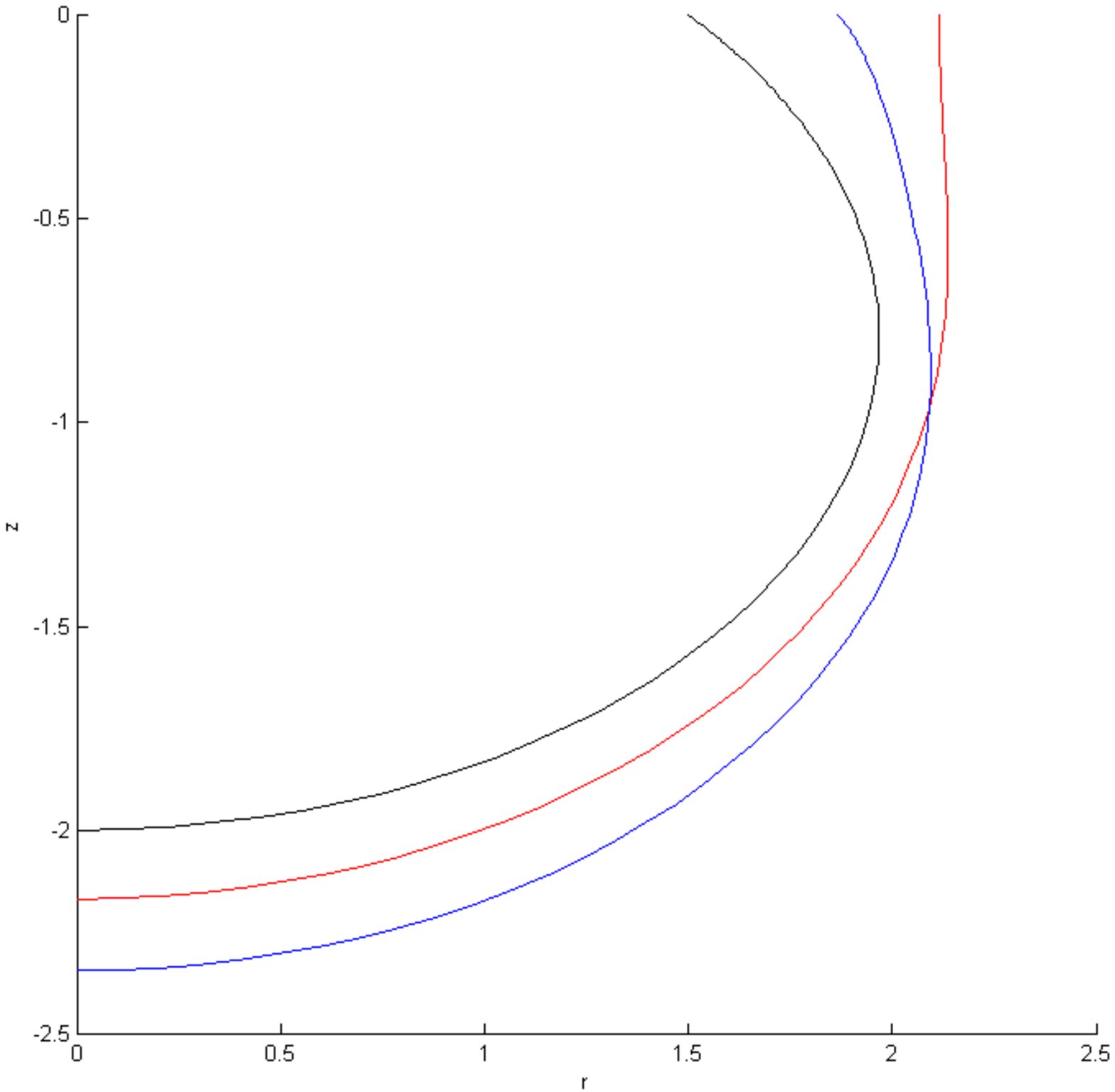}
		\caption{Comparison of the wetting fronts at time $t=1$, with the original front in black.}
		\label{f:front_obtuse_comp}
	\end{subfigure}
	\\
	\begin{subfigure}[t]{0.9\textwidth}
		\centering
		\includegraphics[width=\textwidth]{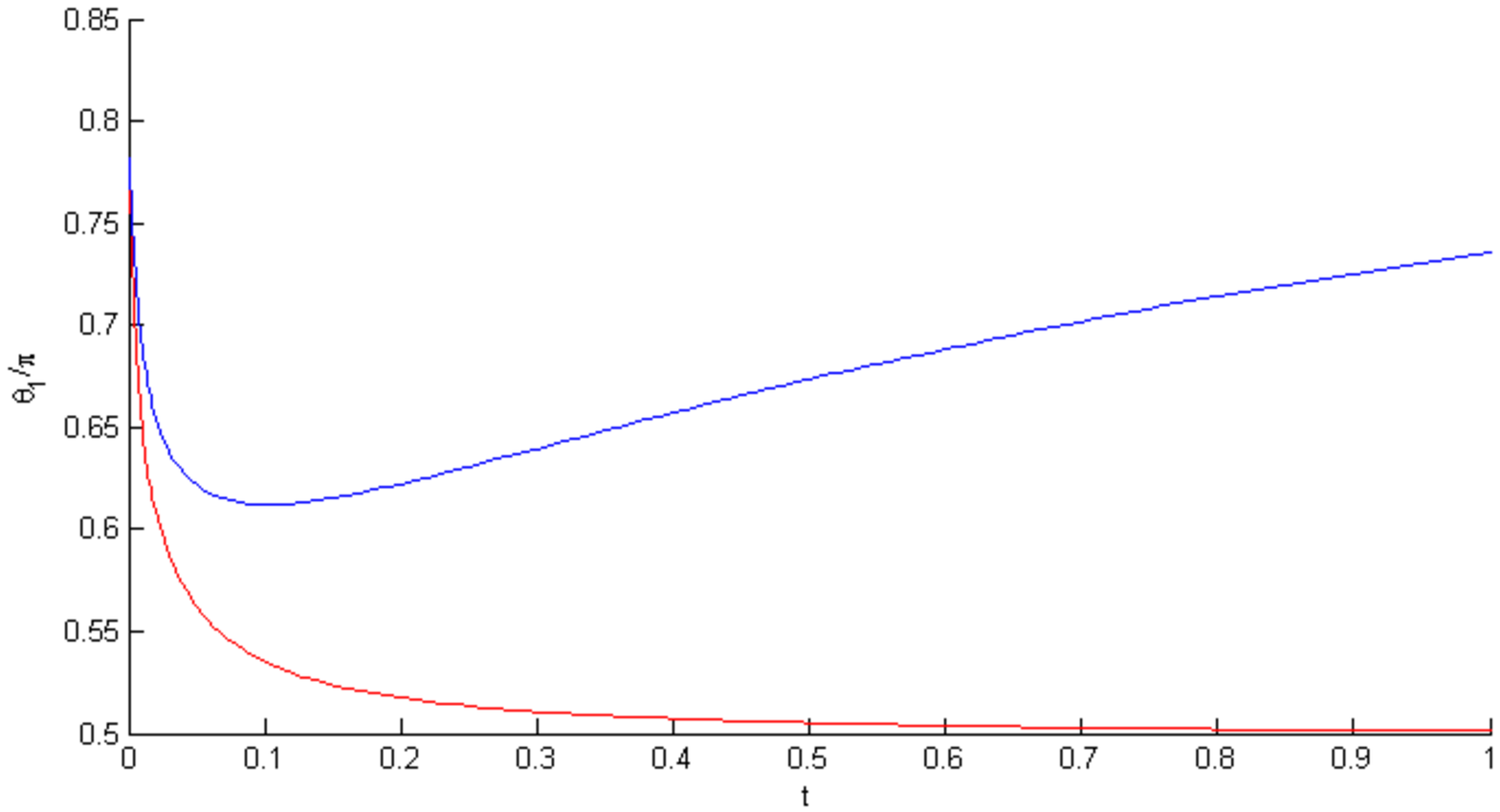}
		\caption{The variation of $\theta_1$ with time.}
		\label{f:front_obtuse_ang}
	\end{subfigure}
	\caption{Plots depicting the dynamics of the wetting front for initial conditions $\theta_1=0.8\pi$, $r_f=1.5$ and $H=2$. Red curves are for without gravity and blue are for with gravity. (a) and (b) include the wetting front at times $t \in \{0,0.1,\ldots,1\}$.}
	\label{sf:obtuse_set}
\end{figure}

\begin{figure}[ptb]
	\centering
	\begin{subfigure}[t]{\textwidth}
		\centering
		\includegraphics[width=0.65\textwidth]{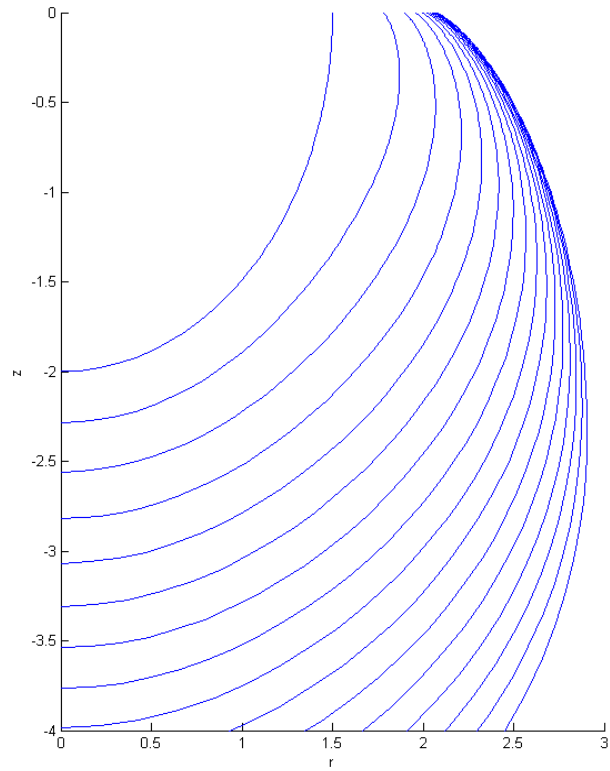}
		\caption{Plot of the wetting front at times $t\in\{0,1,\ldots,10\}$}
		\label{f:front_long}
	\end{subfigure}
	\\
	\begin{subfigure}[t]{\textwidth}
		\centering
		\includegraphics[width=0.7\textwidth]{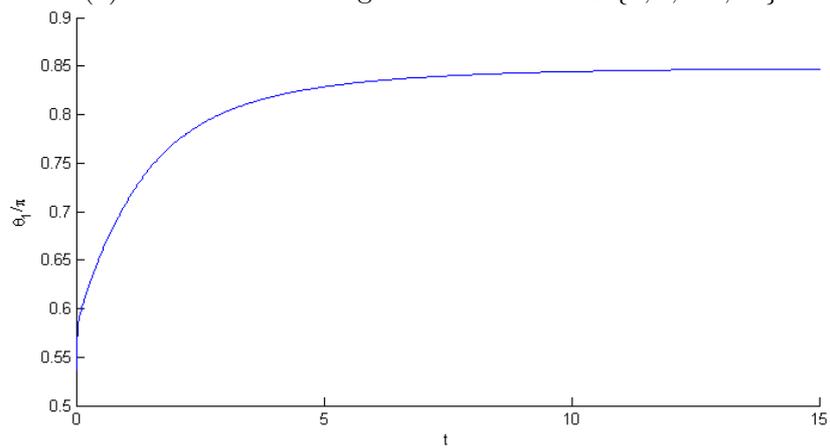}
		\caption{The variation of $\theta_1$ with time.}
		\label{f:front_long_ang}
	\end{subfigure}
	\\
	\begin{subfigure}[t]{\textwidth}
		\centering
		\includegraphics[width=0.7\textwidth]{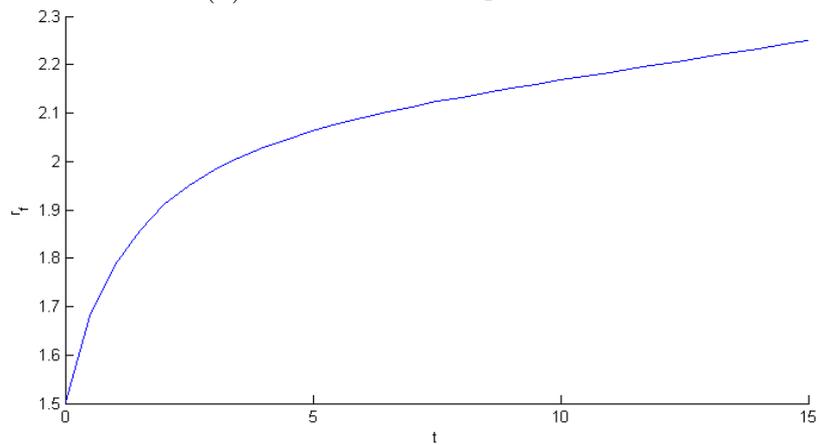}
		\caption{The variation of $r_f$ with time.}
		\label{f:front_long_cl2}
	\end{subfigure}
	\caption{Plots depicting the dynamics of the wetting front for initial conditions $\theta_1=0.5\pi$, $r_f=1.5$ and $H=2$ for $\gamma=0.2$. Shows late times of the same situation as figure \ref{sf:right_set}.}
	\label{sf:long_set}
\end{figure}

\begin{figure}[ptb]
	\centering
	\includegraphics[width=\textwidth]{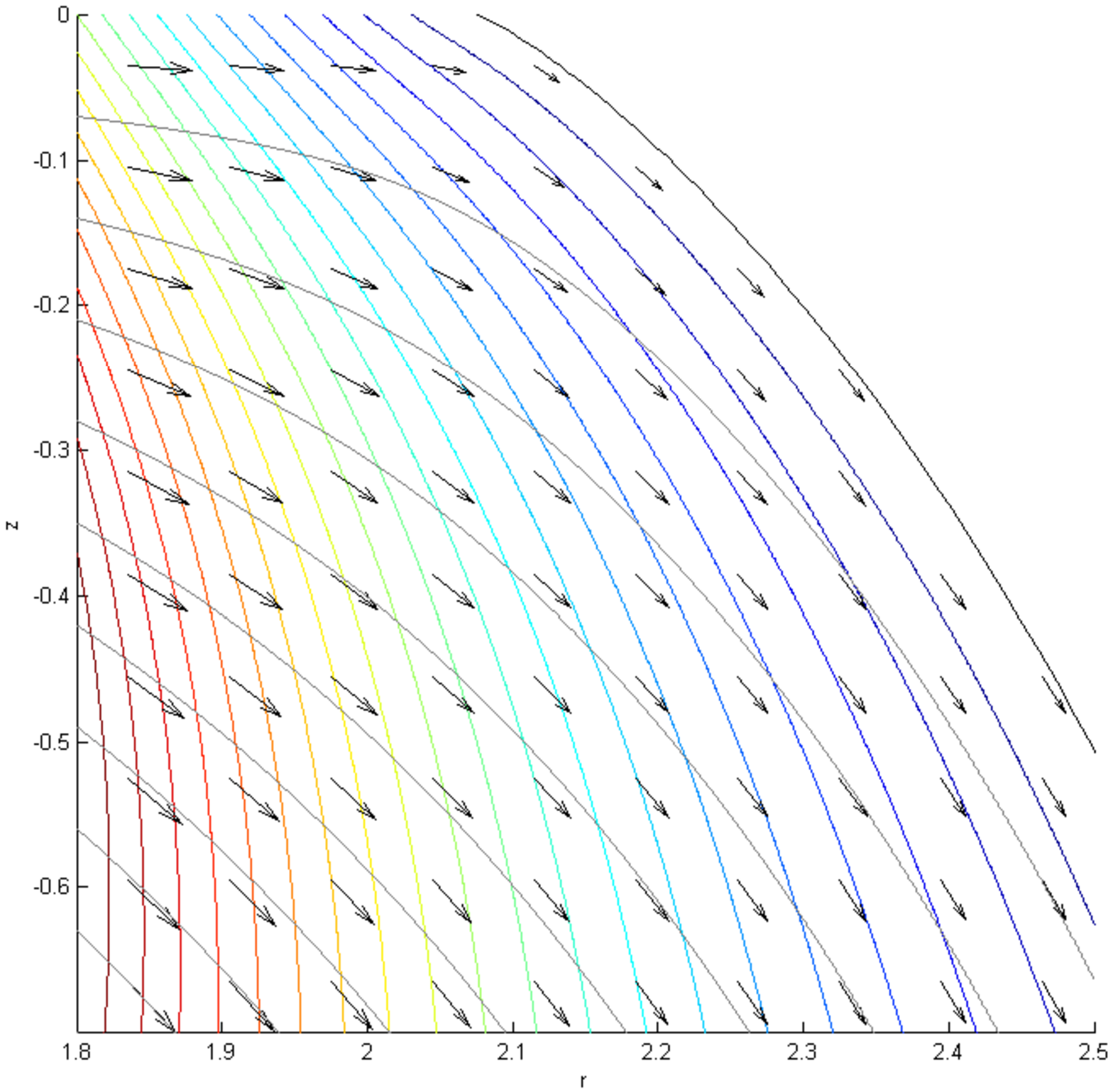}
	\caption{Plot of velocity and pressure locally to the contact line for the dynamics in figure \ref{sf:long_set} at time $t=15$. The front has reached a state where the fluid largely flows along it causing the contact angle to become fixed and the contact line to become slow.}
	\label{f:uvp_z_long}
\end{figure}

The aim of this section is to produce simulations of an already established wetted region to see the contact angle variation and advancement of the wetting front. We shall compare the advancement of the wetting front both without gravity ($\gamma=0$) and with ($\gamma=0.2$). For a typical set of figures see \cref{sf:acute_set}. The plots without gravity are in red and with gravity are in blue. (a) and (b) are plots of the wetting front at uniformly distributed points in time, (c) is a comparison of the wetting front at the latest time simulated and (d) shows the contact angle variation.

This figure (figure \ref{sf:acute_set}) depicts the dynamics for an initially acute contact angle. It shows that the contact line $C_1$ advances much slower with gravity than without, this should be expected from the discussion of the asymptotic analysis in the previous section, where we showed that gravity `pulls' the wetting front back local to $C_1$. Around the bottom of the front, close to $C_0$, gravity can be seen to aid the advancement of the wetting front, this should be no surprise. The contact angle variation is consistent with our asymptotic analysis. Without gravity, the leading order terms in \eqref{eq:asym_vs_right} are linear and quadratic, neither of which cause contact angle variation. Our analysis showed that $\theta_1(t)=\pi/2$ is a a solution, now our numerical result show us that it is stable. With gravity, the contact angle initially increases very rapidly, as we argued that it should for $\theta_1 \approx \pi/2$. It then slows down to what appears to be a linear function of time, this cannot continue since that would result in $\theta_1>\pi$ which is not physical. The behaviour at greater times will be discussed later.

Figure \ref{sf:right_set} is for the wetting front initially perpendicular to the substrate surface, and shows very similar results. The reader should briefly compare figures \ref{f:front_acute_comp} and \ref{f:front_right_comp}. We might naively expect that the initially larger wetted region should remain larger, but this is not the case, the smaller advances faster to catch up producing indistinguishable results.

This is not the case for an initially obtuse contact angle, as depicted in figure \ref{sf:obtuse_set}, although this is likely because the initial wetted region occupies space that the previous two cases do not reach in the times that we consider. It is likely that if we were to run the simulation over perhaps as little as five units of time that the wetted regions reached would be indistinguishable. The other interesting behaviour of this front is that of the contact angle. Without gravity the contact angle converges to $\pi/2$ as always, but with gravity it initially decreases, and then changes to being increasing. Looking at figure \ref{f:front_obtuse_g}, at time $0.1$ the contact line $C_1$ has advanced greatly but the front local to it has not advanced as much. It would seem that this contact line is initially too close to the drawing area, and that during rapid advancements the contact angle $\theta_1$ becomes closer to $\pi/2$. We will see a further example of this in the next section. With regard to our discussion of \eqref{eq:asym_vs_obtuse}, it would seem that $c_0$ does indeed change sign during advancements (see figure \ref{f:front_obtuse_ang}), and that the contact angle does not monotonically tend towards a prescribed value for all time, although it may do so as $t\rightarrow\infty$.

Finally, we consider the large times for the wetting front under the effect of gravity. We impose initial condition $\theta_1=\pi/2$ and simulate. From \ref{f:front_long} we see very clearly that the contact line $C_1$ slows down as it advances, and that the point $C_0$ moves at approximately uniform speed, the effect of gravity dominating the motion. From figure \ref{f:front_long_ang} we see that the contact angle does in fact tend to a constant value. We cannot reach any conclusions about the long time limit of $r_f$ (the radial coordinate of $C_1$) from the data that we have, it may tend towards a constant value, or may continue to increase slowly up to infinity. We also plot velocity and pressure local to $C_1$ in the style of section \ref{ss:pv_dist} in figure \ref{f:uvp_z_long}. It shows that the velocities on the wetting front are almost tangential to it, the fluid falling under gravity, which is why the front is dramatically slower than without gravity where the velocity distribution would be similar to that plotted in figure \ref{f:uvp_z_obtuse_ng}.

In this section we have presented the first set of results for the dynamics of the wetting front, but there is still much to investigate. The most important unresolved issues are how the limit of $\theta_1$ and $r_f$ as $t\rightarrow\infty$ depends on $\gamma$, and whether $r_f$ is even convergent. In addition we discussed how the wetting fronts we produced appear to converge on the same dynamics as time passes. It is conceivable that in the state space of all possible wetting fronts there is a stable manifold that all (or a large subset of) physical initial conditions converge onto and move along as time passes. This stable manifold would have to be the set of wetting fronts produced from the initial condition of $\Gamma_0=\{(r,z) \: : \: r\in[0,1] , z=0\}$, the wetted region of zero volume. It is stressed that, at present, this is only a possibility, although one worth investigation. 

\subsubsection{A Small Initial Wetted Region}

\begin{figure}[ptb]
	\centering
	\begin{tabular}{c c}
		\begin{subfigure}[t]{0.45\textwidth}
			\centering
			\includegraphics[width=\textwidth]{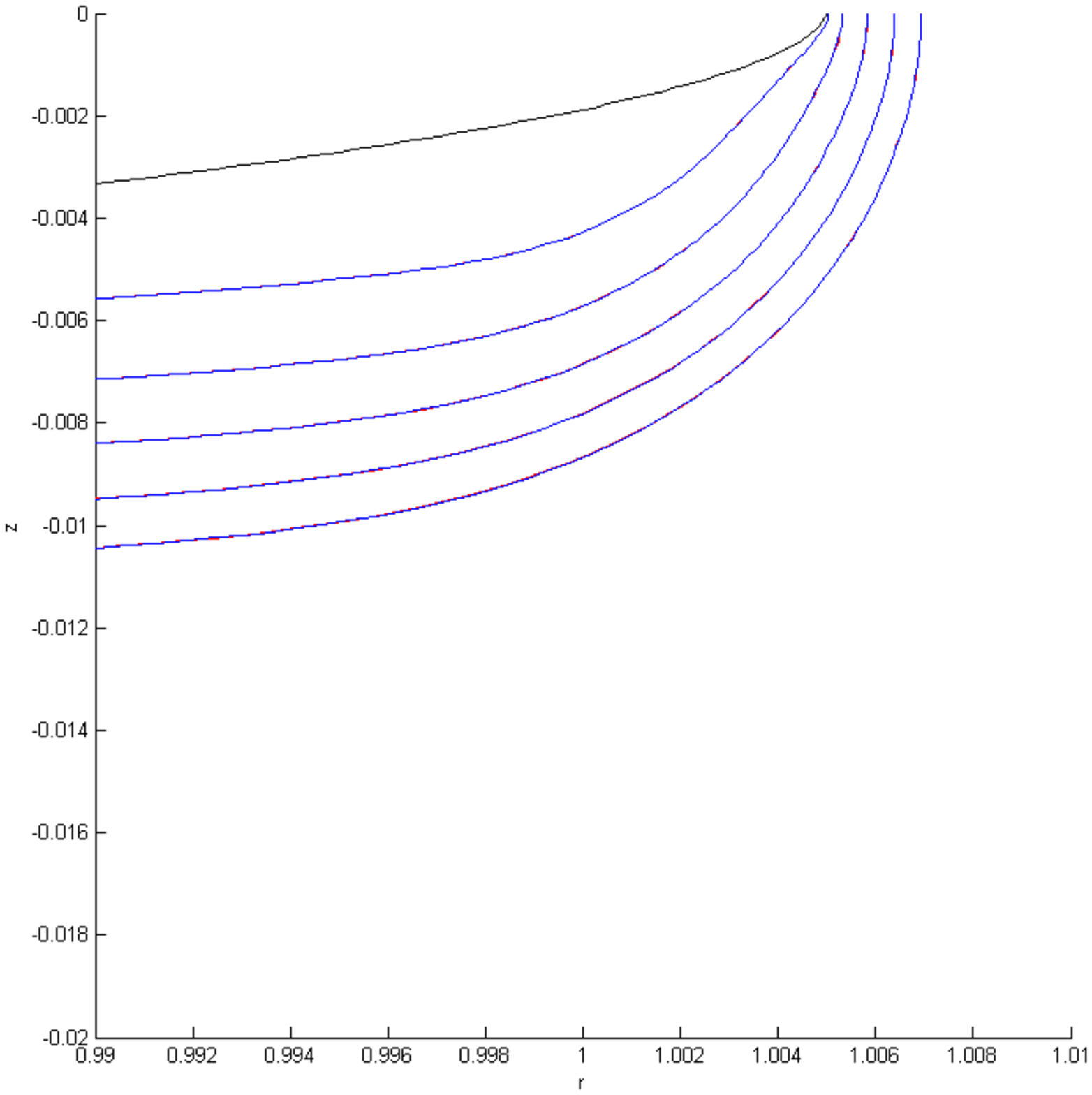}
			\caption{$t\in\{0,1\cdot10^{-5},\ldots,5\cdot10^{-5}\}$}
			\label{f:front_imbib_0}
		\end{subfigure}
		&
		\begin{subfigure}[t]{0.45\textwidth}
			\centering
			\includegraphics[width=\textwidth]{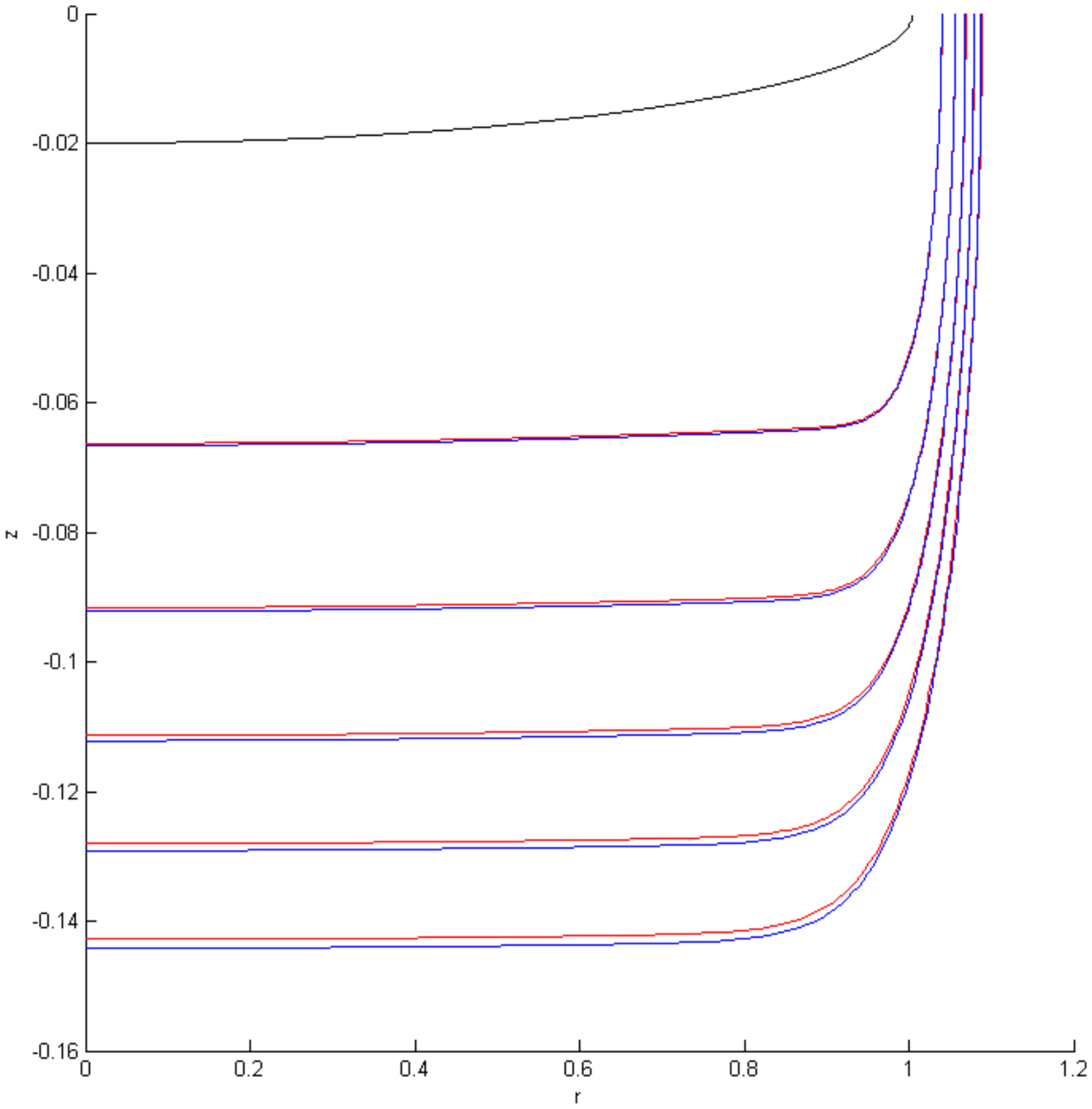}
			\caption{$t\in\{0,2\cdot10^{-3},\ldots,10\cdot10^{-3}\}$}
			\label{f:front_imbib_1}
		\end{subfigure}
	\end{tabular}
	\\
	\begin{subfigure}[t]{\textwidth}
		\centering
		\includegraphics[width=0.75\textwidth]{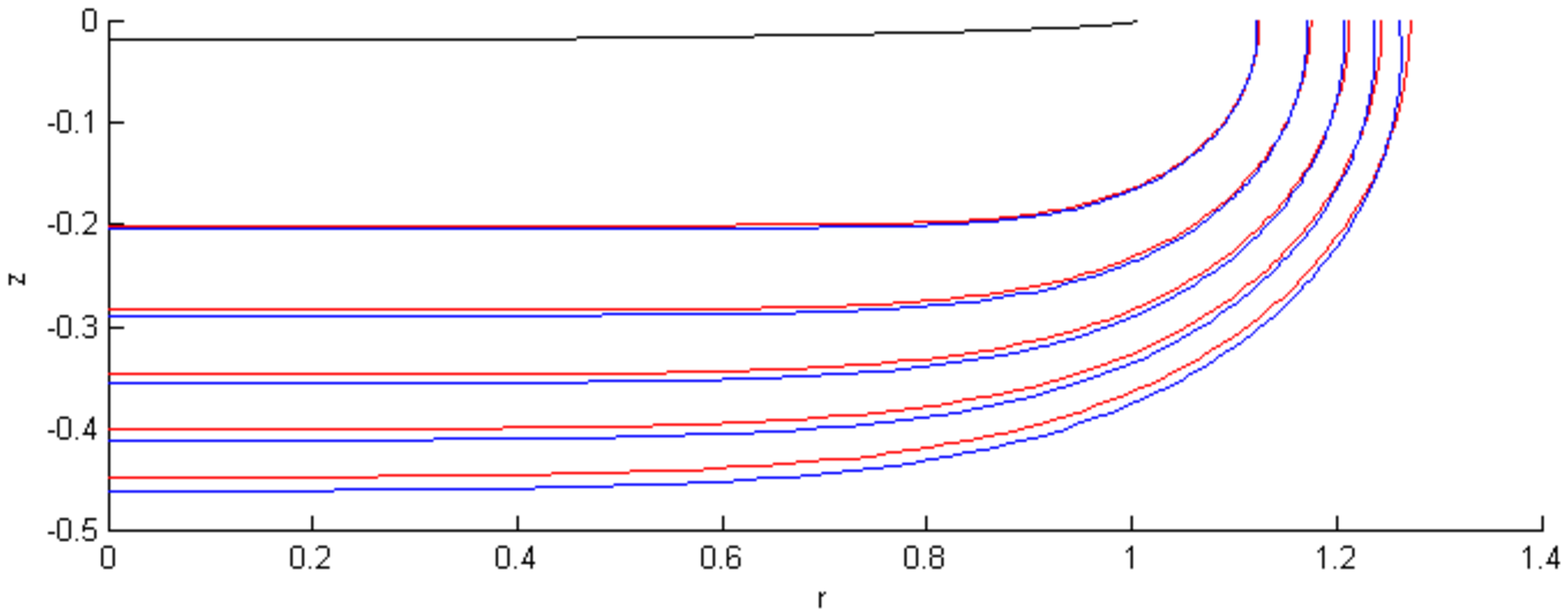}
		\caption{$t\in\{0,2\cdot10^{-2},\ldots,10\cdot10^{-2}\}$}
		\label{f:front_imbib_2}
	\end{subfigure}
	\\
	\begin{subfigure}[t]{\textwidth}
		\centering
		\includegraphics[width=0.75\textwidth]{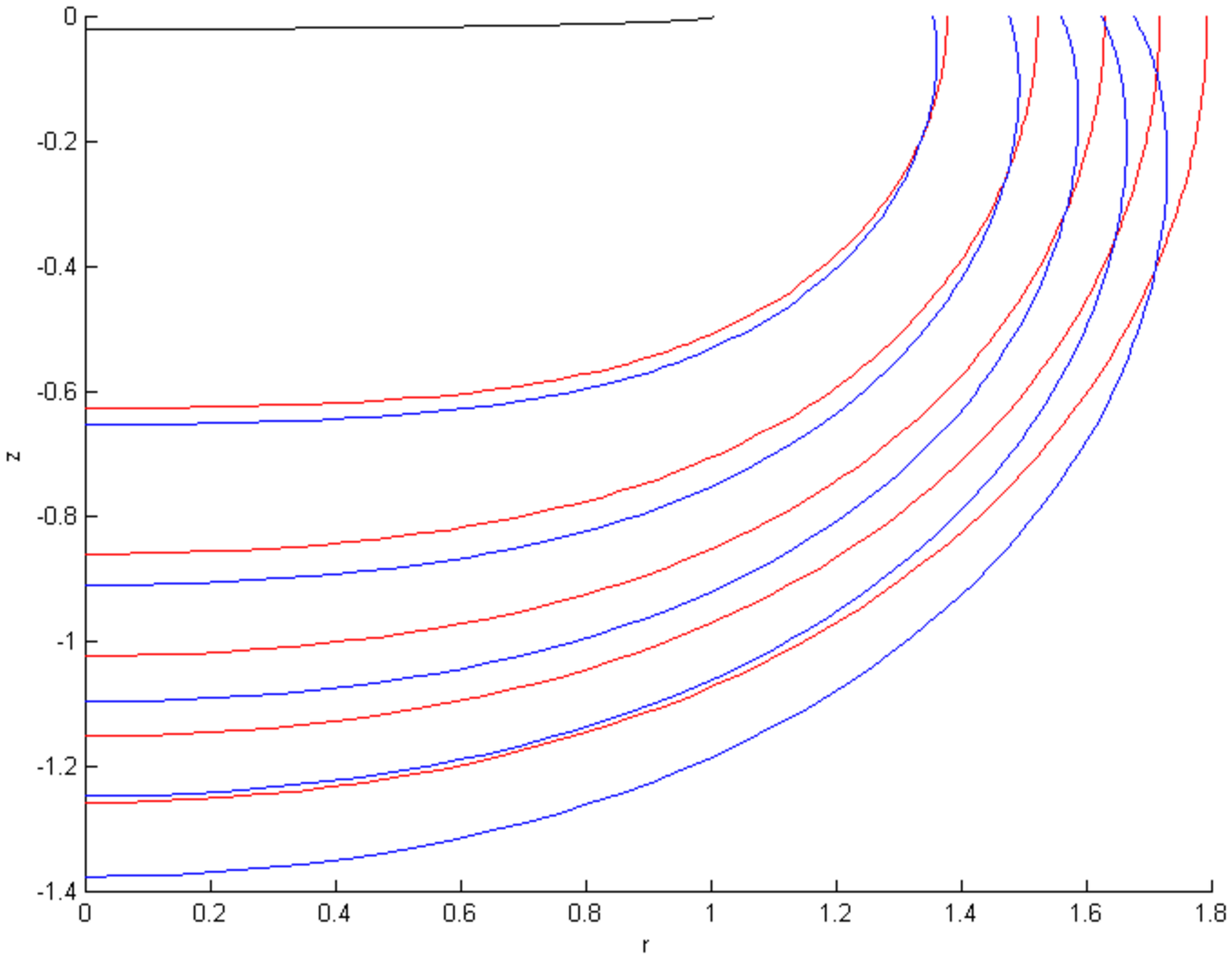}
		\caption{$t\in\{0,2\cdot10^{-1},\ldots,10\cdot10^{-1}\}$}
		\label{f:front_imbib_3}
	\end{subfigure}
	\caption{Plots depicting the dynamics of the wetting front for initial conditions $\theta_1=0.5\pi$, $r_f=1.005$ and $H=0.02$. Red curves are for without gravity and blue are for with gravity ($\gamma=0.2$), the initial front is plotted in black. Notice that (a) only includes part of the domain and (b) has a distorted aspect ratio.}
	\label{sf:imbib_set}
\end{figure}

\begin{figure}[ptb]
	\centering
	\begin{subfigure}[t]{\textwidth}
		\centering
		\includegraphics[width=0.8\textwidth]{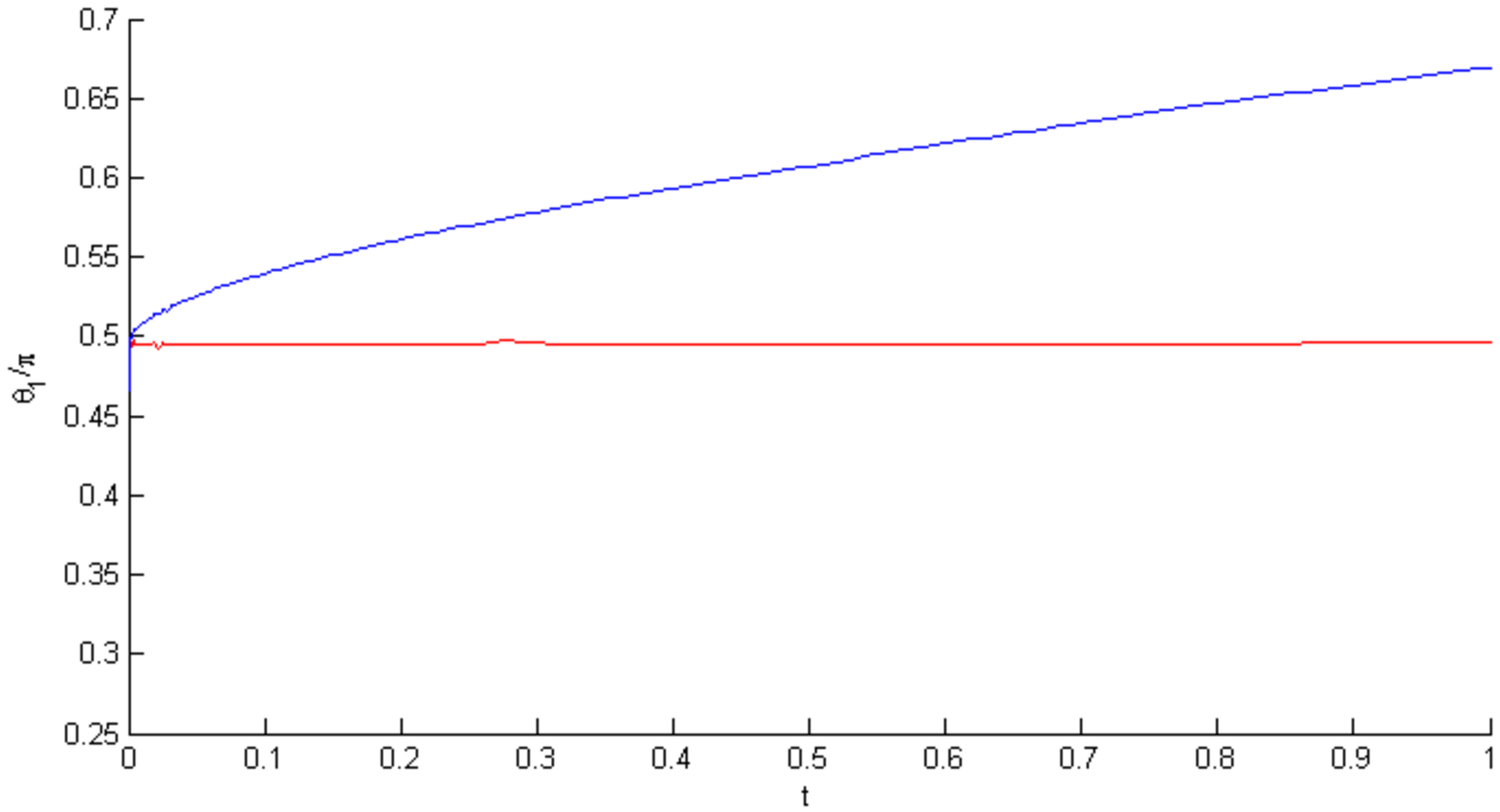}
		\caption{Comparison of the contact angle variation}
		\label{f:front_imbib_ang}
	\end{subfigure}
	\\
	\begin{subfigure}[t]{\textwidth}
		\centering
		\includegraphics[width=0.8\textwidth]{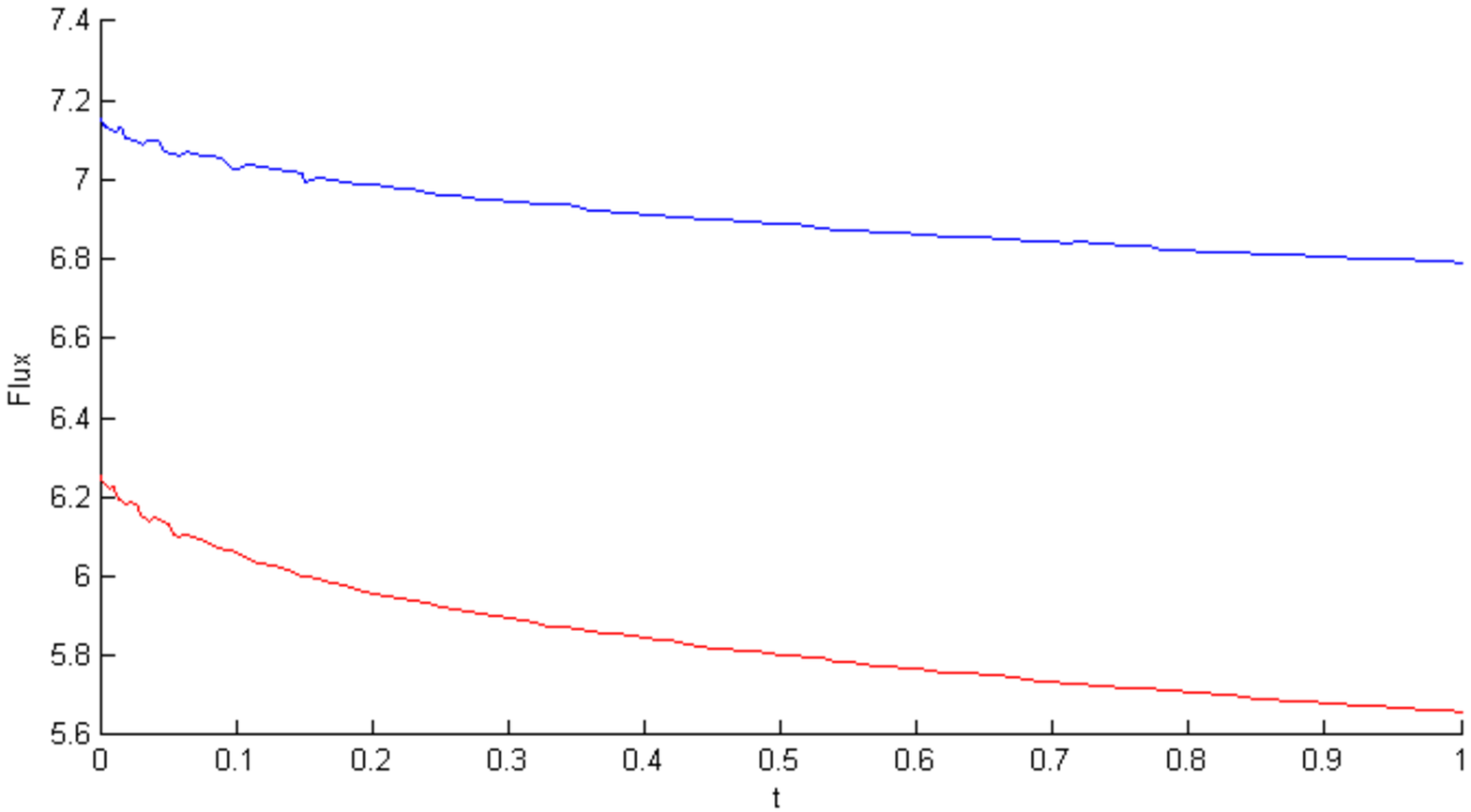}
		\caption{Comparison of the volume flux variation}
		\label{f:front_imbib_flux}
	\end{subfigure}
	\\
	\begin{subfigure}[t]{0.8\textwidth}
		\centering
		\includegraphics[width=\textwidth]{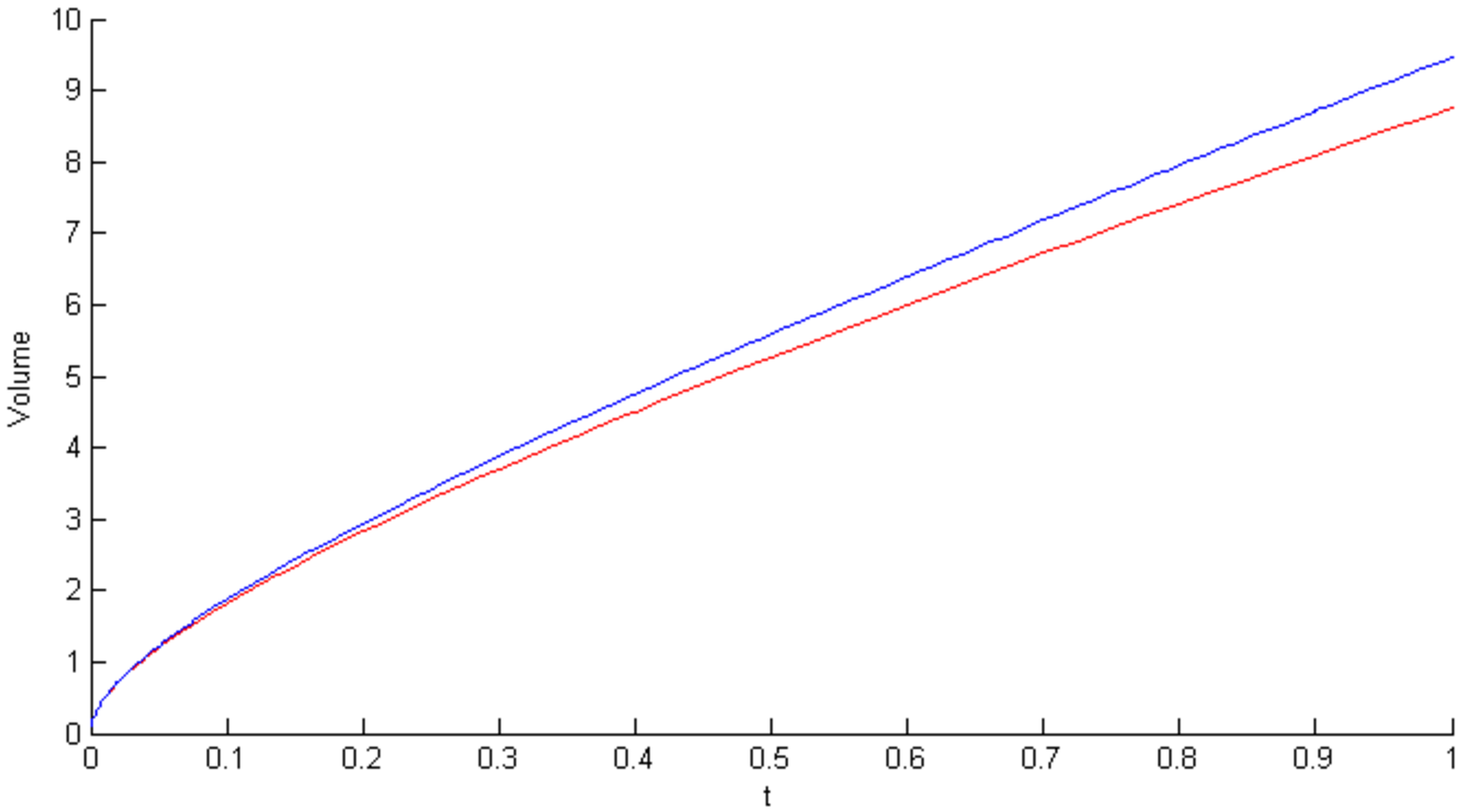}
		\caption{Comparison of the total volume variation}
		\label{f:front_imbib_vol}
	\end{subfigure}
	\caption{Plots of the measured quantities for the initial conditions used in figure \ref{sf:imbib_set}. Red curves are for without gravity and blue are for with gravity ($\gamma=0.2$).}
	\label{sf:imbib_measure}
\end{figure}

In this section we simulate the imbibition from a very small initial wetted region. This is to gain insight into the dynamics that result from imbibing into a porous solid without an initial wetted region, and choose this approach because our numerical scheme cannot solve over a region of zero volume. This is plotted in figure \ref{sf:imbib_set}. Figure \ref{f:front_imbib_0} plots very early times, it is seen that initially the fluid flows mainly in the vertical direction (see $t=10^{-5}$) before advancing in the vertical direction. We propose that this is because the initial condition is not part of the stable manifold in the state space of wetting fronts, and the front is first converging upon it and then propagating along it. Examining the front for times up to as high as $t=2\cdot10^{-2}$ the front has a definite structure, with a flat horizontal profile from the axis of symmetry up to a particular radius, before curving up to meet the surface of the porous substrate approximately at the perpendicular. We assume that this behaviour is exhibited at all times for imbibition into a porous solid without an initial wetted region.

Such behaviour is not what is assumed in \cite{spread_porous}, where lubrication theory is used to examine the imbibition of a thin liquid drop. They assume that, because the drop and wetted region are thin that the radial derivative of pressure, and thus the radial velocity, is small. This is trivially not the case. At early times $\Gamma_1$ and $\Gamma_3$ will be approximately the same length, thus the pressure will change by the same amount over a similar distance and the radial and axial velocities are seen to be comparable.

At later times, $t>10^{-1}$, the wetting front evolves into an arc comparable to those seen in figures \ref{f:front_acute_comp} and \ref{f:front_right_comp}. We therefore propose that the front evolves as seen in figure \ref{f:front_long} for later times (of course we must ignore the plot of the initial condition from figure \ref{f:front_long}).

Figure \ref{sf:imbib_measure} contains plot of the measured quantities. It should be noted that the heuristic used to find volume flux and contact angle are sub-optimal, which is why there are some jumps in the plots. These are not problems with the numerical solution (at least, not more so than has already been discussed), but rather in extracting information from it. In figure \ref{f:front_imbib_ang} we see the usual behaviour of $\theta_1=\pi/2$ being stable without gravity, and with gravity the contact angle increases up to the stable value plotted in \ref{f:front_long_ang}. The volume flux into the wetted region, plotted in figure \ref{f:front_imbib_flux}, is found to be higher with gravity than without, this is because gravity is aiding the advancement of the wetting front causing the volume of the wetted region to increase faster than it does with pressure gradient alone. As time passes the pressure gradient decreases, because the wetted region is larger, and so the fluid imbibes more slowly. These features are seen again in the plot of the total volume of the wetted region, figure \ref{f:front_imbib_vol}.

\section{Summary and Concluding Remarks}	\label{s:conc}

In this section we will overview the discoveries that we have made. Firstly we will discuss the qualitative results that we have produced. After that we will overview those results which are important to the field of flows in porous media, and may affect future research in this area. Finally further investigation that could be performed into the current formulation will be discussed.

We produce solutions for incompressible Darcy imbibition with a wetting front that has a constant contact angle within the pores, as is formulated in section \ref{s:ProbForm}. In section \ref{s:Numerics} we put forward a numerical scheme that is suitable for solving this formulation, and can easily be modified to solve for non-linear boundary conditions, such as those produced by a dynamic contact angle or the modes proposed by Shikhmurzaev and Sprittles in \cite{wetting_dynamics_shk}. This numerical scheme is used to produce the velocity and pressure distributions across the wetted region that are plotted in section \ref{ss:pv_dist}. These reveal that for small domains gravity has little effect, whilst for large domains the fluid can clearly be seen to fall under its action far from the drawing area. In the region around the contact line CL1 (see figure \ref{f:intro_CA_CL}) the velocities are found to be singular, whilst around CL2 the velocity distribution is highly dependent on the contact angle CA2 along with the strength of the gravitational effect. Asymptotic analysis is performed in section \ref{ss:asymp}, guided and confirmed by the numerical results in section \ref{ss:num_asymp}, that reveal the behaviour local to the contact lines. The analysis local to CL2 was then interpreted in relation to the dynamics of the wetting front in section \ref{ss:asym_phys}, giving the different possible behaviours. It was predicted that, for the case without gravity, the contact angle CA2 would have a constant solution $\pi/2$, and for initial conditions of an angle less than $\pi/2$ the angle would converge on $\pi/2$. Also, for the case with gravity, the contact angle would certainly increase to be larger than $\pi/2$. The predictions from our analysis were confirmed by the numerical simulations in section \ref{ss:simulate}, the contact angle converging to $\pi/2$ without gravity and a larger angle with gravity. Gravity also makes the wetted region move faster downwards, which causes the volume of the wetted region to increase faster, and retards the advancement of the contact line CL2. Finally, we observed that the wetted regions evolution seems to be largely independent of the initial conditions, converging on the same dynamics as time passes. 

In our asymptotic analysis, section \ref{ss:asymp}, we obtained singular velocities. In section \ref{ss:asym_phys} we discuss the physical meaning of this, which we conclude must be that Darcy's equation is invalid in these regions, and an improvement is required. Considering another phenomenon, Darcy's equation is used successfully to model capillary rise in porous columns, as discussed in our introduction. However, if this column was tipped on its side during the imbibition then the equation that describes the process would no longer be Darcy's equation, as shown by our analysis. An improvement is required not only for the phenomenon considered here, but for a wide range of phenomena existing in research, engineering and nature. This improvement should, first and foremost, not ignore inertial effects. It is also possible that long range viscous effects will exist due to the enormous velocity gradients present. In any case, an investigation into producing a valid equation for this phenomena is required to advance the field of fluid flows in porous materials.

With regard to the current formulation, that is believed to be qualitatively correct, it has revealed that the value of the contact angle at CL2, CA2 or $\theta_1$, is convergent on different values depending on the strength of the gravitational effect, specified by the value of $\gamma$. It would be of interest to discover how the limiting value of the contact angle depends on $\gamma$. It would also be of informative to see if the contact line CL2 stops moving when it is far from the axis of symmetry, i.e. if it too converges depending on $\gamma$. In addition, we proposed that there may be a stable manifold in the state space of all possible wetting fronts that is converged onto for all physical initial conditions. All of these properties should be investigated.

\bibliographystyle{abbrv}
\bibliography{Bibliography/Bibliography}

\begin{thebibliography}{10}

\bibitem{review_adler}
P.~M. Adler and H.~Brenner.
\newblock Multiphase flow in porous media.
\newblock {\em Annual review of fluid mechanics}, 1988.

\bibitem{disordered_alava}
M.~Alava, M.~Dube, and M.~Rost.
\newblock Imbibition in disordered media.
\newblock {\em Advances in physics}, 2004.

\bibitem{spread_porous}
N.~Alleborn and H.~Raszillier.
\newblock Spreading and sorption of a droplet on a porous substrate.
\newblock {\em Chemical Engineering Science}, 2004.

\bibitem{about_bj}
J.~Auriault.
\newblock About the {B}eavers and {J}oseph boundary condition.
\newblock {\em Transport in porous media}, 2010.

\bibitem{natural_rock}
G.~I. Barenblatt, V.~M. Entov, and V.~M. Ryzhik.
\newblock {\em Theory of Fluid Flows Through Natural Rocks}.
\newblock Kluwer Academic Publishers, 1990.

\bibitem{bc_beavers}
G.~S. Beavers and D.~D. Joseph.
\newblock Boundary conditions at a naturally permeable wall.
\newblock {\em Journal of fluid mechanics}, 1967.

\bibitem{droplet_experiment_chandra}
S.~Chandra and C.~T. Avedisian.
\newblock Observations of droplet impingement on a ceramic porous surface.
\newblock {\em International Journal of Heat and Mass Transfer}, 1992.

\bibitem{interface_pinning}
T.~Delker, D.~B. Pengra, and P.~zen Wong.
\newblock Interface pinning and the dynamics of capillary rise in porous media.
\newblock {\em Physical Review Letters}, 1996.

\bibitem{penetration_denesuk}
M.~Denesuk, G.~L. Smith, B.~J.~J. Zelinski, N.~J. Kreidl, and D.~R. Uhlmann.
\newblock Capillary penetration of liquid droplets into porous materials.
\newblock {\em Journal of colloid and interface science}, 1993.

\bibitem{dynamics_denesuk}
M.~Denesuk, B.~J.~J. Zelinski, N.~J. Kreidl, and D.~R. Uhlmann.
\newblock Dynamics of incomplete wetting on porous materials.
\newblock {\em Journal of colloid and interface science}, 1994.

\bibitem{wetting_quartz}
D.~Diggins, L.~G.~J. Fokkink, and J.~Ralston.
\newblock The wetting of angular quartz particles: Capillary pressure and
  contact angles.
\newblock {\em Colloids and Surfaces,}, 1990.

\bibitem{wetting_fukai}
J.~Fukai, Y.~Shiiba, T.~Yamamoto, O.~Miyatake, D.~Poulikakos, C.~M. Megaridis,
  and Z.~Zhao.
\newblock Wetting effects on the spreading of a liquid droplet colliding with a
  flat surface: Experiment and modeling.
\newblock {\em Physics of Fluids}, 1995.

\bibitem{bc_haber}
S.~Haber and R.~Mauri.
\newblock Boundary conditions for darcy's flow through porous media.
\newblock {\em International journal of multiphase flow}, 1983.

\bibitem{drop_penetration}
K.~P. Hapgood, J.~D. Litster, S.~R. Biggs, and T.~Howes.
\newblock Drop penetration into porous powder beds.
\newblock {\em Journal of colloid and interface science}, 2002.

\bibitem{hilfer_theory}
R.~Hilfer.
\newblock Macroscopic capillarity without a constitutive capillary pressure
  function.
\newblock {\em Physica A}, 2006.

\bibitem{hilfer_numerics}
R.~Hilfer.
\newblock Percolation as a basic concept for macroscopic capillarity.
\newblock {\em Springer}, 2009.

\bibitem{spread_holman}
R.~K. Holman, M.~J. Cima, S.~A. Uhland, and E.~Sachs.
\newblock Spreading and infiltration of inkjet-printed polymer solution
  droplets on a porous substrate.
\newblock {\em Journal of colloid and interface science}, 2002.

\bibitem{darcy_intro}
M.~{King Hubbert}.
\newblock Darcy's law and the field equations of the flow of underground
  fluids.
\newblock {\em Transactions of the american institute of mining and
  metallurgical engineers}, 1956.

\bibitem{spread_sessile_numerics}
B.~Markicevic, T.~G. D’Onofrio, and H.~K. Navaz.
\newblock On spread extent of sessile droplet into porous medium: Numerical
  solution and comparisons with experiments.
\newblock {\em Physics of Fluids}, 2010.

\bibitem{cap_rise_martic}
G.~Martic, J.~D. Coninck, and T.~D. Blake.
\newblock Influence of the dynamic contact angle on the characterization of
  porous media.
\newblock {\em Journal of colloid and interface science}, 2003.

\bibitem{stable_darcy}
A.~Masud and T.~J.~R. Hughes.
\newblock A stabilized mixed finite element method for darcy flow.
\newblock {\em Computer methods in applied mechanics and engineering}, 2002.

\bibitem{popovich_rise_spread}
L.~L. Popovich, D.~L. Feke, and I.~Manas-Zloczower.
\newblock Influence of physical and interfacial characteristics on the wetting
  and spreading of fluids on powders.
\newblock {\em Powder Technology}, 1999.

\bibitem{fluid_poz}
C.~Pozrikidis.
\newblock {\em Fluid dynamics: theory, computation, and numerical simulation}.
\newblock Springer, second edition, 2009.

\bibitem{cap_rise_quere}
D.~Qu\'{e}r\'{e}.
\newblock Inertial capillarity.
\newblock {\em Europhysics Letters}, 1997.

\bibitem{intro_fem}
J.~N. Reddy.
\newblock {\em An introduction to the finite element method}.
\newblock Mc Graw Hill Education, third edition, 2005.

\bibitem{droplet_reis}
N.~C. Reis, R.~F. Griffiths, and J.~M. Santos.
\newblock Numerical simulation of the impact of liquid droplets on porous
  surfaces.
\newblock {\em Journal of Computational Physics}, 2004.

\bibitem{droplet_reis_2}
N.~C. Reis, R.~F. Griffiths, and J.~M. Santos.
\newblock Parametric study of liquid droplets impinging on porous surfaces.
\newblock {\em Applied mathematical modelling}, 2008.

\bibitem{roberts_FEM}
J.~E. Roberts and J.~M. Thomas.
\newblock Mixed and hybrid methods.
\newblock {\em Handbook of Numerical Analysis 2, Finite Element Methods - part
  1}, 1991.

\bibitem{shk_capillary_flows}
Y.~D. Shikhmurzaev.
\newblock {\em Capillary flows with forming interfaces}.
\newblock Chapman \& Hall/CRC, 2008.

\bibitem{cap_rise_shk}
Y.~D. Shikhmurzaev and J.~E. Sprittles.
\newblock Anomalous dynamics of capillary rise in porous media.
\newblock {\em Physical Review E}, 2012.

\bibitem{wetting_dynamics_shk}
Y.~D. Shikhmurzaev and J.~E. Sprittles.
\newblock Wetting front dynamics in an isotropic porous medium.
\newblock {\em Journal of fluid mechanics}, 2012.

\bibitem{dynam_angle_shk}
Y.~D. Shikhmurzaev and J.~E. Sprittles.
\newblock Dynamic contact angle of a liquid spreading on an unsaturated
  wettable porous substrate.
\newblock {\em Journal of fluid mechanics}, 2013.

\bibitem{extended_washburn}
K.~S. Sorbie, Y.~Z. Wu, and S.~R. McDougall.
\newblock The extended washburn equation and its application to the oil/water
  pore doublet problem.
\newblock {\em Journal of colloid and interface science}, 1995.

\bibitem{fem_framework_shk}
J.~E. Sprittles and Y.~D. Shikhmurzaev.
\newblock Finite element framework for describing dynamic wetting phenomena.
\newblock {\em International journal for numerical methods in fluids}, 2012.

\bibitem{washburn_applicability}
J.~Szekely, A.~W. Neumann, and Y.~K. Chuang.
\newblock The rate of capillary penetration and the applicability of the
  washburn equation.
\newblock {\em Journal of colloid and interface science}, 1970.

\bibitem{washburn_original}
E.~W. Washburn.
\newblock The dynamics of capillary flow.
\newblock {\em Physical Review}, 1921.

\bibitem{darcy_whitaker}
S.~Whitaker.
\newblock A theoretical derication of darcy's law.
\newblock {\em Transport in porous media}, 1986.

\bibitem{forcheimer_derivation}
S.~Whitaker.
\newblock The forchheimer equation: A theoreticaldevelopment.
\newblock {\em Transport in Porous Media}, 1996.

\bibitem{cap_rise_powders}
L.~R. White.
\newblock Capillary rise in powders.
\newblock {\em Journal of colloid and interface science}, 1982.

\end{thebibliography}

\end{document}